\documentclass[useAMS,usenatbib]{mn2e}

\usepackage{amsmath}
\usepackage{amssymb}
\usepackage[dvipsnames]{xcolor}
\usepackage{graphicx}
\usepackage{enumitem}
\usepackage{graphics}
\usepackage{hyperref}
\hypersetup{
     colorlinks   = true,
     citecolor    = blue
}

\def\apj{{ ApJ}}
\def\apjl{{ApJL}}

\def\apss{{ Ap\&SS}}
\def\aap{{ A\&A}}

\def\mnras{{ MNRAS}}
\def\araa{{ ARA\&A}}

\def\nat {{ Nature}}
\def\pasj{{ PASJ}}
\def\pasa{{ PASA}}
\def\ssr{{ Space Sci. Rev.}}

\def\physrep{{ Physics Reports}}

\def\prd{{ Phys. Rev. D}}

\def\physscr{{Physica Scripta}}

\def\mr{\mathrm}

\def\mc{\mathcal}
\def\d{\mathrm{d}}
\def\para{\parallel}
\def\b{\boldsymbol}
\def\t{\widetilde}
\def\hx{\hat{\b{x}}}

\def\hz{\hat{\b{z}}}

\def\me{m_{\rm e}}
\def\eps{\epsilon}
\def\gp{\gamma_{\rm p}}
\def\bp{\beta_{\rm p}}
\def\gb{\gamma_{\rm b}}
\def\bb{\beta_{\rm b}}
\def\gbar{\bar{\gamma}}
\def\tp{\theta'}

\def\op{\omega_{\rm p}'}
\def\oB{\omega_{\rm B}'}
\def\ops{\omega_{\rm p,s}}

\def\mue{\mu_{\rm e}}
\def\bphi{\beta_{\phi}}
\def\gphi{\gamma_{\phi}}

\makeatletter
\newcommand{\lambdabar}{{\mathchoice
  {\smash@bar\textfont\displaystyle{0.25}{1.2}\lambda}
  {\smash@bar\textfont\textstyle{0.25}{1.2}\lambda}
  {\smash@bar\scriptfont\scriptstyle{0.25}{1.2}\lambda}
  {\smash@bar\scriptscriptfont\scriptscriptstyle{0.25}{1.2}\lambda}
}}
\newcommand{\smash@bar}[4]{%
  \smash{\rlap{\raisebox{-#3\fontdimen5#10}{$\m@th#2\mkern#4mu\mathchar'26$}}}%
}
\makeatother

\newcommand{\myemail}{wenbinlu@astro.as.utexas.edu}

\title{On the radiation mechanism of repeating fast radio bursts}
\author[Lu \& Kumar]
  {Wenbin Lu$^1$\thanks{\myemail},
  Pawan Kumar$^1$\thanks{pk@astro.as.utexas.edu}\\
  $^1$Department of Astronomy, University of Texas at Austin, Austin,
TX 78712, USA}
\date{\today}

\pagerange{\pageref{firstpage}--\pageref{lastpage}} \pubyear{2018}
\def\LaTeX{L\kern-.36em\raise.3ex\hbox{a}\kern-.15em
    T\kern-.1667em\lower.7ex\hbox{E}\kern-.125emX}

\begin{document}
\label{firstpage}
\maketitle

\begin{abstract}
Recent observations show that fast radio bursts (FRBs) are energetic
but probably non-catastrophic events occurring at cosmological
distances. The properties of their progenitors are largely unknown in
spite of many attempts to determine them using the event rate,
duration and energetics. Understanding the radiation mechanism for
FRBs should provide the 
missing insights regarding their progenitors, which is investigated 
in this paper. The high brightness temperatures ($\gtrsim$$10^{35}\,$K)
of FRBs mean that the 
emission process must be coherent. Two general classes of coherent
radiation mechanisms are considered --- maser and the antenna
mechanism. We use the observed properties of the repeater FRB 121102
to constrain the plasma conditions needed for these 
two mechanisms. We have looked into a wide variety of maser mechanisms
operating in either vacuum or plasma and find that none of them can
explain the high luminosity of FRBs without invoking unrealistic or
fine-tuned plasma conditions. The most favorable mechanism
is antenna curvature emission by coherent charge bunches where the
burst is powered by magnetic reconnection near the surface of a
magnetar ($B\gtrsim10^{14}\,$G). We show that the plasma in the twisted 
magnetosphere of a magnetar may be clumpy due to two-stream instability. 
When magnetic reconnection occurs, the pre-existing density clumps may
provide charge bunches for the antenna mechanism to operate. This
model should be applicable to all FRBs that have multiple outbursts
like FRB 121102.
\end{abstract}


\section{Introduction}\label{sec:intro}
The first fast radio burst (FRB) was reported about a decade ago from
analyzing the archival data of the Parkes
radio telescope \citep{2007Sci...318..777L}. This so-called ``Lorimer
burst'' (FRB 010724) had a peak flux density of $>$$30\,$Jy at $1.4\,$GHz and
duration of $\sim$$5\,$ms. The dispersion measure 
$DM=375\mr{\,pc\,cm^{-3}}$, i.e. the column density of free electrons
integrated along the line of sight, exceeds the contribution from 
the interstellar medium of the Milky 
Way by a factor of $\sim$$10$. Thus, it was inferred that FRBs are from
cosmological distances $\sim$Gpc with DM dominated by the extremely
dilute intergalactic medium (IGM). Indeed, no H$\alpha$ filaments or H II regions
that could explain the large DM were found in archival images
\citep{2007Sci...318..777L, 2014ApJ...797...70K}. Follow-up
observations for $\sim$$100\,$hours did not find any more bursts at this
location, which implies that it may be a catastrophic event such as
coalescence of relativistic objects. However, this single event offered
limited clue for understanding its nature. Later on, four more FRBs
with similar properties as the ``Lorimer burst'' were discovered by the
High Time Resolution Universe survey designed to detect such short
timescale radio transients \citep{2013Sci...341...53T}. Thus, FRBs are
established as a new type of astrophysical phenomenon. Since then, more FRBs
have been discovered and their all-sky rate is estimated to be
$\sim10^3$ to $10^4\,\d^{-1}$ at $\sim1\,$GHz above
$\sim1\mr{\,Jy\,ms}$ \citep{2013Sci...341...53T, 2015MNRAS.447..246P,
  2016MNRAS.455.2207R, 2016MNRAS.460L..30C}.

The breakthrough came
when one burst originally discovered by the Arecibo telescope, FRB 121102,
was found to repeat 
\citep{2014ApJ...790..101S, 2016Natur.531..202S}. It not only showed
that as least this FRB is not a catastrophic event but also
allowed interferometric follow-up observations to
determined the precise location to an accuracy of $\sim3\,$mas
\citep{2017Natur.541...58C, 2017ApJ...834L...8M}. The location of this
FRB is found to be 
in the star-forming region of a low-metallicity 
dwarf galaxy at redshift\footnote{Throughout this paper, we assume
  {\it Planck} best-fit cosmology \citep{2016A&A...594A..13P} and this
  redshift corresponds to luminosity distance $0.97\,$Gpc and
  angular-diameter distance $0.68\,$Gpc.} $z = 0.193$
\citep{2017ApJ...834L...7T, 2017ApJ...843L...8B}, similar to the
environment of hydrogen-poor superluminous supernovae and long
gamma-ray bursts \citep{2017ApJ...841...14M}.

Confirmation of cosmological origin means that the bursts from FRB
121102 are quite energetic. If the FRB sources are isotropic (the
effect of anisotropy will be included later on), then the luminosity is
\begin{equation}
  \label{eq:100}
  L_{\rm iso} = 4\pi D_{\rm L}^2 \mc{S}_{\nu}\Delta \nu \simeq
  (1.2\times10^{42}\mr{\,erg\,s^{-1}}){\mc{S}_{\nu}\over\mr{Jy}}
  \left( {D_{\rm L}\over\mr{Gpc}} \right)^2\Delta \nu_9,
\end{equation}
where $\mc{S}$ is the flux density, $D_{\rm L}$ is the luminosity
distance and $\Delta\nu = \Delta\nu_9\,$GHz is the width of the FRB
spectrum. We define the \textit{apparent} brightness temperature by using the 
maximum transverse area of the emitting region for a
non-relativistic source $\pi (c t_{\rm FRB})^2$,
\begin{equation}
  \label{eq:63}
  \begin{split}
      T_{\rm B} = {\mc{S}_{\nu} D_{\rm A}^2\over 2\pi 
    t_{\rm FRB}^2\nu^2k_{\rm B}} \simeq (1.1\times10^{35}\mr{\,K})\
  {\mc{S}_{\nu}\over\mr{Jy}} \left( {D_{\rm A}\over\mr{Gpc}} \right)^2 t_{\rm
    FRB,-3}^{-2}\nu_9^{-2},
  \end{split}
\end{equation}
where $t_{\rm FRB} = t_{\rm FRB,-3}\,$ms is the burst duration,
$\nu=\nu_9\,$GHz is the observational frequency, $D_{\rm A}$ is the
angular-diameter 
distance, $c$ is the speed of light in vacuum and $k_{\rm B}$ is the
Boltzmann constant. Note that eq. (\ref{eq:63}) is only a lower
limit on the true brightness temperature, which is unknown.
If the source is moving toward the Earth at Lorentz factor $\Gamma$,
then the transverse area of the emitting region may be $\pi (\Gamma
ct_{\rm FRB})^2$, and in this case the true brightness temperature in
the lab frame is smaller than that in eq. (\ref{eq:63}) by a factor
of $\Gamma^2$. From Lorentz transformation, the true brightness
temperature in the source's comoving frame is even smaller by another
factor of $\Gamma$. Still, for any reasonable Lorentz factor, a
coherent radiation mechanism is required \citep{2014PhRvD..89j3009K}.

Although the DMs of the repeating bursts from FRB 121102 stay constant
with time (within the measurement error $\sim5\mr{\,pc\,cm^{-3}}$),
their fluences/fluxes vary by a factor of $\sim10^3$ and the durations vary
from $\lesssim1\,$ms to $\sim10\,$ms \citep{
2016Natur.531..202S, 2016ApJ...833..177S, 2017ApJ...850...76L}. 
The isotropic equivalent energy distribution function is a
single power-law with $\d N/\d E_{\rm iso}\propto E_{\rm iso}^{-1.7}$
spanning $E_{\rm iso}\in$$(10^{37.3}, 
10^{40})\,$erg with no evidence of a cut-off at either the low- or
high-energy end \citep{2017ApJ...850...76L}. None of these events show
evidence of frequency-dependent asymmetric pulse broadening as
observed in Galactic pulsars \citep{2016Natur.531..202S}. Thus, their
durations are likely 
intrinsic, if the cosmological time dilation is neglected. We note that
about half of the other (so-far) non-repeating FRBs show pulse
broadening with width $W\propto \nu^{\sim-4}$, which is consistent with
scattering by inhomogeneities of the circumstellar/interstellar medium
in the host galaxy along the line of sight \citep{2014ApJ...785L..26L,
  2015Natur.528..523M, 2016arXiv160505890C, 2016ApJ...832..199X}. The
other half do not show any evidence of scattering broadening and their durations
(from $\ll$$1\,$ms to $\sim$$10\,$ms) are consistent with being intrinsic
\citep{2017arXiv171008026R}.

The distances of most FRBs and hence their
luminosities are unknown. Their DMs are too large to fit in the
empirical scaling laws between DM and scattering broadening for
Galactic pulsars. This suggests that a large fraction of the DMs may be
due to the IGM whose contribution to scattering broadening is
negligible at low redshifts \citep{2013ApJ...776..125M,
  2016arXiv160505890C}. If the IGM contributes a large portion of the
DMs, one can estimate the luminosity distances of known FRBs to be $D_{\rm
  L}\in$(1, 10) Gpc, 
minimum\footnote{Note that, except for FRB 121102 whose 
precise location is known, the reported peak fluxes from other
FRBs are only lower limits inferred based on the assumption that they
occurred at the nearest beam center, see \citet{2017arXiv171008026R}.}
peak isotropic luminosities $L_{\rm 
iso,min}\in$$(10^{42.5}, 10^{44})\mr{\,erg\,s^{-1}}$, and minimum
isotropic energies $E_{\rm iso,min}\in$$(10^{39.5},
10^{42})\mr{\,erg}$ \citep[see the FRB catalog by][]{2016PASA...33...45P}.
It has been shown that the energy 
distribution function of FRB 121102 is so far consistent with being
representative of all FRBs \citep{2016MNRAS.461L.122L}. One can also see
that those (so-far) non-repeating FRBs have much larger
luminosities/energies than the events from the repeater. This is most
likely a selection effect because FRB 121102 has better localization and can be
observed by more sensitive telescopes. Very recently, a very bright burst
from FRB 121102 was discovered by the Apertif Radio Transient System
\citep{2014htu..conf...79V} with peak flux density $\mc{S}_{\nu}\sim
24\,$Jy and duration $\Delta t_{\rm FRB}\sim 1.3\,$ms
\citep{2017ATel10693....1O}, corresponding 
to isotropic luminosity $L_{\rm iso}\sim
2.6\times10^{43}\mr{\,erg\,s^{-1}}$, isotropic energy $E_{\rm iso}\sim 
3\times10^{40}\,$erg and apparent brightness temperature $T_{\rm
  B}\sim 3\times10^{35}\,$K. These 
energetics are comparable to the (so-far) 
non-repeaters, further suggesting that the repeater may not be a
special member of the FRB family. 

In this paper, we mainly focus on the radiation mechanism
of the repeater FRB 121102, although our analysis should be applicable to
other repeating FRBs as well. Our general
guide line is that, any FRB 
model must explain not only the {\it typical} isotropic luminosity
$L_{\rm iso}\sim 10^{43}\mr{\,erg\,s^{-1}}$, 
energy $E_{\rm iso}\sim 10^{40}\,$erg, apparent brightness temperature
$T_{\rm B}\sim 10^{35}\,K$ and duration $t_{\rm
  FRB}\sim1\,$ms but also the large {\it variations} of these
quantities at a given frequency ($\sim$GHz). 

This paper is organized as follows. In \S 2, we discuss the
constraints on the nature of FRB progenitors from the event rate,
duration and energy budget. In \S 3, we describe two general
classes of coherent emission mechanisms --- maser and the antenna
mechanism. The goal of this paper is to test each of these coherent
emission mechanisms and see whether they are consistent
with the basic properties of FRBs. In \S 4, we discuss various maser
mechanisms operating inside the corotating magnetosphere of a neutron star.
In \S 5, we discuss the possibility of maser emission
powered by the dissipation of a relativistic outflow at large distances
from the central object. Then in \S 6, we discuss the antenna mechanism. 
Conclusions are drawn at the end of each section. In \S 7, we discuss
the differences between the mechanisms of FRBs and pulsar radio
emission. A summary of the paper is provided in \S 8. Throughout the
paper, the convention $Q = 10^nQ_n$ and CGS units are used.

\section{General considerations of FRB progenitors}

We first summarize the general constraints on FRB progenitor
models from the event rate, duration and energy budget. Then we
review various FRB progenitor models proposed in the 
literature. We show that these lowest-order constraints (from the
event rate, duration and energy) are not sufficient to prove or falsify many 
of these models. Thus, one is forced to take one step further and
consider the radiation mechanisms and the required plasma
conditions, which will be described in later sections.

\subsection{Event rate, duration and energy budget}

There may be two classes of FRBs: repeating and non-repeating. In the
following, we provide order-of-magnitude estimates of the birth rate
of FRB progenitors $R_0$ (in unit $\rm Gpc^{-3}\,yr^{-1}$) based on
the repeating or non-repeating hypothesis. The
all-sky detection rate above $\sim1\rm\,Jy\,ms$ at $\sim1.4\,$GHz is
denoted as $R_{\rm det}= 10^{3.5} R_{\rm det,3.5}\rm\,d^{-1}$. A typical FRB 
with isotropic energy $E_{\rm iso} \simeq 10^{40}A^{-1}\,$erg within redshift
$z\simeq 0.5$ (corresponding to an IGM DM of $\simeq
400\rm\,pc\,cm^{-3}$) will have fluence $\gtrsim A^{-1}\rm\,Jy\,ms$,
where $A < 1$ is the attenuation factor due to the off-center position
of the FRB in the beam of detection. The
comoving volume out to this redshift is $V\simeq 30 
\rm\,Gpc^{3}$. We define a beaming factor $f_{\rm
  b}$ for each burst as the solid angle of the radiation beaming cone
divided by $4\pi$.

{\it If} the majority of FRBs are non-repeating, then the birth rate
of FRB progenitors $R_0$, averaged within $z\simeq 0.5$, can be
estimated by the projected all-sky detection rate
\begin{equation}
  \label{eq:101}
  R_{0} \simeq {(1+z) R_{\rm det} \over f_{\rm b}V}
  \simeq 6\times 10^4 {R_{\rm det, 3.5}\over f_{\rm b}} \rm\,Gpc^{-3}\,yr^{-1}.
\end{equation}
Note that we have assumed that all FRBs have similar isotropic energy
of $\sim10^{40}A^{-1}\rm\,erg$. There are potentially more undetected FRBs
with much smaller energies $E_{\rm iso}\ll 10^{40}A^{-1}\rm\,erg$ (as seen from
the repeater FRB 121102), so eq. (\ref{eq:101}) should be considered as a
stringent lower limit. The volumetric rate of 
core-collapse supernovae (CCSNe) at $z\simeq 0.5$ is $R_{\rm CCSN}\simeq
3\times10^{5}\rm\,Gpc^{-3}\,yr^{-1}$ \citep{2014ARA&A..52..415M}. The beaming
factors of FRBs are unknown, but all known examples of coherent radio
emission (e.g. pulsar radio emission) show strong beaming
\citep[see][for a review]{2017RvMPP...1....5M}.
% This is essentially
% because photons with phase coherence must travel nearly parallel to
% each other and their separation in the longitudinal direction must be
% smaller than $\lambda/2\pi$ ($\lambda$ being the
% wavelength).
Therefore, a small beaming factor $f_{\rm b}\ll 1$ posts
severe challenge to non-repeating FRB models that are based on black
holes or neutron stars \citep[e.g.][]{2013PASJ...65L..12T, 2013ApJ...776L..39K,
  2014A&A...562A.137F, 2014ApJ...780L..21Z, 2015MNRAS.450L..71F},
because the progenitors' birth rate in this case is greater or at least
comparable to the rate of CCSNe. This has been pointed out by
\citet{2016MNRAS.457..232C}.

% Taking an average IGM contribution of $DM_{\rm IGM}\sim 300\rm\ pc\
% cm^{-3}$, we obtain an average redshift of $\bar{z}\sim 0.4$, and the comoving
% volume within this redshift is $V\sim 7\rm\ Gpc^3$. The all-sky FRB
% detection rate above $\mathcal{F}_0\sim 2\rm\ Jy.ms$ is $\dot{N}_{\rm det}\simeq
% 3\times10^3 \rm\ d^{-1} \simeq 10^6\rm\ yr^{-1}$. As an
% order-of-magnitude estimate, let us assume that all
% FRBs are from roughly the same redshift $z \sim 0.3$. 
% If most FRBs are non-repeating, then the source birth rate per
% unit volume $\dot{N}$ equals to the FRB detection rate divided by the
% total volume $V\sim 7\rm\ Gpc^3$, i.e. $\dot{N} \sim
% 10^5\rm\ Gpc^{-3}\ yr^{-1}$, which is comparable to the core-collapse
% supernova (CCSN) rate in the local 
% universe.

A more natural way of explaining the high observed FRB rate is that
they are from repeating sources. First, it is a fine-tuned coincidence
that the first (and only) FRB found by Arecibo is from a new, repeating class,
while all the other $\sim$$30$ FRBs belong to a different non-repeating
class. Second, the hypothesis that all FRBs are repeating with a
similar energy distribution function as FRB 121102 is so far
consistent with all observations \citep{2016MNRAS.461L.122L}. In fact,
if FRB 121102 had a location error similar to those found by the
Parkes telescope, the true location may fall into the low-sensitivity
gaps between beams during follow-up observations and
perhaps none of the subsequent bursts could have been
detected. Based on the assumption of a universal energy distribution
function (i.e. the same repetition rate at any given isotropic energy),
\citet{2016MNRAS.461L.122L} derived that the ratio between 
the birth rate of FRB progenitors and CCSN rate is in the range $(10^{-5},
10^{-3})(\tau_{\rm active}/30\rm\, yr)^{-1}A^{0.7}f_{\rm b,tot}^{-1}$ with 3$\sigma$
confidence, where $\tau_{\rm active}$ is the duration of the bursting
activity per progenitor, $A<1$ is the typical off-beam-center
attenuation factor for Parkes FRBs, and $f_{\rm b,tot}$ is the total
beaming factor (the combined solid angle of all bursts from the same
progenitor divided by $4\pi$). We note that this 
rate ratio is $\sim2\times10^{-4}$ for hydrogen-poor superluminous supernovae
\citep{2013MNRAS.431..912Q} and $\sim10^{-3}$ for
gamma-ray bursts \citep[GRBs,][here we have assumed a typical GRB
beaming factor of $10^{-2}$]{2010MNRAS.406.1944W}. Thus, FRB
progenitors may be rare objects in the Universe. We note that a similar
conclusion was drawn by \citet{2017ApJ...843...84N} and \citet{
2017ApJ...850...76L}.

Although the existence of a (small) population of non-repeating FRBs
cannot be ruled out, they are not the main focus of this
paper. Instead, we restrict ourselves solely to the repeater FRB 121102,
which has accumulated a large amount of data from extensive
observations. Hereafter, unless specially noted with
``non-repeating'', an ``FRB'' means one of the bursts from the 
repeating source FRB 121102, and ``the (FRB)
progenitor'' means the central object responsible for the many bursts
from FRB 121102.

In the following, we consider general constraints on the
progenitor from FRB durations and the total energy reservoir. And
then, for the neutron-star (NS) progenitor model, we derive
constraints on the basic properties of the NS such as surface B-field 
strength and rotation period.

Durations of FRBs are likely controlled by the
dynamical time of the system. For instance, the free-fall time near
the surface of a star of radius $R$ and mass $M$ is
$t_{\rm ff} \sim (R^3/GM)^{1/2}$, which is $\sim 0.1\,$ms for a NS or
stellar-mass black hole (BH). On the other hand, a white dwarf has
free-fall time $t_{\rm ff}\sim 10\,$s, which is much longer than FRB
durations. Even the light-crossing time for a white dwarf $R/c\sim
30(R/10^9\mr{\,cm})\,$ms is too long to be consistent with FRB
durations. The typical timescale for a 
sudden accretion of a block of gas is either given by the dynamical
time at the mass feeding end or the viscous time of the accretion
disk. For a binary system where an object is accreting mass from the
compact/non-compact companion \citep[e.g. as in the model
of][]{2016ApJ...823L..28G}, the dynamical timescale of the system 
is much longer than FRB durations.

There is also the possibility that the emitting plasma is moving
towards the observer at Lorentz factor $\Gamma\gg 1$. Such a
relativistic plasma can only be launched from relativistic
compact object (a NS or BH). In this case, 
the plasma can dissipate its free energy via internal dissipations
(e.g. magnetic reconnection or internal shocks) or external shocks
(when the plasma interacts with the surrounding medium).
The distance between the dissipation location and the center of the
progenitor star can be much larger than $ct_{\rm FRB} =
3\times10^7t_{\rm FRB,-3}\,$cm by a factor of $\sim 
2\Gamma^2$. We conclude that the constraint from FRB durations
leave NSs or BHs\footnote{If one only considers FRB durations,
  intermediate-mass or supermassive 
BHs are viable progenitors because the outflow Lorentz factor $\Gamma$ may
be large and the dissipation 
region could be much smaller than the size of the causally-connected
region $\sim R/\Gamma^2$ \citep[such as in the models
of][]{2016PhRvD..93b3001R, 2017MNRAS.471L..92K}. Our discussion is
applicable to these 
high-mass progenitors as well.
% Other exotic/unknown sources proposed in
% the literature, e.g. cosmic string and Thompson's EM explosion, 
% do not have a sound theoretical description of the plasma conditions
% in the radiating region. We will not discuss these models, although the
% general arguments for coherent radiation mechanisms proposed in the
% current paper are also applicable to them.
} as the most possible progenitors.

The repetition pattern of FRB 121102 is sporadic and non-Poissonic
\citep{2017arXiv170504881O}. Adding up the 
isotropic equivalent energy of all the bursts detected by the
Arecibo campaign \citep[$4.3\times10^{39}\rm\,erg$ up to
Feb. 2016,][]{2016ApJ...833..177S} and
then dividing it by the total on-source time 15.8 hr, we obtain a
long-time averaged luminosity
\begin{equation}
  \label{eq:60}
  \langle L_{\rm frb}\rangle_{\rm Arecibo}\simeq 8\times10^{34}
  (f_{\rm b,tot}/f_{\rm r})\rm\,erg\,s^{-1},
\end{equation}
where $f_{\rm r}$ is the radio emission efficiency and $f_{\rm b,tot}$
is the total beaming factor (the combined solid angle of all bursts,
including those beamed away from the Earth, divided by $4\pi$). Note
that this is only a lower limit 
because bursts with much higher fluences than the observed ones
require a long 
monitoring time and bursts with much lower fluences are not
observable. The energy distribution function $\d N/\d E_{\rm
  iso}\propto E_{\rm iso}^{-1.7}$ implies that most of the energy is
near the high-energy end $E_{\rm iso,max}$, which is currently unknown.
For Very Large Array (VLA) observations at 3 GHz 
by \citet{2017ApJ...850...76L}, the total 
burst energy is $1.9\times10^{40}\rm\,erg$ and the total on-source
time is $\sim 60\,$hr, so the time-averaged luminosity is $\langle
L_{\rm frb}\rangle_{\rm VLA}\simeq 9\times10^{34}f_{\rm
 r}^{-1}f_{\rm b,tot}\rm\,erg\,s^{-1}$. The same analysis with the
Green Bank Telescope observations by \citet{2016ApJ...833..177S}
gives a similar result.
% This also agrees with the (cumulative) energy
% distribution function $\dot{N}(>E_{\rm iso}) \simeq 3\,E_{\rm
%   iso,39}^{-0.7} \mr{\,d^{-1}}$ estimated by
% \citet{2017ApJ...850...76L}, although 
% most of the energy is at the high-energy end $E_{\rm iso,max}$, which
% is currently unknown.
In the following, we take $\langle L_{\rm
  frb}\rangle_{\rm 
 Arecibo}$ as a lower limit and obtain the energy reservoir required
to supply the bursting activity for a duration $\tau_{\rm active}$
\begin{equation}
  \label{eq:61}
  E_{\rm tot} \gtrsim (7.5\times10^{43}\mr{\,erg})\ (f_{\rm
    b,tot}/f_{\rm r}) (\tau_{\rm active}/30\rm\,yr).
\end{equation}

If FRBs are powered by accretion onto BHs, the minimum accretion rate
is $\dot{M}_{\rm min}\gtrsim \langle L_{\rm frb}\rangle_{\rm Arecibo}/c^2 \sim
10^{-13}\, (f_{\rm
    b,tot}/f_{\rm r})\mr{\, M_{\rm \sun}\, yr^{-1}}$, which can 
be satisfied by many known accreting systems. If FRBs are produced by
magnetic dissipation in the magnetosphere of a 
NS and the B-field energy is not replenished by differential rotation
on timescales $\ll \tau_{\rm active}$, the minimum surface B-field
strength is given by $B_*^2R_*^3/6 
\gtrsim E_{\rm tot}$ ($R_*\approx10\,$km being the NS radius), i.e.
\begin{equation}
  \label{eq:58}
  B_*\gtrsim (2.1\times10^{13}\mr{\,G})\ (f_{\rm b,tot}/f_{\rm r})^{1/2}
  (\tau_{\rm active}/30\rm\,yr)^{1/2}.
\end{equation}
FRB 121102 has been repeating since discovery in 2011
\citep{2014ApJ...790..101S}. To avoid the chance of coincidence, the
true active duration $\tau_{\rm active}\gg 6\,$yr. The radiation
efficiency $f_{\rm r}$ and (total) beaming factor $f_{\rm b,tot}$ are
poorly constrained, but these two factors tend to cancel each other in
a square root term in eq. (\ref{eq:58}), so we obtain a rough estimate of
the surface B-field strength $B_*\gtrsim$a few$\times10^{13}\,$G in
the NS progenitor scenario.

If the NS has radius $R_* \approx 10\,$km, surface dipole B-field near
the polar cap $B_* = 10^{14}B_{*,14}\,$G, and spin period $P =
0.1P_{-1}\,$s, the spin-down 
luminosity is \citep{2006ApJ...648L..51S}
\begin{equation}
  \label{eq:53}
  L_{\rm sd}= {6\pi^4 B_*^2 R_*^6 \over P^4c^3}\simeq
  (2.2\times10^{39}\mr{\,erg\,s^{-1}})\ B_{*,14}^2P_{-1}^{-4},
\end{equation}
where we have assumed a magnetic inclination angle of $
45^{\rm o}$ (not sensitive).
% Multiwavelength upper limits above
% constrains\footnote{If the system is optically thick to X-rays
% due to bound-free absorption \citep[as mentioned
% by][]{2017ApJ...841...14M}, the absorbed energy will be reprocessed
% into UV-optical emission. It is also unrealistic to have the PWN
% emission peaking above 10 keV. Thus, we consider
% $10^{41}\rm\,erg\,s^{-1}$ as the upper limit of the bolometric
% luminosity of the PWN surrounding the NS.} $L_{\rm sd}\lesssim
% 10^{41}\rm\,erg\,s^{-1}$, i.e. 
% \begin{equation}
%   \label{eq:56}
% P\gtrsim 38B_{*,14}^{1/2}\rm\ ms.
% \end{equation}
The spin-down timescale is given by the total rotational energy
divided by the spin-down luminosity
\begin{equation}
  \label{eq:54}
  t_{\rm sd} = {2\pi^2 I_*/P^2 \over L_{\rm sd}}\simeq (29\mr{\,yr})\
  B_{*,14}^{-2} P_{-1}^2,
\end{equation}
where $I_*\approx 10^{45}\mr{\,g\,cm^2}$ is the moment of inertia of
the NS. The NS was born with a supernova remnant\footnote{If the NS is
  born in a ``dark'' stellar collapse without a supernova ejecta
  \citep[e.g.][]{2017MNRAS.469L.104K}, then the constraint on the age
  is weaker $t_{\rm age}\gtrsim 6\,$yr and hence the lower limit on
  the spin period in eq. (\ref{eq:59}) will be weaker by a factor of
  $\sim$$2$.}, which is currently expanding. Non-detection of a time 
derivative of DM from the 
repeater over the past $\sim 6$ yrs means that the age
of the system $t_{\rm age}\gtrsim 30\,$yr
\citep{2016ApJ...824L..32P, 2016MNRAS.461.1498M}. The absence of
free-free absorption of GHz waves also gives a similar constraint on
the age. The fact that the 
spin-down time must be longer than the age means $t_{\rm sd}\gtrsim
30\rm\,yr$, which constrains the current rotation
period\footnote{If the radiation has beaming angle $\Delta \theta$ and
  the beaming cone corotates with the NS, then the cone sweeps 
across the observer's line of sight in a time $\simeq 16P_{-1}(\Delta
\theta/1\mr{\,rad})\rm\ ms$. We take the conservative limit
$\Delta\theta<1\mr{\,rad}$ and then the longest burst from FRB 121102
($\Delta t_{\rm FRB}\sim8\,$ms, assuming intrinsic) gives $P\gtrsim
50\,$ms.}
\begin{equation}
  \label{eq:59}
  P\gtrsim 0.1 B_{*,14}\rm\,s.
\end{equation}
% If the NS's initial rotation period was much shorter than the current
% value, then the current spin-down time roughly equals to the system's
% age $t_{\rm sd}\simeq t_{\rm age}$, which means
% \begin{equation}
%   \label{eq:55}
%   P \simeq (0.1\mr{\,sec})\ B_{*,14}(t_{\rm age}/30\rm\,yr)^{1/2}.
% \end{equation}
% We insert eq. (\ref{eq:55}) into eq. (\ref{eq:56}) and obtain a
% constraint on the surface dipole B-field strength at the poles
% \begin{equation}
%   \label{eq:57}
%   B_*\gtrsim (3.8 \times10^{13}\mr{\,G})\ (t_{\rm age}/30\rm\,yr)^{-1},
% \end{equation}
% which is consistent with the energetic consideration
% (eq. \ref{eq:58}).
We conclude from eqs. (\ref{eq:58}) and (\ref{eq:59}) that
observations are consistent with a slowly rotating high-B NS as the
FRB progenitor. On the other hand, a BH with a small accretion rate
$\gtrsim 10^{-13}\rm\,M_{\odot}\, yr^{-1}$ is also possible.

\subsection{No easy answer to the progenitor puzzle}

From \S 2.1, we see that NS or BH progenitors can comfortably meet the
requirements of short durations and relatively small
energy budget. The low progenitor birth rate simply requires a special
subgroup of NSs or BHs. Many models that satisfy these
constraints have been proposed in
the literature \citep[see a review
by][]{2016MPLA...3130013K}. They fall into two general categories in
terms of emission locations: (1) emission
within the magnetosphere of NSs such as hyper-energetic
giant pulses \citep{2015ApJ...807..179P, 2016MNRAS.457..232C, 
  2016MNRAS.462..941L}, emission accompanying magnetar 
flares \citep{2010vaoa.conf..129P, 2014ApJ...797...70K,
  2016ApJ...826..226K}, emission due to B-field disturbance by
infalling gas/bodies \citep{2015ApJ...809...24G, 2016ApJ...829...27D,
  2017ApJ...836L..32Z}, or small-scale magnetic reconnection events
near the NS surface \citep{2017MNRAS.468.2726K}; (2) emission from a
relativistic outflow 
which is undergoing internal dissipation or interacting with the
surrounding medium at large distances from the central object
\citep{2014MNRAS.442L...9L, 2016PhRvD..93b3001R, 2016MNRAS.461.1498M,
  2017ApJ...842...34W, 2017ApJ...843L..26B, 2017MNRAS.471L..92K}.

We see that there is no easy answer to the progenitor puzzle, if one
only considers the lowest-order constraints from FRB event rate, durations and
energetics.

In this paper, we take one step further and explore all possible
coherent radiation mechanisms operating in the radio wavelengths. We
carefully study the plasma 
conditions needed to reproduce the basic observational properties of
FRBs. These aspects have not been considered in most of the previous
works \citep[except][]{2017MNRAS.468.2726K}, mainly because no
consensus has been reached over the (coherent) pulsar radio emission
mechanism, despite decades of 
hard work and debates \citep[see the review by][]{2016JPlPh..82c6302E,
  2017RvMPP...1....5M}.
However, FRBs are drastically different from normal pulsar radio
emission in that they are much brighter (by a factor of
$\sim10^{10}$) and only last for a brief period of time. As we show
later in \S 4, \S 5 and \S 6, much more stringent constraints can be put on the
source plasma of FRBs, and therefore most of the
viable options for pulsar radio emission can actually be ruled out.

\section{possible radiation mechanisms}\label{sec:rad_mech}
There are generally two classes of coherent emission mechanisms: maser
and the antenna mechanism \citep{1969Ap&SS...4..464G}. The first one
requires special particle distribution function (population inversion)
so that incoming radiation from certain direction has negative
absorption coefficient; an example is vacuum synchrotron
maser which occurs in the direction near the edge of the $\gamma^{-1}$
beaming cone around the momentum vector of the emitting particle
\citep{1991MNRAS.252..313G}. The antenna mechanism involves 
phase-coordinated time-dependent  
currents; a widely discussed case under the conditions of the
NS magnetosphere is curvature emission by charge bunches of size
$\lesssim$ the wavelength of emission
\citep{1975ApJ...196...51R}. 

A special case of maser mechanism is collective plasma emission
(hereafter plasma maser). In this process, plasma waves are excited
and exponentially amplified by certain plasma instabilities and then the energy in 
plasma waves is transformed into escaping modes of electromagnetic (EM)
radiation. For example, the most widely discussed pulsar radio emission
mechanism describes that a fast beam of particles runs into a
slowly-moving plasma, and the free energy associated with the
relative motion is transferred to plasma waves due to an
instability, and then these waves are converted to escaping radio
modes \citep{2017RvMPP...1....5M}. One important 
point, which will be useful later, is that in any plasma maser
mechanism, the energy of the observed EM radiation comes from
particles' kinetic energy.

If the emission region is at radius $r$ from the center of the
progenitor, the strength of the electromagnetic fields associated with FRB
radiation in the source region is
\begin{equation}
  \label{eq:5}
  E_{\rm \perp EM} \simeq B_{\rm \perp EM}\geq \sqrt{L_{\rm iso}\over
    r^2 c}\simeq (1.8\times 10^{10}\ \mr{G})L_{\rm
      iso,43}^{1/2} r_6^{-1}, 
\end{equation}
where ``$\perp$'' means the fields are perpendicular to the line of
sight and we have ``$>$'' when the local curvature of the emitting surface
is smaller than $r$. This EM wave is very intense in that the
dimensionless non-linearity parameter
\begin{equation}
  \label{eq:6}
 a_0={eE_{\rm \perp EM} \over m_{\rm e}\omega c}\geq 5.0\times 10^7 L_{\rm
      iso,43}^{1/2} r_6^{-1} \nu_9^{-1},
\end{equation}
where $\omega = 2\pi \nu$ is the angular frequency. This means that
free electrons along the line of sight at distance $r\lesssim 
5\times 10^{13}\rm\ cm$ from the progenitor will be accelerated
to relativistic speeds. In the regime of non-linear optics $a_0\gg 1$,
particles' effective mass and hence the effective plasma frequency
depend on the wave amplitude \citep{2006RvMP...78..309M}.

On the other hand, in the case where the EM waves are generated in
(and/or propagate through) a medium with strong B-field perpendicular
to the wave electric field, the acceleration due to 
$E_{\rm \perp EM}$ only lasts for a gyration time
$\omega_{\rm B}^{-1}$ [$\omega_{\rm B} = eB/(m_{\rm e}c)$], and hence
the non-linearity parameter becomes $a_0=E_{\rm \perp
  EM}/B$. As long as $E_{\rm \perp
  EM}/B\ll 1$, the propagation characteristics
of the wave can be treated by linear approximation. This will be the
base of the discussion on plasma dispersion relation in \S 4.2.

To make our discussion as general as possible, in the following
sections, we consider NSs or 
BHs as viable FRB progenitors and the emitting plasma could
either be within the corotating magnetosphere of a NS or a relativistic
outflow launched from a NS or BH. We discuss possible maser
mechanisms in \S 4 and \S 5, and then the antenna mechanism is
discussed in \S 6.

% All possible (maser and
% antenna) mechanisms are discussed for both locations.

\section{Masers inside the neutron star magnetosphere} 
Based on the assumption that the source plasma is confined by the
B-field of a NS, we first constrain the source location and the
B-field strength in the source region from energetic 
requirements. And then in \S 4.1, we show that the traditional
magnetosphere of a rotating NS or magnetar cannot provide enough
particle kinetic energy to power FRBs. Then, under the assumption that
some explosive process injects a large amount of particle kinetic
energy in the magnetosphere, we derive the basic requirements on
particles' distribution function for various maser mechanisms,
including plasma maser (\S 4.2) and masers in vacuum (\S
4.3). Conclusions are drawn at the end of each subsection.

We assume that the B-field configuration at radius $r\gg R_*$ is dipolar
\begin{equation}
  \label{eq:16}
  B(r) = B_*(r/R_*)^{-3} \simeq (10^{11}\ \mathrm{G}) B_{*,14} r_7^{-3}.
\end{equation}
The energy density of the FRB EM waves at a distance
$r$ from the center of the star is given by 
\begin{equation}
  \label{eq:121}
  U_{\rm EM} = {L_{\rm iso}\over 4\pi r^2c} \simeq (2.6\times10^{17}
\mr{\,erg\,cm^{-3}}) \, L_{\rm iso,43} r_7^{-2}.
\end{equation}
We require that the total energy density of the source plasma to be a factor of
$f_{\rm r}^{-1}$ ($f_{\rm r}$ being the radiation efficiency in the
radio band) higher than $U_{\rm EM}$. Moreover, for the external
B-field to confine the plasma, the magnetic energy density
$B^2/8\pi$ must exceed that of the plasma by at least a factor of
$\zeta\gg 1$,
\begin{equation}
  \label{eq:122}
    {B(r)^2/8\pi} \gtrsim \zeta {U_{\rm EM}/f_{\rm r}}.
\end{equation}
Combining eqs. (\ref{eq:16}) and (\ref{eq:122}), we obtain
\begin{equation}
  \label{eq:129}
    r \lesssim (3.5\times10^{7}\mr{\,cm})\, B_{*,14}^{1/2} \zeta_{1}^{-1/4}
  \left({L_{\rm iso,43}\over f_{\rm r}}\right)^{-1/4},
\end{equation}
and 
\begin{equation}
  \label{eq:131}
  B \gtrsim (2.4\times10^9\mr{\,G})\, B_{*,14}^{-1/2} \zeta_{1}^{3/4}
  \left({L_{\rm iso,43}\over f_{\rm r}}\right)^{3/4}, 
\end{equation}
where we have used $\zeta_1= \zeta/10$.
Therefore, the radiation process occurs much below the light cylinder
$R_{\rm LC} = Pc/2\pi\simeq 4.8\times10^8P_{-1}\,$cm and the emission
region has a strong B-field $B\gtrsim 10^9\,$G. The transverse
momentum of an electron or positron is lost in a short time
$\sim 10^{-9}\gamma (B/10^9\rm\,G)^{-2}\,s$ due
to cyclotron/synchrotron cooling ($\gamma$ being the Lorentz factor),
so particles are forced to stay 
at the lowest Landau level and only move along the B-field lines
(unless there is a mechanism that keeps exciting them to higher
Landau levels). For later usage, we also note that the ratio between
cyclotron frequency $\omega_{\rm B} \equiv eB/\me c$ and the frequency
of the observed radio waves $\omega \equiv 2\pi \nu$ has a lower limit
\begin{equation}
  \label{eq:151}
  \omega_{\rm B}/\omega \gtrsim 2.2\times10^6\,\nu_9^{-1}B_{*,14}^{-1/2}
  \zeta_{1}^{3/4}\left({L_{\rm iso,43}\over f_{\rm r}}\right)^{3/4}.
\end{equation}

\subsection{Explosive particle injection needed for masers}
In this subsection, we show that, in the conventional picture,
particles above the polar cap of a rotating NS \citep{1969ApJ...157..869G,
  1971ApJ...164..529S, 1975ApJ...196...51R, 1979ApJ...231..854A,
  1982RvMP...54....1M} or in the twisted magnetosphere of a magnetar
\citep{2002ApJ...574..332T} do not have enough kinetic energy
to power FRBs. Some explosive particle injection process
is needed for any maser mechanism powered by particles' kinetic
energy. 

Throughout this paper, particle number density is often expressed
as a multiplication factor $\mc{M}$ times the
Goldreich-Julian (G-J) density \citep{1969ApJ...157..869G} 
\begin{equation}
  \label{eq:27}
  n_{\rm GJ}\simeq {B\over ecP} \simeq (6.9\times10^{13}\,
  \mr{cm^{-3}})\, B_{14} P_{-1}^{-1}.
\end{equation}
When the density of the pair plasma falls below either the G-J density or
the minimum density required to support the induced current $J =
(c/4\pi)|\nabla\times \b{B}|$, the region becomes charge starved and
the E-field parallel to the B-field cannot be screened. Thus,
particles are accelerated by the unscreened E-field to high Lorentz
factors and then produce 
$\gamma$-rays which may turn into pairs via B-assisted photon decay
$\gamma+\mr{B}\rightarrow \mr{e}^{+} + \mr{e}^{-}$ 
\citep{2006RPPh...69.2631H}. When the B-field strength is
super-critical $B\gtrsim B_{\rm QED} = 4.4\times10^{13}\,$G, some $\gamma$
photons with polarization perpendicular to the osculating plane of the
B-field may split into two photons (with polarization parallel to the
osculating plane) before the B-assisted photon decay. The energies of
the two daughter photons are only slightly smaller (by a factor of
$\sim$2), so they may still turn into pairs. Two photon annihilation
process $\gamma + \gamma \rightarrow e^{+} + e^{-}$ may also occur
but is subdominant \citep{2001ApJ...554..624H}. The secondary
$\mr{e}^{\pm}$ pairs will produce more $\gamma$-rays and then more
pairs. Such pair cascade proceeds until the number density of pairs is
high enough to screen the parallel E-field. 

Above the polar cap of a rotating NS, the initial
$\gamma$-rays are produced by curvature radiation
\citep{1975ApJ...196...51R}. The maximum
kinetic energy density of the plasma (before or after pair
cascade) in the open field line region is given by $\gamma_{\rm th}n_{\rm
  GJ}\me c^2$, where $\gamma_{\rm 
  th}$ is the threshold Lorentz factor for the initiation of pair
cascade. This is because primary particles accelerated by the parallel
E-field to $\gamma\gtrsim \gamma_{\rm th}$ will produce copious pairs,
and then the parallel E-field is quickly shielded when the number
density reaches $n\gtrsim n_{\rm GJ}$. Below, we provide a rough
estimate of the threshold Lorentz factor $\gamma_{\rm th}$ above the
polar cap, following \citet{2010MNRAS.406.1379M}.

Curvature radiation (CR) has characteristic energy
\begin{equation}
  \label{eq:97}
  {\eps_{\rm CR}\over \me c^2} \simeq {\gamma^3 hc\over 2\pi \rho \me
    c^2} = 3.9\times10^3\, \gamma_7^3 \rho_7^{-1},
\end{equation}
where we have normalized the energy by the electron rest mass
energy, $\gamma$ is the Lorentz factor of the electron (or positron),
$h$ is the Planck constant, and $\rho$ is the local curvature radius.
The electron (or positron) is accelerated by the parallel
E-field within the charge-deficit ($n\ll n_{\rm GJ}$) gap above the
polar cap of the NS
\begin{equation}
  \label{eq:197}
  \gamma \me c^2 = e\Phi_{\rm gap} \simeq 2\pi n_{\rm GJ}e^2 H^2,
\end{equation}
where $\Phi_{\rm gap}$ is the voltage drop across the gap and $H$ is
the gap height. Note that, in eq. (\ref{eq:197}), we have assumed $H$
to be much smaller than the size of the polar cap $\sim
R_*^{3/2}R_{\rm LC}^{-1/2}$, this is because, as we will show later,
$\Phi_{\rm gap}$ is much smaller than the total
voltage drop across the polar cap \citep{1975ApJ...196...51R}
\begin{equation}
  \label{eq:104}
  \Phi_{\rm pc} \simeq 2\pi^2 R_*^3 B_*/(P^2 c^2) \simeq
  (6.6\times10^{16}\,\mr{V})\, B_{*,14} P_{-1}^{-2}.
\end{equation}
Then, $\gamma$-ray photons with energy $\eps_{\rm CR}\gg \me c^2$ will undergo
B-assisted decay into pairs within a propagation length  $\rho \me
c^2/\eps_{\rm CR}$ (when the angle between the photon's momentum vector and the
B-field becomes $\sim\me c^2/\eps_{\rm CR}$). Thus, the height of the
charge-deficit gap is given by
\begin{equation}
  \label{eq:198}
  H\simeq \rho\me c^2/\eps_{\rm CR}\simeq
  (2.6\times10^{3}\mr{\,cm})\, \rho_7^2 
  \gamma_7^{-3}.
\end{equation}
We combine eqs. (\ref{eq:197}) and (\ref{eq:198}) and obtain the
threshold Lorentz factor
\begin{equation}
  \label{eq:108}
  \gamma_{\rm th}\simeq 1.9\times10^7\,\rho_7^{4/7}
  (B_{*,14}/P_{\rm -1})^{1/7}. 
\end{equation}
The voltage drop across the gap is $\Phi_{\rm gap}=\gamma_{\rm th}\me
c^2/e\simeq  (1\times 10^{13}\mr{\,eV})\, \rho_7^{4/7}\,
  (B_{*,14}/P_{\rm -1})^{1/7}\ll \Phi_{\rm pc}$.
Therefore, the maximum (instantaneous) isotropic luminosity powered by
particles accelerated above the polar cap can be
estimated

% \footnote{A pile-up of plasma kinetic energy is possible when
% the group velocity of the EM waves is smaller than the bulk velocity
% of the plasma. Let's think about this point. However, when the
% plasma's kinetic energy density is smaller than that of the EM waves,
% the EM waves stop growing. Another way of effectively piling up plasma
% kinetic energy is that the plasma has a very large density
% contrast, i.e. the peak density is much higher than the average density.}
\begin{equation}
  \label{eq:90}
  \begin{split}
      L_{\rm iso,max}^{\rm (pc)} &\simeq 4\pi R_*^2 \gamma_{\rm th}n_{\rm
    GJ}\me c^3 \\
  &\simeq (4.1\times10^{38}\mr{\,erg\,s^{-1}})\, \rho_7^{4/7}
  (B_{*,14}/P_{-1})^{8/7}.
  \end{split}
\end{equation}
We note that the giant pulses from e.g. the Crab pulsar
can reach an instantaneous isotropic luminosity of $\sim
10^{37}\mr{\,erg\, s^{-1}}$ \citep{2007ApJ...670..693H}, which is
consistent with the constraints 
from eq. (\ref{eq:90}), considering that the B-field
curvature radius above the polar cap may be $\rho\gtrsim 10^7\,$cm for a
dipolar geometry. However, we see that traditional pair creation
processes above the polar cap of a rotating NS cannot produce FRB
isotropic luminosities $\sim10^{43}\mr{\,erg\, s^{-1}}$.

On the other hand, the magnetosphere of a magnetar is believed to be
different from normal NSs in that the evolution of the
ultrastrong ($\gtrsim 10^{14}\,$G) B-field anchored on the active
stellar crust leads to a twisted external magnetosphere with strong
persistent currents \citep{2002ApJ...574..332T}. For a large twist angle
$\sim1$ radian, the current flowing along B-field lines at radius $r$
can be estimated
\begin{equation}
  \label{eq:77}
  J = {c\over 4\pi}|\nabla\times\b{B}| \sim {cB\over 4\pi r},
\end{equation}
which corresponds to a {\it minimum} plasma density
\begin{equation}
  \label{eq:1}
  n_{\rm min} = J/(ec)\sim (1.7\times10^{16}\mr{\,cm^{-3}})
  B_{14} r_6^{-1}.
\end{equation}
Pair cascade is initiated by $\gamma$-rays produced by
resonant inverse-Compton scattering of electrons (or positrons) off 
ambient X-ray photons \citep{2007ApJ...657..967B}. In this case, the
threshold Lorentz factor is given by the resonance condition 
$\gamma_{\rm th}\simeq \gamma_{\rm res}$, when the energy of X-ray
photons in the electron's comoving 
frame equals to the Landau energy increment
\begin{equation}
  \label{eq:102}
  \gamma_{\rm res}\eps_{\rm x} \simeq (B/B_{\rm QED})\me c^2,
\end{equation}
where $\eps_{\rm x}$ is the X-ray photon's energy in
the NS frame and $B_{\rm QED}=4.4\times10^{13}\,$G is the critical
B-field strength. The X-ray spectra
of Galactic magnetars usually show a power-law component extending
beyond 10$\,$keV, which may be due to atmospheric heating or
Comptonization of low-energy thermal X-rays from the surface
\citep{2017ARA&A..55..261K}. The mean energy of the scattered photons is
\begin{equation}
  \label{eq:103}
  \begin{split}
      {\eps_{\rm IC}\over \me c^2} &\simeq \gamma_{\rm res}\, \mr{min}\left(1,
  {B\over B_{\rm QED}} \right)\\
&\simeq 1.2\times10^2\, \left({\eps_{\rm x}\over 10\mr{\,keV}}
  \right)^{-1} B_{\rm 14}\, \mr{min}\left(1, 
  {B\over B_{\rm QED}} \right).
  \end{split}
\end{equation}
If $B\gtrsim B_{\rm QED}$, the scattered photons, initially at an
angle $\lesssim \gamma_{\rm res}^{-1}$ wrt the B-field, undergo B-assisted
decay into pairs within a propagation length\footnote{If $B\gg B_{\rm QED}$,
  a fraction of the photons scattered at an angle $\gg
  \gamma_{\rm res}^{-1}$ wrt the B-field may turn into pairs right away
  \citep{2007ApJ...657..967B}.} $\rho \me c^2/\eps_{\rm IC}\ll
\rho$. In this way, the voltage drop in the corona loops 
above a magnetar is maintained near the $e^{\pm}$ production
threshold. The maximum (instantaneous) isotropic luminosity powered by 
particles in these current-threading corona loops near the surface of a
magnetar is 
\begin{equation}
  \label{eq:112}
  \begin{split}
  L_{\rm iso,max}^{\rm (magnetar)} &\sim 4\pi R_*^2 \gamma_{\rm res}n_{\rm
    min}\me c^3 \\
  &\sim (6.1\times10^{35}\mr{\,erg\,s^{-1}})\, \left({\eps_{\rm
        x}\over 10\mr{\,keV}} \right)^{-1} B_{*,14}^2.
  \end{split}
\end{equation}
We see that FRBs with $L_{\rm iso}\sim
10^{43}\rm \,erg\,^{-1}$ cannot be produced by particles' kinetic
energy stored in the magnetosphere of magnetars, unless the B-field
strength is unrealistically high $B_*\gtrsim 4\times 10^{17} (\eps_{\rm
  x}/10\mr{\,keV})^{1/2}\,$G (which means the magnetic energy
is comparable to the gravitational binding energy of the NS).

We conclude that some explosive particle injection process
is needed for any maser mechanism powered by particles' kinetic
energy (but not for antenna mechanism powered by E-fields, as shown
later in \S 6). We note
that pairs may be injected by magnetic reconnection processes
near the surface of magnetars. A catastrophic example of explosive
pair production is magnetar giant flares \citep{2006MNRAS.367.1594L}, although
here the magnetic fields undergo large-scale reconfiguration and the
resultant plasma is highly optically thick to photons from radio up to
$\gamma$-ray wavelengths.

In the following two subsections, we discuss plasma maser (\S 4.2) and then
masers in vacuum (\S 4.3), under the assumption that a large number
of pairs are suddenly injected by some other process. General
constraints on the particle distribution function are derived in each case.

\subsection{Plasma maser (collective plasma
  emission)}\label{sec:plasma_maser} 
We consider the situation where a beam of particles runs into a target
plasma and subluminal waves are excited and then amplified due to
a certain beam instability. Two instabilities are considered:
cyclotron-Cherenkov (or anomalous Doppler) instability and Cherenkov
instability.

In \S 4.1, we have shown that some explosive particle injection
process is needed for any maser mechanism powered by particles'
kinetic energy. Thus, we assume both the beam and target
plasma to be made of electrons 
and positrons with number densities $n_{\rm b} = n_{\rm b-}+ n_{\rm
  b+}\gg n_{\rm GJ}$ and $n = n_{-}+n_{+}\gg n_{\rm GJ}$,
respectively. To maintain the charge balance in the magnetosphere, we
require $|n_{\rm b-} - n_{\rm b+}|\sim  n_{\rm GJ}$ and $|n_- -
n_+|\sim n_{\rm GJ}$. 

We take the B-field to be along
the $\hz$ direction ($\b{B} = B\hz$), which points towards the
observer. The (1-dimensional) distribution functions of 
the beam and target plasma in the NS frame are denoted as $f_{\rm
  b}(u_{\rm b})$ and $f(u)$, where $u_{\rm b}$ and $u$ are
4-velocities in the $\hz$ direction and we have normalized $\int
f_{\rm b} \d u_{\rm b} = \int f \d u = 1$. We assume that beam
particles are moving towards the observer at relativistic speeds, thus
$u_{\rm b}\gg 1$. To make our discussion as general as possible, we
allow particles in the target plasma to be moving towards ($u>0$) or
away from ($u<0$) the observer. Particle velocities divided by speed
of light are denoted as $\beta_{\rm b}$ and $\beta$ and their Lorentz
factors are $\gamma_{\rm b}$ and $\gamma$. Our goal is to explore the
general constraints on these two distribution functions in order for
plasma maser mechanism to produce FRBs. 

We introduce another inertial frame (hereafter the {\it plasma frame}) 
which is moving at velocity $\beta_{\rm p}c$ (Lorentz factor
$\gamma_{\rm p}$) wrt the NS frame. Note that both $\beta_{\rm p}>0$
and $\beta_{\rm p}\leq 0$ are possible. All quantities in the plasma
frame are denoted with a prime ($'$) and unprimed quantities are
measured in the NS frame. We also denote the 
mean value of a quantity $K'$ as $\langle 
K'\rangle  \equiv \int K' f' \d u'$. The relative Lorentz factor
$\gamma_{\rm p}$ between the two frames is defined such that the mean
4-velocity is zero $\langle u' \rangle = \langle \gamma'\beta' \rangle
= 0$ in the plasma frame. The mean Lorentz factor in the plasma frame
can be considered as the temperature of the plasma, so we denote
$T'\equiv \langle \gamma' 
\rangle$. In the NS frame, a large variation of particle Lorentz factors 
cannot be avoided during pair creation, so the
plasma is at least mildly hot ($T'\gtrsim 
2$) and most likely extremely hot ($T'\gg 1$). The discussion in this
subsection is applicable for both cases.

The emission is powered by the free energy associated with the {\it
  relative} motion between the two plasmas. The EM waves strongly
interact and exchange energy/momentum with the target plasma during
wave excitation and amplification. In the plasma frame, the incoming
beam is decelerated due to the pressure of the target plasma $\sim$$T'
n'\me c^2$. We define $f_{\rm r}<1$ as the fraction of the momentum loss
from the beam that goes to the momentum flux of the FRB EM waves, so we
obtain $ T' n'\me c^2 f_{\rm r} \simeq U_{\rm EM}'$, where $U_{\rm
  EM}'$ is the energy density of the EM waves of the FRB in the plasma
frame. Another way of understanding is that the inertia of the target plasma
$\sim$$T'n'\me$ needs to be large enough to extract
momentum from the beam at the rate of $U_{\rm EM}'c$. 
When the target plasma is moving towards the observer
($\beta_{\rm p}>0$), we have $U_{\rm EM}' = U_{\rm EM}/\gamma_{\rm
  p}^2$ (where $U_{\rm EM}$ is given by
eq. \ref{eq:121}), which means
\begin{equation}
  \label{eq:18}
  \gp^2T' n' \simeq (3.2\times10^{23}\ \mr{cm}^{-3}) (L_{\rm iso, 43}/f_{\rm
    r})  r_7^{-2}.
\end{equation}
When the target plasma is at rest or moving away
from the observer ($\beta_{\rm   p}\leq 0$), the energy density in the
plasma frame is $U_{\rm EM}' = \gamma_{\rm p}^2U_{\rm EM}$. Since the
following derivations only require $f_{\rm r}<1$, we keep using
eq. (\ref{eq:18}) for simplicity but one should keep 
in mind that $f_{\rm r}$ should be substituted by $f_{\rm
  r}/\gamma_{\rm p}^4$ in the case of counter-streaming beam and
target plasma.

The number density can be expressed 
in unit of the G-J density (we assume a dipolar B-field
$B=B_*(r/R_*)^{-3}$ for $r\gg R_*$)
\begin{equation}
  \label{eq:26}
  \mc{M}={\gp n'\over n_{\rm GJ}} \simeq 4.6\times10^{10} {L_{\rm
      iso, 43}\over \gamma_{\rm p,2} T' f_{\rm r}}B_{*,14}^{-1} P_{-1} r_7.
\end{equation}
We define the non-relativistic plasma
frequency $\op$ and cyclotron frequency $\oB$ (note that $\omega_{\rm
  B}' = \omega_{\rm B}$) as 
\begin{equation}
  \label{eq:123}
  \begin{split}
      \op &= \sqrt{4\pi e^2 n'\over \me} \simeq
  (3.2\times10^{14}\mr{\,s^{-1}})\, {\left(L_{\rm iso,43}/f_{\rm
        r}\right)^{1/2} \over r_7\gamma_{\rm p,2} T'^{1/2}},\\
\oB &= {eB\over \me c} = (1.8\times10^{18}\mr{\,s^{-1}})\,B_{11}.
  \end{split}
\end{equation}
We assume that the dispersion relation of the interesting wave-mode with
frequency $\omega'$ and wave-vector $\b{k}'$ is purely determined by
the target plasma, which is reasonable if $n'\gg
n_{\rm b}'$. Instabilities due to the existence of the beam plasma will
be considered as a perturbation. Without loss of generality, we take
the wave vector to be in the $x'$-$z'$
plane at an angle $\tp$ wrt the $z'$-axis, 
i.e. $\b{k}' = k'(\sin\tp \hx' + \cos\tp\hz')$. The refractive index
is defined as $\t{n}' \equiv k'c/\omega'$ and the
phase velocity along the $\hz'$ direction is $\beta_{\phi}' \equiv
\omega'/(k'c\cos\tp) = (\t{n}'\cos\tp)^{-1}$. Only subluminal waves
($|\bphi'|<1$) can be excited, so we can define a Lorentz factor
corresponding to the phase velocity $\gphi' = (1 - \bphi'^2)^{-1/2}$.
The unprimed version of these symbols have the same
meanings but in the NS frame.

To avoid severe Landau damping, we only consider waves with
$\gphi'\gg T'$ (and hence $\gphi'\gg 1$, $|\bphi'|\approx1$). In the
absence of the beam particles (which will be included later), the
wave-number $k'$ and frequency $\omega'$ are both real positive
numbers. Lorentz transformation of the 4-wavevector $(\omega, \b{k})$
gives
\begin{equation}
  \label{eq:115}
  \omega = \gp\omega' (1 + \bp/\bphi'), k\cos\theta c = \gp\omega'
  (\bp + 1/\bphi').
\end{equation}
% The phase velocity in the lab frame (along the $\hz$ direction) is
% \begin{equation}
%   \label{eq:147}
%   \bphi = {\omega\over k\cos\theta c} = {\bp + \bphi' \over 1 + \bp \bphi'}.
% \end{equation}
We are interested in waves with frequency $\nu
=\omega/(2\pi) = \nu_9\,$GHz in the NS frame. For convenience, we
introduce three more variables 
\begin{equation}
  \label{eq:92}
  \begin{split}
      \Delta_1 &\equiv T'\omega_{\rm p}'^2/\omega_{\rm B}'^2 \lesssim
 {1\over 2\gp^2 \zeta} \ll 1,\\
 \Delta_2 & \equiv {\op^2\over \mc{M}\oB
   \omega'} \simeq  2.0\times10^{-8}\, {\bphi' + \bp\over \bphi' P_{-1}
     \nu_9}\ll 1,\\ 
  \Delta_3&\equiv T'\op^2/\omega'^2 \simeq 2.6\times10^{13}\, (\bphi' +
  \bp)^2 {L_{\rm iso,43}/f_{\rm r}\over r_7^2\nu_9^{2}},
  \end{split}
\end{equation}
where $\zeta\gg 1$ is the minimum ratio between the
energy density of the B-field and that of the target plasma in the NS
frame (according to eqs. \ref{eq:122} and \ref{eq:18}). We also note
that the B-field is invariant ($B = B'$) under 
Lorentz transformation parallel to $\hz$. The fact that $\Delta_1\ll 1$ and
$\Delta_2\ll 1$ greatly simplifies the dispersion relation. In
  order for the wave to be in resonance with the 
beam particles, its phase velocity must be in nearly the same direction ($+z'$
direction) as the beam at a relativistic speed. The phase velocity of
the wave must also be pointing towards the observer ($+z$ direction)
in the NS frame. Thus, we require 
\begin{equation}
  \label{eq:141}
  \bphi'>0,\ \gphi'\gg 1,\ \bphi' + \bp>0.
\end{equation}
An upper limit of the beam (and plasma) Lorentz factor is given by the
requirement that particles do not lose more than half of their
energy due to curvature cooling over a propagation length $\rho$ (curvature 
radius), so we have
\begin{equation}
  \label{eq:125}
  \gamma_{\rm b,max} = \left({3m_{\rm e}c^2\rho\over 4 e^2}
\right)^{1/3}\simeq 3.0\times10^6 \rho_7^{1/3}.
\end{equation}
We also note that two drift velocities associated with the
inhomogeneities of the B-field --- curvature drift and grad-B drift ---
are both extremely small near the NS surface. The curvature drift
velocity in the NS frame is given by
\begin{equation}
  \label{eq:72}
  \begin{split}
      v_{\rm d}/c ={\gb \me \beta_{\rm b}^2c^3\over eB\rho} \simeq
  (1.7\times10^{-9})\, \gamma_{\rm b,6}\rho_7^{-1}B_{11}^{-1}.
  \end{split}
\end{equation}
The grad-B drift is even smaller because it is proportional to
the particle's transverse velocity squared $\beta_{\rm b,\perp}^2$, which is
suppressed by cyclotron/synchrotron cooling. Therefore, the
Cherenkov-drift resonance (at $\omega - \beta_{\rm
  b}kc\cos\theta  - kv_{\rm d}\sin\theta=0$ in the NS frame), whose
growth rate is proportional to the drift velocity
\citep{1999ApJ...512..804L}, can be ignored. Then the 
dispersion relation can be calculated in the uniform B-field
approximation. 

We denote the normalized 1-dimensional distribution
function along the B-field to be $f_s'(u')$
(and $\int f_s'\d u'=1$), where $s=-$ (or $+$) means
electrons (or positrons). Since $|n_-' - n_+'|/n'\simeq \mc{M}^{-1}\ll
1$ (see eq. \ref{eq:26}), the distribution functions of electrons and
positrons are nearly 
identical and are simply denoted as $f'(u')$ without subscript. The
non-relativistic plasma frequencies of these two 
species are nearly the same $\omega_{\rm p,-}'
\approx \omega_{\rm p,+}'$, so we have (following eq. \ref{eq:92})
\begin{equation}
  \label{eq:127}
  \Delta_{1,-} \approx \Delta_{1,+} \approx
T'{\omega_{\rm p,\pm}'^2/\omega_{\rm B,\pm}'^2}\approx \Delta_1/2.
\end{equation}
Since the cyclotron frequency is an odd function of charge
sign, we have
\begin{equation}
  \label{eq:126}
\sum_s {\ops'^2\over\omega_{\rm B,s}' \omega'} \approx {1\over \mc{M}} 
{\omega_{\rm p,\pm}'^2\over\omega_{\rm B}' \omega'} \approx \Delta_{2,-} \approx
\Delta_{2,+}\approx \Delta_2/2.
\end{equation}

Consider a plane-wave ($\propto \mr{e}^{i\b{k}'\cdot
\b{r}'-i\omega't'}$) perturbation with wave-vector $\b{k}' = 
k'(\sin\tp \hx' + \cos\tp \hz')$. We take the Fourier-Laplace
transform of the Maxwell equations
\begin{equation}
  \label{eq:201}
  \begin{split}
      &c\nabla'\times \b{B}' = 4\pi \b{J}' + \partial \b{E}'/\partial t', \b{J}'
  = \overset\leftrightarrow{\sigma} \b{E}',\\
&c\nabla' \times \b{E}' = - \partial \b{B}'/\partial t', 
  \end{split}
\end{equation}
and then obtain
\begin{equation}
  \label{eq:181}
  \b{k}'\times[\b{k}'\times\b{E}'] + \omega'^2 \b{E}'/c^2 + 4\pi i\omega'
\overset\leftrightarrow{\sigma}\cdot \b{E}'/c^2 = 0,
\end{equation}
where $\b{E}'(\omega', \b{k}')$ is the amplitude of the
transformed E-field perturbation and the conductivity tensor
$\overset\leftrightarrow{\sigma}(f', 
\omega',\b{k}')$ depends 
on the distribution function $f'$. Making use of the refractive index
$\t{n}' \equiv \omega'/\b{k}'c$ and dielectric tensor
$\overset\leftrightarrow{\mc{E}} 
(f', \omega', \b{k}') \equiv \overset\leftrightarrow{I} + 4\pi i
\overset\leftrightarrow{\sigma}/\omega'$, we obtain the dispersion 
relation in the plasma frame 
\begin{equation}
  \label{eq:83}
  \mr{det}\left[ \t{n}'^2\left({\b{k}'\b{k}' \over k'^2} -
        \overset\leftrightarrow{I} \right) + \overset\leftrightarrow{\mc{E}}
\right] = 0,
\end{equation}
where $\overset\leftrightarrow{I}$ is a unit tensor and we have
\begin{equation}
  \label{eq:96}
  {\b{k}'\b{k}' \over k'^2} -
        \overset\leftrightarrow{I} = 
  \left[ {\begin{array}{ccc}
   -\cos^2\tp & 0 & \sin\tp\cos\tp \\
   0 & -1 & 0 \\
    \sin\tp\cos\tp & 0 & -\sin^2\tp \\
  \end{array} } \right].
\end{equation}
For wave modes far away from the cyclotron resonance for
{\it target plasma particles} ($\omega_{\rm B}/\gamma'\omega' \gg
|1 - \beta'\t{n}'\cos\tp|$), the dielectric
tensor is \citep[e.g.][]{1998MNRAS.293..447L,
 1999JPlPh..62..233M, 2001MNRAS.325..715G}
\begin{equation}
  \label{eq:84}
  \overset\leftrightarrow{\mc{E}} = 
  \left[ {\begin{array}{ccc}
   1 + \chi_1 & -i\chi_2 & -\chi_4 \\
   i\chi_2 & 1 + \chi_1 & -i\chi_5 \\
    -\chi_4 & i\chi_5 & 1 - \chi_3 \\
  \end{array} } \right],
\end{equation}
where
\begin{equation}
  \label{eq:89}
  \begin{split}
    \chi_1 &= \Delta_1 \langle \gamma' \mu'^2\rangle/T',\
 \chi_2 = -{\Delta_2\over 2} \left\langle {\mu'} \right\rangle,\ \\
\chi_3 &= \Delta_3 \left\langle {1\over
    \gamma'^3 \mu'^2} 
\right\rangle/T' - \Delta_1 \t{n}'^2\sin^2\tp \left\langle {\gamma'
    \beta'^2 \mu'} \right\rangle/T',\\
\chi_4 &= -\Delta_1 \t{n}' \sin\tp \left\langle {\gamma' \beta' \mu'} \right\rangle/T',\ 
\chi_5 = {\Delta_2\over 2} \t{n}' \sin\tp \left\langle {\beta'} \right\rangle,\\
\mu'&\equiv 1 -\beta'\t{n}'\cos\tp,\ \gamma' = \sqrt{u'^2+1}, \\
\beta' &= u'/\sqrt{u'^2 + 1},\ \langle \ldots \rangle \equiv \int (\ldots) f'(u') \d u'.
  \end{split}
\end{equation}
We are interested in waves with $\bphi'\approx 1$
and $\gphi'\gg \langle \gamma'\rangle =T'$ (to avoid Landau damping)
and hence $\mu' = 
1-\beta'/\bphi'\approx 1 - \beta'$. We ignore the terms $\chi_2$ and
$\chi_5$ associated with $\Delta_2\sim 10^{-8}$
(eq. \ref{eq:92}). Making use of $\langle\gamma'\beta'\rangle =
\langle u'\rangle =0$, $\langle 
\gamma'\rangle=T'$, we obtain $  \langle \gamma' \mu'^2\rangle\simeq
2T'$ and $\left\langle {\gamma' \beta' \mu'} \right\rangle\simeq
-T'$, and hence
\begin{equation}
  \label{eq:143}
  \chi_1\simeq 2\Delta_1 = (\gp^2\zeta)^{-1}\ll 1,\ \chi_4\simeq
  \Delta_1\t{n}'\sin\tp \ll 1.
\end{equation}
For waves with phase velocities corresponding to Lorentz factors much
higher than the plasma temperature $\gphi'\gg T'$, we have
$\left\langle {\gamma'^{-3} 
    \mu'^{-2}} \right\rangle\simeq T'$ and hence $\chi_3\simeq \Delta_3$.
Therefore, the dispersion relation (eq. \ref{eq:83}) gives two
branches of solutions for the X-mode ($\b{E}$ perpendicular to the
$\b{k}$-$\b{B}$ plane) and Alfv{\'e}n-mode ($\b{E}$ in the
$\b{k}$-$\b{B}$ plane)
\begin{equation}
  \label{eq:144}
  \t{n}_{\rm X}'^2 = 1 + \chi_1 = 1 + 2\Delta_1,\ \mbox{for X-mode,}
\end{equation}
and 
\begin{equation}
  \label{eq:146}
  \t{n}_{\rm A}'^2\cos^2\tp \approx 1 + 2\Delta_1 + {\sin^2\tp
  \over \Delta_3\cos^2\tp - 1},\ \mbox{for Alv{\'e}n-mode}.
\end{equation}
where higher order terms $\mc{O}(\Delta_1^2)$ have been ignored and we
have made use of $\gphi'\gg 1$ and hence $ \chi_1(\t{n}_{\rm
  A}'\cos\tp-1)\approx \Delta_1/\gphi'^2\ll 1$. 

We are interested in wave growth at either cyclotron-Cherenkov
instability or Cherenkov instability when a beam runs through the
target plasma. There are two possible cases:

\begin{itemize}[leftmargin=0.7cm]
\item [(1)] The target plasma is moving in the
same direction as the beam towards the observer ($\bp>0$). This is
a natural consequence of injection of a high-Lorentz factor beam
(along open or closed B-field lines) which
is capable of initiating pair cascades (as described in \S 4.1).
% In this case, we have $\Delta_3\gg 1$ and hence the dispersion
% relation is simpler.
\item [(2)] The target plasma is moving away from the observer
  ($\bp<0$) in the opposite direction of the beam. This is possible
  if there are two independent particle injection (e.g. magnetic
  reconnection) regions near the NS surface where the feet of the
  closed B-field lines are anchored.
% In this case, the factor $(\bphi' 
% +\bp)^2$ could be extremely small (for instance $\sim\gp^{-4}$) and we
% cannot decide whether $\Delta_3$ is much greater than 1. In principle,
% all waves with $\gphi'\gg T'$ may grow as long as certain resonance
% condition is met. However, 
Since Alfv{\'e}n-mode waves propagate along the
B-field line and hence cannot escape \citep{1986ApJ...302..138B}, we
only consider the excitation and growth of X-mode waves. 
\end{itemize}
% In case (1), we have $\beta_{\phi}'+\beta_{\rm p}\approx 2$ and hence
% (from eq. \ref{eq:92})
% \begin{equation}
%   \label{eq:158}
%   \Delta_3 \simeq 1.0\times10^{14}\, {(L_{\rm iso,43}/f_{\rm r})}
%   r_7^{-2}\nu_9^{-1} \ggg 1.
% \end{equation}
% In case (2), we need to substitute $f_{\rm r}$ by $f_{\rm
%   r}/\gamma_{\rm p}^4$, as required by the minimum inertia of the
% target plasma capable of extracting enough momentum from the beam, and
% hence 
First, we consider the cyclotron-Cherenkov resonance at
\begin{equation}
  \label{eq:148}
    \omega' - \bb' k' \cos\theta' + {\oB\over \gb'} = 0\,
  \Longleftrightarrow\, 1 - \bb'/\bphi' + {\oB\over \gb'\omega'} = 0.
\end{equation}
Such resonance can only occur when $\bb'>\bphi'$. In the limit
$\gb'\gg \gphi'\gg 1$, we obtain
\begin{equation}
  \label{eq:149}
  \omega' = {2\gphi'^2} \oB/\gb'.
\end{equation}
Making use of eq. (\ref{eq:115}) and $\gb' = \gb\gp(1 - \bp)$, we
obtain the beam Lorentz factor in the lab frame
\begin{equation}
  \label{eq:150}
  \begin{split}
    \gb &= {2(\bphi' + \bp) \over 1 - \bp}\gphi'^2{\omega_{\rm B} \over
    \omega}\\
&\gtrsim 2.2\times10^6 {2(\bphi' + \bp) \over 1 - \bp}\gphi'^2
\nu_9^{-1}B_{*,14}^{-1/2} \zeta_{1}^{3/4}\left({L_{\rm iso,43}\over
    f_{\rm r}}\right)^{3/4},
  \end{split}
\end{equation}
where we have used the minimum ratio between the cyclotron frequency
$\omega_{\rm B}$ and the
frequency of the observed radio waves $\omega \equiv 2\pi \nu$ given
by eq. (\ref{eq:151}). For this beam Lorentz factor to be compatible
with the constraint from curvature cooling (eq. \ref{eq:125}), we
require
\begin{equation}
  \label{eq:152}
  0<{2(\bphi' + \bp) \over 1 - \bp}\gphi'^2 \lesssim 1.
\end{equation}
This is not possible for the case (1) where both the beam and target 
plasma are moving towards the observer ($\bp>0, \bphi'\approx 1,$ and $
\gphi'\gg 1$). For case (2) where
the target plasma is moving away from the observer $(\bp\approx -1)$,
this condition gives
\begin{equation}
  \label{eq:153}
  0<\bphi' + \bp \lesssim \gphi'^{-2}\, \Longleftrightarrow\,
\gphi'/\sqrt{3}\lesssim \gp < \gphi'.
\end{equation}
This is not possible with the X-mode, which only allows
$\bphi'\simeq (1 - \Delta_1)/\cos\tp \geq 1 - \Delta_1$, i.e. $\gphi'
\gtrsim \gp \sqrt{\zeta}\gg \gp$. Note that here $\zeta\gg 1$ is the
minimum ratio of 
the magnetic energy density to the kinetic energy density of the
plasma. Therefore, we conclude 
cyclotron-Cherenkov resonance condition cannot be satisfied given that
the B-field needs to be strong enough to confine the
beam and the target plasma.

Next, we consider the Cherenkov instability, which occurs when the
velocity of the beam particles equals to the phase velocity of a
certain wave. In the presence of the beam, the dielectric tensor (eq. \ref{eq:84})
needs to be modified to include both the beam and the target plasma. Since
the cyclotron-Cherenkov resonance of the beam particles can be ignored (it
requires $\gamma_{\rm b}$ to be much greater than the upper limit
given by from curvature cooling), the modification of the dielectric
tensor is done by re-defining the averaged quantities as $\langle
\ldots \rangle \equiv \int (\ldots) [f'(u') \d u' + f_{\rm b}(u_{\rm b}')\d u_{\rm
  b}' ]$ and using different densities $n_{\rm b}'$ and $n'$ 
in the definitions of the plasma frequencies of the
beam $\omega_{\rm pb}'$ and target plasma $\omega_{\rm p}'$. Under
this new definition, all the terms in the
dielectric tensor involving $\chi_1,\chi_2, \chi_4$, and $\chi_5$ are real,
except $\mc{E}_{zz}$. This is because 
$\chi_3=\Delta_3\left\langle  \gamma'^{-3} \mu'^{-2}\right\rangle$ has
significant imaginary part close to the Cherenkov resonance
\begin{equation}
  \label{eq:134}
  \mu_{\rm b}' = 1 - \beta_{\rm b}'\t{n}'_{\rm A}\cos\tp = 0,\
  \mr{i.e.}\ \bb' = \bphi'.
\end{equation}
The beam particles only couple to the z-component
  of the electric field of the wave, so X-mode waves with
  $\b{E}\perp\b{B}$ cannot be excited by the beam particles due to
  this instability. Therefore, we are only interested in the
excitation and growth of Alfv{\'e}n-mode waves in case (1) where both 
the beam and target plasma are moving towards the observer. In this
case, we have 
\begin{equation}
  \label{eq:158}
  \Delta_3 \simeq 1.0\times10^{14}\, {(L_{\rm iso,43}/f_{\rm r})}
  r_7^{-2}\nu_9^{-1} \ggg 1,
\end{equation}
and the dispersion relation of the Alfv{\'e}n-mode is
\begin{equation}
  \label{eq:159}
  \t{n}_{\rm A}'\cos\tp \approx 1 + {\chi_1\over 2} + {\mr{tan}^2\tp\over
    2\chi_3}.
\end{equation}
The growth rate of Alfv{\'e}n-mode waves excited at the Cherenkov
resonance is given by the imaginary part of the complex frequency
$\mr{Im}(\omega')$, which will be calculated below.

We only consider the beam particles  
near the resonance with fractional Lorentz factor spread $\Delta
\gb'/\gb'\lesssim 1$. 
Thus, the Lorentz factor of the beam particles $\gb'\simeq \gphi'$ must
be much greater than the temperature of the target 
plasma $T'$. Therefore, $\chi_1$ and $\chi_4$ are dominated 
by the target plasma and the contribution by the beam particles can be
ignored ($|\mu_{\rm b}'|\lesssim \gb'^{-2}$ is very small). Since
$\t{n}'_{\rm A}\cos\tp\approx 1 + \chi_1/2$ (to the first order), the
resonant beam Lorentz factor is
\begin{equation}
  \label{eq:135}
  \gb' \approx \chi_1^{-1/2} \gtrsim \sqrt{2\zeta}\, \gp
\end{equation}
and the Lorentz factor $\gb$ in the NS frame is a factor of $2\gp$
larger. At a sufficiently large distance from the NS surface
(e.g. $r\sim 10^7\,$cm), this beam Lorentz factor
$2\sqrt{2\zeta}\gp^2$ may not violate the constraint from curvature
cooling in eq. (\ref{eq:125}), so the growth of Cherenkov instability is
possible. From eq. (\ref{eq:159}), the
growth rate of the Cherenkov instability is
\begin{equation}
  \label{eq:137}
  \mr{Im}(\omega')\approx {k'c \cos\tp \sin^2 \tp \over 2}
  {\mr{Im}(\chi_3) \over |\chi_3|^2}\lesssim {k'c \cos\tp \sin^2 \tp
    \over 2|\chi_3|}, 
\end{equation}
where $\chi_3$ has two components given by the beam and
target plasma and we have $|\chi_3|\gtrsim \Delta_3\ggg
1$. Therefore, the growth rate is negligibly small at radio wavelengths.
We note that the growth of Cherenkov instability is faster at higher
frequencies \citep[up to $\omega' \sim \omega_{\rm p}' T'^{1/2}$, as
pointed out by e.g.][]{1999JPlPh..62..233M}. Therefore, even if the
beam-plasma resonance condition is met, most of the 
radiation will be at much higher frequencies than $\sim$GHz.

To summarize the main results of this subsection, we find that it is
unlikely that the plasma maser mechanism is responsible for FRBs,
because of the following inconsistencies: 
(i) in the case where both the beam and target plasma are moving
towards the observer, the cyclotron-Cherenkov resonance condition
requires unrealistically high beam Lorentz factors that are
inconsistent with curvature cooling; 
(ii) in the case where the beam and target plasma are
counter-streaming, the cyclotron-Cherenkov resonance condition
requires the ratio of magnetic energy density and particles' kinetic
energy density to be less than unity;
(iii) curvature drift velocity is negligibly small and hence
Cherenkov-drift instability cannot be important for FRBs; (iv) 
Cherenkov instability associated with Alfv{\'e}n mode has too small a
growth rate to be important for FRBs. These
inconsistencies basically come from the fact that plasma maser is powered by
particles' kinetic energy and the high luminosities of FRBs $L_{\rm iso}\sim
10^{43}\mr{\,erg\,s^{-1}}$ require very large particle number
densities and hence plasma frequencies much higher than $\sim$GHz. Note
that plasma maser may still 
be responsible for normal pulsar radio emission, which is persistent on
timescales much longer than the rotation period and has much lower
luminosities (by a factor of $\sim10^{10}$).

\subsection{Masers in vacuum}
In this subsection, we consider the possibility of negative absorption
in vacuum (refractive index $\tilde{n}= 1$) when particles'
distribution function has population inversion. The
source may be powered by either particles' kinetic energy or field
energy which maintains the population
inversion. \citet{1995MNRAS.276..372L} showed that negative curvature
absorption is possible when the B-field lines have
torsion. \citet{2017MNRAS.465L..30G}
proposed that vacuum synchrotron maser may be responsible for
FRBs. For synchrotron maser to be in the radio band, the B-field
strength should be smaller than $\sim$$10^3\,$G, which means that the
source should operate far away from the NS's surface
perhaps close to or beyond the light-cylinder. Nevertheless, we 
treat synchrotron and curvature emission in a unified way, because 
particles have (locally) helical trajectories and radiate similarly.

\begin{figure}
  \centering
\includegraphics[width = 0.45 \textwidth,
  height=0.25\textheight]{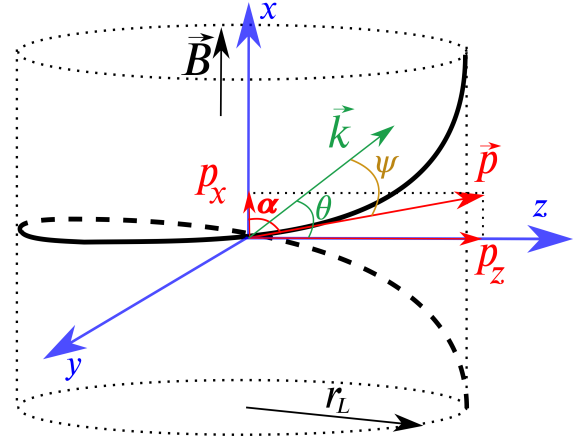}
\caption{A particle in a synchrotron orbit with pitch angle
  $\alpha$ given by $\mr{tan}\,\alpha = p_{\rm z}/p_{\rm x}$. The
  magnetic field is uniform $\b{B} = B\hx$ and the
  Larmor radius $r_{\rm L}$ is given by eq. (\ref{eq:11}). We are
  interested in the emissivity in the direction of the wave vector
  $\b{k}$, which is in the x-z plane at an angle $\theta$ wrt the
  z-axis. The angle between the wave vector $\b{k}$ and the particle's
  momentum vector $\b{p}$ is denoted as $\psi$.
}\label{fig:orbit}
\end{figure}

We consider a particle in
gyro-motion around a uniform B-field with a helical trajectory of
pitch angle $\alpha$, as shown in Fig. (\ref{fig:orbit}). We use
Cartesian coordinates with $\hx$ 
in the direction of the B-field ($\b{B} = B\hx$) and the component of
the particle's momentum parallel to the B-field is $\b{p}_{\para} = p_{\rm
  z}\hx = \gamma \beta_{\rm x}mc\hx$. Here, $\gamma$ is the Lorentz factor,
$\beta_{\rm x} = v_{\rm x}/c$, and $m$ is the particle mass. The
perpendicular momentum component is 
in the direction of the z-axis, i.e. $\b{p}_{\perp} = p_{\rm z}\hz =
\gamma \beta_{\rm z}mc\hz$ ($\beta_{\rm z} = v_{\rm z}/c$). The y-axis
is in the direction of 
$\b{p}\times \b{B}$. Thus, the 4-momentum of the particle can be written in
Cartesian (t, x, y, z) components
\begin{equation}
  \label{eq:8}
  \vec{p} = \gamma mc(1,\ \beta_{\rm x},\  0,\  \beta_{\rm z}).
\end{equation}
The ratio of the transverse momentum and
parallel momentum is given by the pitch angle
\begin{equation}
  \label{eq:73}
  \mr{tan}\,\alpha = {p_{\rm x}\over p_{\rm z}} = {\beta_{\rm x}\over
    \beta_{\rm z}},
\end{equation}
so we have $\beta_{\rm x} = \beta \cos\alpha$ and $\beta_{\rm z} =
\beta \sin\alpha$ (where $\beta = v/c$ is the total velocity).

We introduce another inertial frame which is moving in the $\hx$
direction at velocity $\beta_{\rm x}c$ or Lorentz factor $\gamma_{\rm
  x} = (1 - \beta_{\rm x}^2)^{-1/2}$. The new frame is called
$O'$-frame and all quantities measured in this frame are denoted with
a prime ($'$), while the old frame is called the $O$-frame where all
quantities are unprimed. The x$'$y$'$z$'$-axes in the $O'$-frame are
parallel to the xyz-axes in the $O$-frame. It is easy to show
that the 4-momentum of the particle in the
$O'$-frame is $\vec{p'} = (\gamma/\gamma_{\rm x}) mc(1,\ 0,\ 0,\
\gamma_{\rm x}\beta_{\rm z})$. Thus, in the $O'$-frame, the particle has
effective mass $\gamma'm = (\gamma/\gamma_{\rm x})m$ and is in a
circular orbit with velocity $\beta_{\rm z}' = \gamma_{\rm
  x}\beta_{\rm z}$. The radius of the circular orbit in the $O'$-frame
is the same as the Larmor radius in the $O$-frame, i.e.
\begin{equation}
  \label{eq:11}
  r_{\rm L} = {\gamma\beta_{\rm z} mc^2\over e B} =
  {\gamma\beta mc^2\sin\alpha \over e B }.
\end{equation}
The B-field is invariant under Lorentz transformation in the $\hx$
direction. 

In the limit $\gamma'=\gamma/\gamma_{\rm x}\gg 1$, the particle's
gyrophase-averaged emissivity in the $O'$-frame at an angle
$\theta'\ll1$ wrt the particle's momentum vector 
is given by \citep[e.g.][]{Jackson1975, 1979rpa..book.....R}
\begin{equation}
  \label{eq:12}
  \begin{split}
      j_{\nu'}'(\gamma', \theta') =&\ {4e^2 r_{\rm L} \over 3c^2}
  {\nu'^2\over \gamma'^4}(1 + \gamma'^2\theta'^2)\times\\
&\ \left[\gamma'^2
    \theta'^2K_{1\over3}^2(y') + (1 +
    \gamma'^2\theta'^2)K_{2\over3}^2(y')\right],\\
y' =\ {\nu'\over 2\nu_{\rm c}'}& (1 + \gamma'^2\theta'^2),\ 
\nu_{\rm c}' = {3c\over 4\pi r_{\rm L}}\gamma'^3,
  \end{split}
\end{equation}
where $K_{a}$ is the modified Bessel function of order $a$. Since
$\beta\approx 1$, we have $\gamma_{\rm x} = (1 - 
\beta^2\cos^2\alpha)^{-1/2}\approx (\sin\alpha)^{-1}$ and $\gamma' \approx
\gamma \sin\alpha$. We are interested in the emissivity in the
$O$-frame at frequency $\nu$ in the $\b{k}$ (wavevector) direction at
an angle $\theta$ wrt the z-axis in 
the x-z plane, and it can be obtained by Lorentz transformations
\begin{equation}
  \label{eq:34}
  \begin{split}
        &\  j_{\nu}(\gamma, \theta) =j_{\nu'}'(\gamma', \theta')
      (\nu/\nu')^2,\\
      &\ \nu = {\nu'\over \gamma_{\rm x}(1 -
        \beta_{\rm x}\sin\theta)},\ 
       \sin\theta' = {\sin\theta - \beta_{\rm x}\over 1 -
          \beta_{\rm x}\sin\theta}.
  \end{split}
\end{equation}
We define the angle between the wavevector $\b{k}$ and the particle's
momentum $\b{p}$ to be $\psi$, i.e.
\begin{equation}
  \label{eq:36}
  \cos \psi = \b{k}\cdot\b{p}/kp,
\end{equation}
so we have $\psi = \theta - (\pi - \alpha)$ and $\sin\theta =
\sin\psi\sin\alpha + \cos\psi\cos\alpha$. We are interested in the regime $\theta'\ll
1$, $\gamma' \gg 1$ and $\psi\ll 1$, and it is straightforward
to show that
\begin{equation}
  \label{eq:50}
  \gamma'\theta' \approx \gamma\psi \left(1 + {\mr{cot}\,\alpha \over
      2\gamma^2\psi}\right),\ \mr{and}\ 1 + \gamma'^2\theta'^2\approx
  1 + \gamma^2\psi^2,
\end{equation}
where higher order terms smaller by a factor of $\sim$$\gamma^{-2}$
or $\sim$$\psi^2$ have been ignored. We have also assumed that the pitch
angle $\alpha$ is not much smaller than 1 (otherwise the particle is
moving almost in a straight line with very little emission) but
$\alpha$ can be arbitrarily close to $90^{\rm o}$.
Then, we make use of $\nu'\approx \nu \sin\alpha$ and calculate the
emissivity in the $O$-frame in the $\b{k}$ direction (at an 
angle $\psi\ll1$ wrt the particle's momentum vector)
\begin{equation}
  \label{eq:52}
  \begin{split}
&\  j_{\nu}(\gamma, \psi) = {4e^2r_{\rm L}\over 3c^2}{\nu^2\over
  \gamma^4\sin^4\alpha} (1 + \gamma^2\psi^2)\times \\
&\ \left[\gamma^2\psi^2\left(1 + {\mr{cot}\,\alpha \over
      2\gamma^2\psi}\right)^2K_{1\over3}^2(y) + (1 + \gamma^2\psi^2)K_{2\over3}^2(y)
\right], \\ 
&\ y = {\nu\over 2\nu_{\rm c}\sin^2\alpha} (1+\gamma^2\psi^2)^{3/2},\
\nu_{\rm c}= {3c\over 4\pi r_{\rm L}}\gamma^3.
  \end{split}
\end{equation}
The first term [$\propto K_{1/3}^2(y)$] is polarized in the
$(\hat{\b{B}}\times \hat{\b{k}})\times\hat{\b{k}}$ direction and is
hence called O-mode. The second
term [$\propto K_{2/3}^2(y)$] is polarized in the
$\hat{\b{B}}\times\hat{\b{k}}$ direction and is hence called
X-mode. We note that the X-mode emissivity is symmetric about $\psi$
but the O-mode is asymmetric due to the
$\mr{cot}\alpha/(2\gamma^2\psi)$ term. This asymmetry vanishes at
 pitch angle $\alpha = 90^{\rm o}$ when the trajectory is confined
in a plane.

We also note that the emissivity in eq. (\ref{eq:52}) is also valid
for curvature radiation as the particle follows the infinitely
strong B-field lines with or without torsion (corresponding to
$\alpha\neq 90^{\rm o}$ and $\alpha=90^{\rm o}$ respectively), as long
as the Larmor radius is replaced by the curvature radius $\rho$. In the
curvature radiation scenario (assuming infinitely strong B-field), we
have $\hat{\b{p}}\para\hat{\b{B}}$, and then the polarization of the
first term [$\propto K_{1/3}^2(y)$] in eq. (\ref{eq:52}) is still in
the $(\hat{\b{B}}\times \hat{\b{k}})\times\hat{\b{k}}$ direction and
the second term in the $\hat{\b{B}}\times\hat{\b{k}}$ direction, as
long as $\psi\neq 0$. The X-mode/O-mode characteristics are the same
as in synchrotron radiation scenario. 

In the limit $\gamma\gg1$, the net absorption cross-section per
particle in the $\hat{\b{k}}$ direction at frequency $\nu$
is directly related to the emissivity and is given by
\citep[][and references therein]{1991MNRAS.252..313G}
\begin{equation}
  \label{eq:68}
  \begin{split}
      \sigma_{\rm abs}(\nu, \gamma, \psi) \approx {1\over
        2m\nu^2}{1\over \gamma^2}{\partial 
        \over \partial\gamma}[\gamma^2 j_{\nu}(\gamma, \psi)],
  \end{split}
\end{equation}
which is valid for any classical radiating particle (as long as the
correct emissivity is used). If the particles encountered by a certain
light ray in the $\b{k}$ direction have Lorentz factor
distribution $N_{\gamma} = \d N/\d \gamma$ (in unit cm$^{-3}$), the absorption
coefficient in this direction is
\begin{equation}
  \label{eq:71}
  \begin{split}
      \mu_{\rm abs} &= {1\over
        2m\nu^2} \int_1^\infty\d \gamma {N_{\gamma}\over
        \gamma^2} {\partial \over \partial\gamma}[\gamma^2
      j_{\nu}(\gamma, \psi)] \\
 &= -{1\over 2m\nu^2} \int_1^\infty\d \gamma {\partial
   \over \partial\gamma} \left( {N_{\gamma}\over \gamma^2}
\right) \gamma^2  j_{\nu}(\gamma, \psi),
  \end{split}
\end{equation}
and the absorption optical depth is $\tau_{\rm abs} \sim \mu_{\rm
  abs}\times \ell$, where $\ell$ 
is the propagation length. Therefore, two necessary conditions for
negative absorption are: (i) $\partial[\gamma^2
j_{\nu}(\gamma, \psi)]/\partial\gamma < 0$ at least for some $\gamma$
and $\psi$; (ii) $\partial (\gamma^{-2}N_{\gamma})/\partial\gamma>0$ for the same
$\gamma$ as in condition (i). The second condition means population
inversion. 

In order for a particle to radiate significantly at $\sim\,$GHz
frequencies, we require $\nu_{\rm c}\simeq \gamma^3c/(2\pi r_{\rm L})\sim
1\,$GHz, i.e. $\gamma\sim 10^2r_{\rm L,7}^{1/3}$. For synchrotron
radiation $r_{\rm L}\sim \gamma mc^2/(eB)\sim
2\times10^{-4}\mr{\,cm}\,\gamma_2 (m/m_{\rm
  e})\,(B/10^{9}\rm\,G)^{-1}$. The characteristic synchrotron
frequency is much higher than GHz in the NS magnetosphere near the
surface, so synchrotron maser can be rule out for this case (later in
\S5, we will consider synchrotron maser due to dissipation of free
energy in an outflow at large distances where the B-field is much
weaker). In the following, we only consider  
curvature radiation by particles with Lorentz factors not too far from
$\gamma\sim 10^2$. The detailed particle injection physics is likely
complicated, so we simply {\it assume} that population inversion $\partial
(\gamma^{-2}N_{\gamma})/\partial\gamma>0$ is {\it achieved and
  maintained} by some unknown mechanism near Lorentz factor
$\gamma\sim10^2$.

\begin{figure}
  \centering
\includegraphics[width = 0.45 \textwidth,
  height=0.22\textheight]{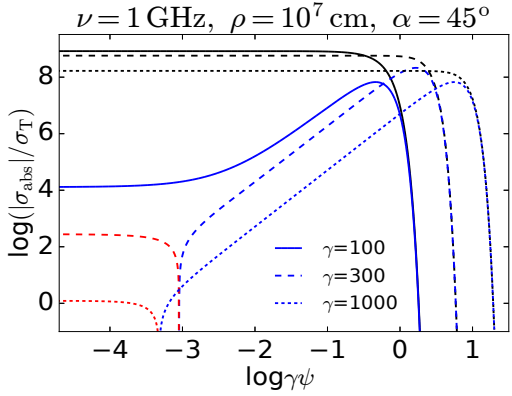}
\caption{Net curvature absorption cross-section (absorption minus stimulated
  emission) at $\nu = 1\,$GHz for curvature radius $\rho = 10^7\,$cm
  and B-field torsion angle $\alpha =45^{\rm o}$. The blue
  and red curves are for the O-mode, and the black curves are for
  the X-mode. We can see that negative absorption (red curves) only
  occurs in the O-mode at very small angles $\psi$ wrt the cone occupied by
  the particle's gyrating momentum vector.
}\label{fig:curv_abs}
\end{figure}

It is widely known that the net absorption cross-section for
curvature radiation is positive for any frequency or angle if the
particle's trajectory is confined in a plane
\citep[corresponding to $\alpha = 90^{\rm o}$ in our
notation,][]{1975MNRAS.170..551B}. From eq. (\ref{eq:72}), we know that 
curvature drift velocity is very small in that the pitch angle of the
helical orbit is extremely close to $90^{\rm o}$: $\cos\alpha\approx v_{\rm
  d}/c\lesssim 10^{-11}(B/10^{9}\mr{\,G})^{-1}$ (for $\gamma\sim
10^2$). The plasma is nearly 
charge-neutral when 
its density is much higher than the G-J density ($\mc{M}\gg1$). Note that
electrons and positrons drift in the opposite directions, so
the negative absorption coefficient is further suppressed by a factor
of $\mc{M}^{-1}$. Thus, the drift-induced
curvature maser scenario \citep{1992MNRAS.258..616L} can be 
ignored. The B-field configuration in the NS magnetosphere may have
torsion due to the existence of high-order multipoles, and the field
line may be locally helical. As shown by \citet{1995MNRAS.276..372L}
and in Fig. (\ref{fig:curv_abs}) 
in this paper, a {\it necessary} condition for the absorption cross-section
to be negative is $\psi\lesssim \mr{cot}\,\alpha/2\gamma^2$ for the
O-mode (and the absorption is always positive for the X-mode).

To produce the observed high brightness temperatures by curvature
maser, an absorption optical depth of $\tau_{\rm abs}\lesssim -30$ is
needed, which means 
\begin{equation}
  \label{eq:74}
  \gamma N_{\gamma} |\sigma_{\rm abs}| {\rho\over\gamma} \gtrsim 30,\ {\rm or}\
  \gamma N_{\gamma} \gtrsim 5\times 10^{17}\,{\rm cm^{-3}}\,
  {\gamma_{2}\over \rho_7}
  {10^3\sigma_{\rm T}\over |\sigma_{\rm abs}|}.
\end{equation}
Such a high density may be achieved near the NS surface ($r\sim10^6\rm\
cm$) with particle number density a factor of $\mc{M}\sim 10^4$ above
the G-J density (eq. \ref{eq:27}).

However, it can be seen in Fig. (\ref{fig:curv_abs}) that, even if
the absorption cross-section is negative at $\psi\lesssim
(2\gamma^2)^{-1}$, it becomes positive with absolute value
a factor of $\gtrsim 10^5$ greater at larger angles
$\psi\sim\gamma^{-1}$. Due to the B-field 
curvature, photon trajectories will unavoidably intersect with other B-field
lines at larger angles before escaping, and hence torsion-induced
curvature maser requires extreme fine-tuned conditions where particle
number density drops by a factor of $\gtrsim10^{5}$ immediately
outside the maser region. 

The above argument applies regardless of whether the maser is powered
by particles' kinetic energy or field energy. For a kinetic-energy
powered maser, an additional issue is that the minimum
number density (given by eq. \ref{eq:26}) must be a factor of
$\mc{M}\gtrsim 10^{10}$ greater than G-J density. It is unclear how
these particles are injected\footnote{
Many theoretical studies of pair production along
open field lines above the polar cap of pulsars
show that $\mc{M}$ ranges between $10^2$ and $10^5$
\citep{1982ApJ...252..337D, 2001ApJ...554..624H, 2010MNRAS.406.1379M,
  2015ApJ...810..144T}, although detailed dynamical processes of pair
production under different (non-dipolar) B-field
geometries still have large uncertainties. Observational studies
of non-thermal emission from pulsar wind nebulae (PWN) arrive at
similar conclusion \citep{1984ApJ...283..710K, 1996MNRAS.278..525A,
  1996ApJ...457..253D}. We note that these PWN studies only focus on
optical to $\gamma$-ray frequencies and that accounting for the radio
emitting electrons/positrons in the Crab Nebula require $\mc{M}\sim
10^6$. These radio emitting particles (with Lorentz factors a
few$\times10^2$) only contributes to $\sim1\%$ of the total energy
of the nebula and their origin has been an unsolved puzzle for decades
\citep{1974MNRAS.167....1R, 2012SSRv..173..341A}. These low energy
particles, with their large number and long synchrotron cooling time, could
be relics from the wind injection in the past \citep[the pulsar spin-down
luminosity could be much higher in the past,][]{1999A&A...346L..49A},
or acceleration of electrons from the Rayleigh-Taylor 
filaments penetrating the shocked wind region
\citep{2013MNRAS.428.2459K, 2017ApJ...841...78T}.}. Also, we can see
from Fig. (\ref{fig:curv_abs}) 
that such a high density will lead to strong positive curvature
self-absorption and the radiation cannot escape.

Under the assumption that the source plasma is in the corotating
magnetosphere of a NS, we conclude that neither plasma maser nor
masers in vacuum are consistent with the basic properties of
FRBs. In the next section, we discuss the possibility that the
source plasma is inside a relativistic outflow launched from a NS or
BH.

\section{Masers in an outflow with internal/external dissipation}
In \S 4, we have discussed various maser mechanisms under the
assumption that the source plasma is confined by the B-field in the
corotating magnetosphere of a NS. In this section, we consider that
FRBs are produced by maser 
mechanisms in an outflow launched from either a BH or NS. The emission
is powered by the dissipation of  free energy in the outflow at large
distances from the central object.

For a BH progenitor, we assume that, due to sudden accretion,
an outflow is launched from the inner disk 
near the event horizon. For a NS progenitor, we assume such 
an outflow is originally launched from near the stellar surface,
based on the fact that most of the magnetic energy in the 
magnetosphere is concentrated near the surface. In both cases,
dissipation of free energy in the outflow may occur due to
external shocks when it runs into some dense clouds or accumulates
enough circum-stellar medium in the forward shock region, or
internal dissipation processes such as magnetic reconnection and
collisions between shells of different speeds. 

We consider that the outflow is moving towards the observer with Lorentz
factor $\Gamma$. At a distance $r$ from the central object, the maser
formation length is limited by the dynamical time and speed of
  light to be $\lesssim r/(2\Gamma^2)$, which constrains the FRB 
duration to be
\begin{equation}
  \label{eq:22}
  t_{\rm FRB} \lesssim {r\over 2\Gamma^2c}
  \simeq (3\,\mr{ms})\, r_{12}\Gamma_2^{-2},
\end{equation}
The variations of $\Gamma$ and $r$ may cause different FRB  
durations. Note that eq. (\ref{eq:22}) also takes into account
(through the ``$<$" sign) the possibility 
that the FRB source plasma is only a small local
patch\footnote{\citet{2016PhRvD..93b3001R} proposed that FRBs may be
  produced by modulational instability induced cavitons under strong plasma
turbulence during the interaction of a relativistic 
lepton beam and a target plasma. In the astrophysical context, they
consider electrons and positrons in a relativistic leptonic jet {\it
  passing through} a cloud of size
$R_{\rm c}\simeq \Gamma^2ct_{\rm FRB}\simeq
(3\times10^{13}\mr{\,cm})\, \Gamma_3^2t_{\rm FRB,-3}$, which could be
much smaller than the causally-connected region of the jet at large
distances from the central engine. The frequency of the escaping
radiation is near (but above) the plasma frequency of the
cloud. However, the dynamics of the jet-cloud interaction is
different from the laboratory beam-plasma interaction they referred to
\citep[e.g.][]{1997RvMP...69..507R}, because astrophysical plasmas are
magnetized and particles' Larmor radii are many orders magnitude
smaller than the cloud size. Therefore, collisionless shocks form
and a contact discontinuity at the two-fluid interface prevents particles in the
jet (or ``beam'') from penetrating through 
the cloud (or ``plasma'').} in the causally-connected region that meets the
maser condition, although the radiation efficiency and the maser
amplification length are reduced when $t_{\rm FRB}\ll r/(2\Gamma^2
c)$.

In the external shock scenario, the emission radius $r$ is roughly given 
by the deceleration radius $r_{\rm dec}$ (defined as where the Lorentz factor of the
outflow drops by a factor of $\sim2$, see eq. \ref{eq:30} in Appendix
B), and the duration is roughly given by $t_{\rm FRB}\sim r_{\rm
  dec}/(2\Gamma^2 c)$. In Appendix B, we consider the dynamics of two
consecutive outflows and show that the second outflow should have a much
longer deceleration time than the first one (because it propagates inside the
cavity opened by the first one). This is in contradiction with the
observation that two consecutive FRBs separated by $\sim40\,$ms have
very similar durations \citep[$\sim2\,$ms,][]{2017ApJ...846...80S}.

Still, FRBs could be powered by internal dissipations in the outflow due to
e.g. magnetic reconnection or collisions between different ejected
shells. In the following two subsections, we discuss the physical
conditions required to produce FRBs by synchrotron maser mechanism in
vacuum \citep{2017MNRAS.465L..30G} in \S 5.1 and 
in plasma \citep{2017ApJ...842...34W} in \S 5.2. In our general
discussion, we assume that, after the onset of internal dissipation
(e.g. propagation of a shock front or
magnetic reconnection trigger), particles have random gyration phases.

We also note that bunching in the gyration phase can occur in the case
of quasi-perpendicular shocks at high magnetization,
as a result of  coherent reflection of (cold) upstream particles by
the shock-compressed B-field \citep{1988PhFl...31..839A,
  1992ApJ...391...73G, 1992ApJ...390..454H}. In this situation, 
coherent gyration of incoming particles generates an X-mode
EM wave precursor ahead of the main shock near the
gyrofrequency.
% \citet{2014MNRAS.442L...9L} and \citet{2017ApJ...843L..26B} 
% proposed that FRBs may be produced by this mechanism when a magnetar 
% outflow drives a shock into the magnetized circumstellar
% medium (the wind nebula). Although this external shock scenario is
% inconsistent with the durations of closely-separated burst pairs as
% shown above, the same radiation mechanism may operate in internal
% shocks between colliding shells as well. 
We discuss this possibility
in \S 5.3.

% Although the models by are based on the shock driven by a magnetar
% outflow propagating into the circumstellar medium as in the
% external shock scenario, the same radiation mechanism may operate
% in internal shocks between different ejected shells as well. The maser
% mechanism in their models is due to bunching in the  
% gyration phase occurring as a result of the reflection of (cold) upstream
% particles by the shock-compressed B-field in the case of magnetized
% quasi-perpendicular shocks \citep{1988PhFl...31..839A}. 

% in the comoving frame of the unshocked upstream
% plasma, where $\nu_{\rm p}' = (4\pi n' e^2/m_{\rm e})^{1/2}$, $\sigma$
% is the upstream magnetization, and $n'$ and $B'$ are the number
% density and B-field strength in the upstream proper frame. 

% We also note that, in the case of
% magnetized quasi-perpendicular shocks (i.e. upstream B-field roughly
% perpendicular to the direction of shock propagation), 

\begin{figure}
  \centering
\includegraphics[width = 0.45 \textwidth,
  height=0.23\textheight]{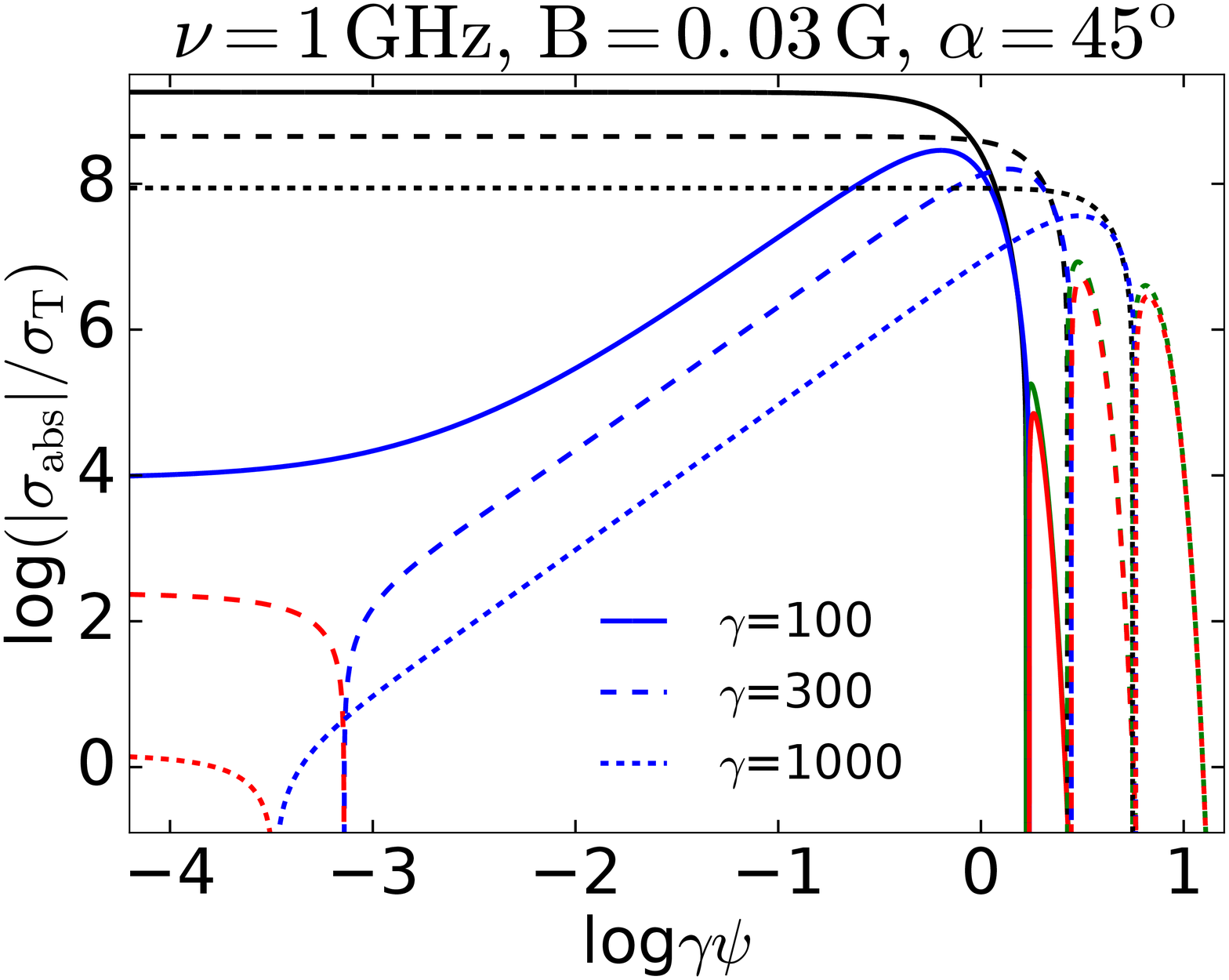}
\includegraphics[width = 0.45 \textwidth,
  height=0.23\textheight]{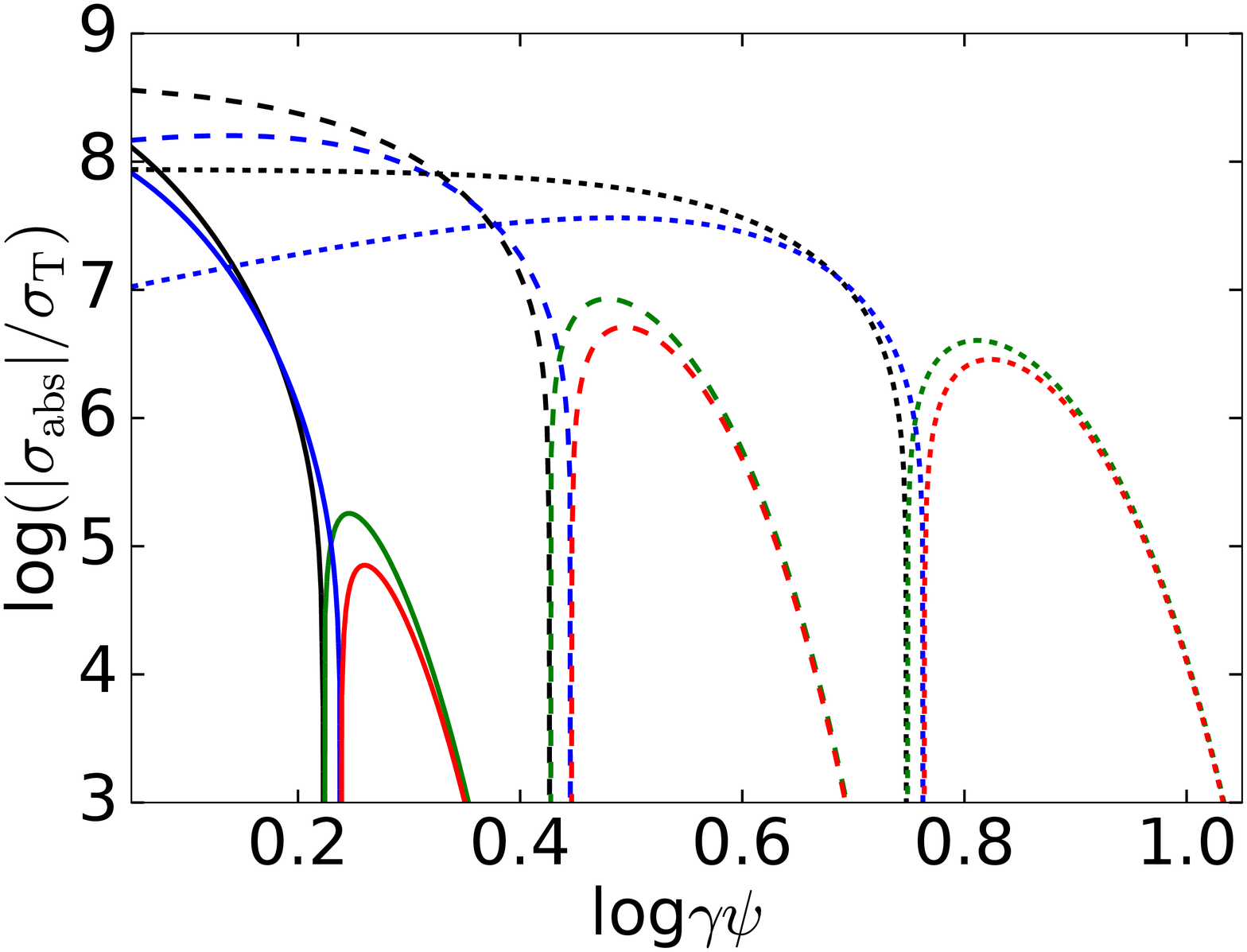}
\caption{\textit{Upper Panel:} Net synchrotron absorption
  cross-section (absorption minus 
  stimulated emission) at frequency $\nu = 
  1\,$GHz for an electron moving 
  in a uniform B-field $B = 0.03\,$G with pitch angle $\alpha =
  45^{\rm o}$. The blue and red curves are for the O-mode, and the black
  and green curves are for the X-mode. The cross-sections are negative
  for the red and green curves. The angle between the line of sight
  and the electron's momentum vector is denoted as
  $\psi$. \textit{Lower Panel:} Zoom-in of the negative
  cross-section regions immediately outside the $\gamma^{-1}$ beaming
  cone.
}\label{fig:syn_abs}
\end{figure}

\subsection{Synchrotron maser in vacuum}
Based on eqs. (\ref{eq:52}) and (\ref{eq:68}), we calculate the
single-particle synchrotron absorption cross-section for a given
frequency $\nu$ at an arbitrary angle $\psi$ away from the particle's
momentum vector. In Fig. (\ref{fig:syn_abs}) we show one case where an
electron is in a helical synchrotron orbit with pitch angle $\alpha =
45^{\mr{o}}$ in a uniform B-field $B = 0.03\,$G. We see that vacuum
synchrotron maser is possible because the absorption
cross-section is negative at angles $\psi\gtrsim$a 
few$\times\gamma^{-1}$ for both O-mode 
and X-mode. However, the absorption cross-section at smaller angles on
the order of $\gamma^{-1}$ is
positive with a much larger absolute value. There is also a very
narrow negative absorption region at $\psi\lesssim \gamma^{-2}$ for
the O-mode if the pitch angle $\alpha\neq 90^{\rm o}$, but the
cross-section is very small.

We can see that, other than population inversion $\partial
(N_\gamma/\gamma^2)/\partial\gamma > 0$, vacuum synchrotron maser
requires the following two extra conditions: (i) the B-field is nearly
uniform to within an angle of 
$\lesssim \gamma^{-1}$; (ii) particles' pitch angle distribution is
narrowly peaked with spread $\Delta \alpha \lesssim \gamma^{-1}$,
where $\gamma$ is the typical Lorentz factor of the
radiating particles in the comoving frame of the source plasma. 
The maximum efficiency of vacuum synchrotron maser is
$\sim\gamma^{-1}$ because particles' pitch angles are not allowed to
change by more than $\sim\gamma^{-1}$ as they cool.

Keeping these points in mind, we consider the emitting plasma to be in
a relativistic outflow moving towards the observer with arbitrary Lorentz
factor $\Gamma$. For electrons with Lorentz factor $\gamma'$ in the
comoving frame, the synchrotron frequency in the observer's frame is
\begin{equation}
  \label{eq:187}
  \nu\simeq \Gamma \gamma'^2\nu_{\rm B}'\sin\alpha',
\end{equation}
where $\nu_{\rm B}' = eB'/(2\pi \me c)$ is the cyclotron frequency,
$\alpha'$ is the pitch angle, and $B'$ is the B-field
strength in the plasma comoving frame.
The outflow kinetic power must exceed the FRB luminosity $L_{\rm iso}$,
so we have
\begin{equation}
  \label{eq:188}
  4\pi r^2 (B'^2/8\pi)\Gamma^2 c\eps_{\rm B}^{-1} \gtrsim L_{\rm iso},
\end{equation}
where $\eps_{\rm B}<1$ is the fraction of outflow power in magnetic energy.
Combining eqs. (\ref{eq:187}) and (\ref{eq:188}) above, we obtain a
lower limit on the radius of emission
\begin{equation}
  \label{eq:189}
  r\gtrsim (7.2\times10^{13}\mr{\,cm})\, \eps_{\rm
    B}^{1/2}\gamma'^2\sin\alpha' L_{\rm 
    iso,43}^{1/2} \nu_9^{-1}.
\end{equation}
When the outflow undergoes dissipation (shocks or magnetic
reconnection) at such a large distance from the central object,
the extremely fine-tuned requirements on the
B-field and pitch-angle distribution are unlikely to be realized. Moreover, even if
population inversion $\partial (N_\gamma/\gamma^2)/\partial\gamma > 0$
is initially achieved, since high-energy electrons radiate faster than
lower-energy ones, the population inversion may be quickly destroyed.

\subsection{Synchrotron maser in plasma}
The synchrotron emissivity (eq. \ref{eq:52}) is strongly modified at frequencies
below the Razin frequency (the plasma frequency multiplied by the
Lorentz factor of the particle). This is because the
Li{\'e}nard-Wiechert potential of a particle undergoing acceleration
is significantly modified by wave dispersion at 
these frequencies. In this subsection, we consider synchrotron
maser in a plasma which is moving towards the observer with bulk
Lorentz factor $\Gamma$. Quantities in the comoving frame of the
radiating plasma are denoted with a prime ($'$) and unprimed
quantities are in the rest frame of the central object.

For a weakly magnetized plasma with randomly oriented B-field, under the
{\it assumption} that the
population-inversion condition $\partial
(N'_{\gamma'}/\gamma'^2)/\partial\gamma' > 0$ is satisfied, 
synchrotron maser may operate below the effective Razin frequency
$\nu_{\rm R*}'=\nu_{\rm p}'\mr{min}(\gamma_{\rm e}', \sqrt{\nu_{\rm p}'/\nu_{\rm
    B}'})$ \citep{2002ApJ...574..861S}, where $\gamma_{\rm e}'$ is the
Lorentz factor of the radiating electrons and the
plasma frequency and gyrofrequency are defined as
\begin{equation}
  \label{eq:33}
  \nu_{\rm p}' = {1\over 2\pi}\left({4\pi n' e^2 \over \bar{\gamma}' m_{\rm
        e}}\right)^{1/2},\ \nu_{\rm B}'=   {e B'\over 2\pi \bar{\gamma}'
    m_{\rm e}c}, 
\end{equation}
and $\bar{\gamma}'$ is the mean Lorentz factor of electrons in the
comoving frame of the emitting plasma. We allow $\bar{\gamma}'$ to be
a free parameter and take $\bar{\gamma}' = 10^2 \bar{\gamma}_2'$ as
our fiducial value (constraints on $\bar{\gamma}'$ will be shown
later). The ratio of these two frequencies can be expressed as
$\nu_{\rm p}'/\nu_{\rm B}' \simeq (\epsilon_{\rm e}/2\epsilon_{\rm
  B})^{1/2}$, where $\epsilon_{\rm e}$ and $\epsilon_{\rm B}$
are the fractions of the energy density in electrons and B-field in the emitting
region. Note that, since EM waves below the plasma frequency cannot
propagate, synchrotron maser near the effective Razin frequency can only
operate when  $\epsilon_{\rm e}/\epsilon_{\rm B}\gg 1$, which means
the radiating plasma is not dominated by magnetic energy.

In the following, we discuss the propagation of the coherent radiation
through the source plasma. To make our discussion as general as
possible, we parameterize the frequency of the emitted radiation in the
comoving frame of the plasma as $\nu'=\xi \nu_{\rm p}'$. For instance, in
the model of \citet{2017ApJ...842...34W}, we have $\xi = 
\sqrt{\nu_{\rm p}'/\nu_{\rm B}'} = (\epsilon_{\rm e}/2\epsilon_{\rm
  B})^{1/4}$. For other maser-type collective plasma emission in a
weakly magnetized plasma, the radiating frequency may be near the plasma
frequency $\nu'\sim \nu_{\rm p}'$ and $\xi\sim 1$.

The frequency in the observer's frame is $\nu = \nu_9\rm \ GHz\simeq
\Gamma \xi\nu_p'$, so we obtain
\begin{equation}
  \label{eq:46}
  \Gamma^2n'\simeq (1.1\times10^{13}\,\mr{cm^{-3}})\,\gbar_2'
\left({\nu_9/\xi} \right)^{2},
\end{equation}
If the emission occurs at a distance $r$ from the central
engine, the isotropic equivalent kinetic luminosity of electrons is
\begin{equation}
  \label{eq:51}
  \begin{split}
      L_{\rm ke, iso} \simeq 4\pi r^2 \Gamma^2 n' \bar{\gamma}' \me c^3.
  \end{split}
\end{equation}
Since $\eps_{\rm e}/\eps_{\rm B}\gg 1$, the total luminosity of the
outflow is a factor $\eps_{\rm e}^{-1}$ higher (a large fraction of
the energy may be in protons). The radiative efficiency in the radio band is
denoted as $f_{\rm r}<1$, so the isotropic FRB luminosity is $L_{\rm
  iso} =L_{\rm 
  ke,iso}\eps_{\rm e}^{-1} f_{\rm r}$. We eliminate $\Gamma^2n'$ with
eqs. (\ref{eq:46}) and (\ref{eq:51}) and then obtain the emission
radius
\begin{equation}
  \label{eq:48}
  \begin{split}
  r\simeq (1.7\times10^{11}\,\mr{cm})\, \gbar_2'  \left({L_{\rm
        iso,43}\eps_{\rm e}\over
      f_{\rm r}}\right)^{1/2} \left({\nu_9\over\xi} \right)^{-1},
  \end{split}
\end{equation}
which has no direct dependence on the bulk Lorentz factor.
The emission radius is related to the FRB duration through
$t_{\rm FRB}\lesssim r/(2\Gamma^2 c)$ (eq. \ref{eq:22}), so we can
constrain the bulk Lorentz factor
\begin{equation}
  \label{eq:76}
  \Gamma \lesssim 53\,\gbar_2'^{-1/2} t_{\rm
    FRB,-3}^{-1/2}\left({L_{\rm iso,43}\eps_{\rm e}\over
      f_{\rm r}}\right)^{1/4} \left({\nu_9\over\xi} \right)^{-1/2}.
\end{equation}
The isotropic equivalent number of electrons in the
causally-connected region is
\begin{equation}
  \label{eq:49}
  \begin{split}
 N &\simeq 4\pi r^3 n'/\Gamma\\
&\gtrsim 4.6\times10^{42}\,\gbar_2'^{-1/2}t_{\rm
  FRB,-3}^{3/2}\left({L_{\rm 
    iso,43} \eps_{\rm e}\over f_{\rm r}}\right)^{3/4} \left({\nu_9\over\xi}
\right)^{1/2},
  \end{split}
\end{equation}
where the upper bound of $\Gamma$ in eq. (\ref{eq:76}) has been used.
For a NS progenitor, the total number of electrons and positrons in
the magnetosphere is roughly given by
\begin{equation}
  \label{eq:43}
  {4\pi R_*^3\mathcal{M} n_{\rm GJ}\over 3}\simeq
  (2.9\times10^{38})\, \mathcal{M}_6 B_{*,14}P_{-1}^{-1}.
\end{equation}
Thus, any model based on outflows from a NS must explain how this
extremely large amount of particles $\mc{M}\gg 10^6$ are created and
ejected (for instance, the outflow has to break the magnetic confinement
if it is launched in the closed field line region). 

The entire outflow has Thomson optical depth
\begin{equation}
  \label{eq:75}
    \begin{split}
\tau_{\rm T} &= {n'\sigma_{\rm T}r/\Gamma} = {L_{\rm iso}
  f_{\rm r}^{-1}\eps_{\rm e} \sigma_{\rm T} \over 4\pi r \Gamma^3
  \gbar' \me c^3} \\ 
&\gtrsim 8.4\times10^{-6}\, \gbar_2'^{3/2} t_{
\rm FRB,-3}^{3/2} \left({L_{\rm
      iso,43}\eps_{\rm e}\over f_{\rm 
      r}}\right)^{-1/4}\left({\nu_9\over\xi} \right)^{5/2}.
  \end{split}
\end{equation}
We note that induced (or stimulated) Compton scattering may be
important even when the Thomson optical depth is small\footnote{The
wavelength of the FRB EM wave is much shorter than the Debye length in the
source region, the strong E-fields associated to the EM wave cannot be
screened, so an electron can interact with many photons at the same
time. Even for wavelength much longer than the Debye length, induced
Raman scattering creating Langmuir waves may become an important
obstacle hindering the propagation of coherent waves.}.
If the source radiates isotropically in the comoving frame and the
spectrum is moderately broad with $\Delta\nu/\nu\gtrsim 1$, the
brightness temperature $T_{\rm B}'$ in the plasma comoving frame is limited
by (see Appendix C) 
\begin{equation}
  \label{eq:64}
  {kT_{\rm B}'\over \bar{\gamma}'^5m_{\rm e}c^2 }  \tau_{\rm
    T}\lesssim 1.
\end{equation}
An intuitive understanding of the $\bar{\gamma}'^{-5}$ factor
  is as follows. For the conservative case with spectrum width $\Delta
  \nu \sim \nu$, induced-Compton scattering is only important when the
  fractional frequency change before and after the scattering is of
  order unity. Thus, the incident photon only interacts with a
  fraction $\sim$$\gbar'^{-2}$ of electrons
  whose momentum vectors are nearly
  parallel to the wavevector (to within an angle
  $\sim$$\gbar'^{-1}$). Therefore, the effective Thomson optical depth is
  $\sim$$\gbar'^{-2}n'\sigma_{\rm T}r/(\gbar'^2\Gamma) =
  \tau_{\rm T}/\gbar'^{4}$. Another factor of $\gbar'^{-1}$
  comes from Lorentz transformation that the brightness temperature in the
  electron' comoving frame is $T_{\rm B}'/\gbar'$.
When the source is relativistic, the relationship between $T_{\rm B}'$
(the true brightness temperature in the comoving frame)
and the \textit{apparent} brightness temperature $T_{\rm B}$ given
by eq. (\ref{eq:63}) is $T_{\rm B}' \simeq T_{\rm B}/\Gamma^3$.
Thus, eq. (\ref{eq:64}) constrains the \textit{apparent} brightness temperature
\begin{equation}
  \label{eq:66}
  T_{\rm B} \lesssim (1.1\times10^{30}\mr{\,K})\, \gbar_2'^{2} t_{\rm FRB,-3}^{-3} 
{L_{\rm iso,43} \eps_{\rm e} \over f_{\rm r}} \left({\nu_9\over\xi} \right)^{-4}.
\end{equation}
Note that the constraint on $T_{\rm B}$ in eq. (\ref{eq:66}) will be
stronger if the FRB spectrum is broader than $\Delta \nu/\nu\sim 1$ or
if the Lorentz factor distribution below $\bar{\gamma}'$ is flatter
than $N_{\gamma'}'\propto \gamma'^4$ (see Appendix C).

Violation of eq. (\ref{eq:64}) or (\ref{eq:66}) causes exponential
loss of photon energy with time and the energy goes back to the
plasma. Note that this constraint is only correct in the linear regime (when the
non-linearity parameter $a_0< 1$), which is valid for a source powered
by particle kinetic energy. It can be shown by inserting the effective
particle mass $\bar{\gamma}'\me$ into eq. (\ref{eq:6}) that the
non-linearity parameter becomes $a_0\simeq (f_{\rm
 r}/\eps_{\rm e})^{1/2}\xi^{-1}$. Since $f_{\rm r}/\eps_{\rm e}< 1$
(only a fraction of the electrons' energy is radiated in the radio
band) and $\xi\gtrsim1$ (for waves to escape), we obtain $a_0\lesssim
1$.

We compare eq. (\ref{eq:66}) with the apparent brightness temperature
in eq. (\ref{eq:63}) and obtain a constraint on the radiative efficiency
\begin{equation}
  \label{eq:174}
  f_{\rm r} \lesssim 2.3\times10^{-6} \xi^4 \eps_{\rm e}  \gbar_2'^2 t_{\rm
    FRB,-3}^{-1}
  \nu_9^{-2} \Delta \nu_{9},
\end{equation}
where $\Delta \nu_9$ is the width of the FRB spectrum.
The maximum mean Lorentz factor $\gbar_{\rm max}'$ corresponds
to the condition when the synchrotron/inverse-Compton cooling time $t_{\rm
  cool}'$ equals to the dynamical time $t_{\rm dy}' =
r/\Gamma c$ (smaller $\gbar'$ corresponds to $t_{\rm cool}'>t_{\rm
  dy}'$), and we obtain
\begin{equation}
  \label{eq:183}
  {r\over \Gamma c} = {3\gbar_{\rm max}' \me c^2 \over 4\sigma_{\rm T}
    c\gbar_{\rm max}'^2 \mr{max}(U_{\rm B}', U_{\rm rad}')},
\end{equation}
where the magnetic and radiation energy densities are
\begin{equation}
  \label{eq:182}
  \mr{max}(U_{\rm B}', U_{\rm rad}') = {L_{\rm iso}/f_{\rm
      r} \over 4\pi r^2 \Gamma^2 c} \mr{max}(\eps_{\rm B}, \eps_{\rm
    e}). 
\end{equation}
Then making use of the expressions for $r$ and $\Gamma$ in
eqs. (\ref{eq:48}) and (\ref{eq:76}) and $\eps_{\rm B}\ll \eps_{\rm
  e}$, we obtain 
\begin{equation}
  \label{eq:185}
 \gbar_{\rm max,2}'^2\simeq 3.5 \left( {L_{\rm iso,43}\eps_{\rm e}
   \over f_{\rm r}}\right)^{1/7} \left( {\nu_9\over \xi
 }\right)^{-10/7} t_{\rm FRB,-3}^{-6/7}.
\end{equation}
Thus, we combine eqs. (\ref{eq:174}) and (\ref{eq:185})
and obtain
\begin{equation}
  \label{eq:186}
  f_{\rm r} \lesssim  3.5\times10^{-5} \xi^{19/4}\eps_{\rm e} L_{\rm
    iso,43}^{1/8} t_{\rm FRB,-3}^{-13/8} \nu_9^{-3} \Delta\nu_9^{7/8}.
\end{equation}
Note that $f_{\rm r}$ is defined as the radiative efficiency in the radio
band. When $\gbar'\simeq\gbar_{\rm max}'$, the source plasma is
radiatively efficient and electrons radiate almost all their kinetic energy
at frequencies $\gg$GHz through 
multiple scattering, since the Compton-Y parameter $Y\sim
\gbar'^2\tau_{\rm T}$ is of order unity.
For synchrotron maser in a relativistic plasma, we have $\xi\simeq
(\epsilon_{\rm e}/2\epsilon_{\rm B})^{1/4}$, and then the constraint in
eq. (\ref{eq:186}) becomes 
\begin{equation}
  \label{eq:176}
  f_{\rm r}\eps_{\rm B}^{19/16}\lesssim 1.5\times10^{-5} 
\eps_{\rm e}^{35/16} L_{\rm iso,43}^{1/8} t_{\rm FRB,-3}^{-13/8}
\nu_9^{-3} \Delta\nu_9^{7/8}.
\end{equation}
Hereafter, we consider the recent detections of bursts from FRB 121102
at $\sim$6 GHz by \citep{2017ATel10675....1G} and take $\Delta \nu_9
\simeq 1$ and $t_{\rm FRB,-3}\simeq 1$.
The electron energy fraction
has been shown to be near equipartition value $\eps_{\rm e}\sim
0.3$ observationally \citep[from the afterglows of 
gamma-ray bursts, ][]{2001ApJ...560L..49P} and theoretically
\citep[from particle-in-cell simulations,
][]{2015SSRv..191..519S}. Thus, the radiative efficiency in the radio
band must be very 
low $f_{\rm r}\lesssim 5\times10^{-9}\eps_{\rm B}^{-19/16}$, as long
as $\eps_{\rm B}$ is not much smaller than $\sim 10^{-5}$. On the
other hand, for possible plasma masers in weakly 
magnetized plasma $\xi\sim 1$, the apparent brightness temperature
implies extremely low radiative efficiency in the radio band $f_{\rm
  r}\lesssim 5\times 10^{-8}$.

\subsection{Bunching in the gyration phase}

It has been proposed that FRBs may be produced by a maser mechanism
due to bunching in gyration phase at quasi-perpendicular
shocks with high magnetization when the upstream particles are
coherently reflected by the 
shock-compressed B-field \citep{2014MNRAS.442L...9L,
  2017ApJ...843L..26B}. These authors considered the external shock
scenario where a magnetar outflow drives a shock into the magnetized
circumstellar medium (the wind nebula). Although this scenario is
inconsistent with the durations of closely-separated burst pairs (as
shown in Appendix B), the same maser mechanism may operate in
internal shocks between colliding shells as well, which is discussed
in this subsection.

Bunching in gyration phase due to coherent reflection of incoming
particles at quasi-perpendicular shocks has been well studied in 1D
simulations with magnetization $\sigma\in (10^{-2}, 5)$
\citep{1988PhRvL..61..779L, 1992ApJ...391...73G, 1992ApJ...390..454H},
where a fraction of $\lesssim10^{-1}$ of the incoming particles'
kinetic energy is converted to a coherent EM precursor. Note that the
magnetization parameter $\sigma$ is defined as
the ratio between the upstream Poynting flux and particles' kinetic flux.
Later 2D simulations with $\sigma=0.1$ by \citet{2009ApJ...698.1523S}
show that the precursor gets increasingly weaker with time as the
coherence of particle reflections between different locations along
the shock 
surface is lost, and the authors speculated that the maser emission
may disappear in sufficiently long simulations or in 3D. More recent 2D
simulations with higher resolutions by \citet{2017ApJ...840...52I}
show that the precursor persists until the end of the simulations (at
time $\sim$$10^3\nu_{\rm p}'^{-1}$) and that the efficiency appears to
converge to $\sim$$10^{-2}$ when $\sigma\in(0.1, 1)$. In this
paper, we take the results from the above 1D simulations as upper
limits but note that longer simulations at higher magnetization in 2D
and 3D are needed to draw a firm conclusion on the efficiency of the
coherent precursor.

Consider two consecutive shells ejected from
the central engine colliding at certain radius $r$ and then the coherent
EM waves propagates through the upstream plasma of the first (slower)
shell whose bulk Lorentz factor is $\Gamma$. The emission frequency in 
the comoving frame of the first shell is $\sim
\mc{R}^{1/4}\nu_{\rm B}'$, where $\mc{R}$ is the ratio between the
luminosities of the second and first shells 
and $\mc{R}^{1/4}>1$ is the relative Lorentz factor between the
upstream and downstream. The magnetization parameter of the first
shell is $\sigma=\nu_{\rm B}'^2/\nu_{\rm p}'^2$, so the
emission frequency can be rewritten 
as $\nu\sim \mc{R}^{1/4}\sigma^{1/2}\nu_{\rm p}'$. We also note that coherent
reflection of incoming particles is only possible when the upstream
plasma in the first shell is initially cold, i.e. electrons' thermal Lorentz
factor $\bar{\gamma}'\ll \mc{R}^{1/4}$.

We put $\xi\sim \mc{R}\sigma^{1/2}$ and
$\bar{\gamma}'< \mc{R}^{1/4}/2$ into eq. (\ref{eq:174}), which comes
from the constraint on the brightness temperature imposed by
induced-Compton scattering, and hence the radiative efficiency (in the
radio band) has an upper limit
\begin{equation}
  \label{eq:203}
  f_{\rm r}\lesssim 6\times 10^{-11}\, \mc{R}^{3/2} \sigma^2 \eps_{\rm e}
  t_{\rm FRB,-3}^{-1} \nu_9^{-2} \Delta \nu_9.
\end{equation}
When the first shell is highly magnetized with $\sigma\gg 1$, we have
$\eps_{\rm e}\lesssim \sigma^{-1}$. We further put in $\nu\sim
6\rm\,GHz$, $\Delta \nu_9\simeq 1$ and
$t_{\rm FRB,-3}\simeq 1$ \citep{2017ATel10675....1G} and 
obtain a stringent constraint on the radiative efficiency $f_{\rm
  r}\lesssim 1.6\times10^{-12}\mc{R}^{3/2} \sigma$. Another constraint comes
from the above mentioned particle-in-cell simulations,
which showed that only a small fraction $\lesssim 10^{-1}$
of the incoming \textit{particles' kinetic energy} can be converted into
coherent EM wave ahead of the main shock, so we also have
$f_{\rm r}\lesssim 10^{-1}\sigma^{-1}$ when $\sigma\gg 1$. Combining
the constraint from induced-Compton scattering with that from
numerical simulations, we obtain $f_{\rm r}\lesssim 4\times10^{-7}
\mc{R}^{3/4}$ (the maximum is reached when $\sigma\simeq
2.5\times10^5 \mc{R}^{-3/4}$). Therefore, as long as the luminosity
ratio $\mc{R}$ between 
two consecutive shells ejected from the central engine is not much
greater than $\sim$$10^{4.5}$, the maser due to bunching in gyration phase at
quasi-perpendicular shocks must have a very low radiative efficiency.

\vspace{1cm}

To summarize the main results of this section, we find that FRB models
based on masers powered by 
internal/external dissipation of the free energy of an outflow suffer
from the following potential inconsistencies: (i) external shock model
cannot reproduce the durations of some closely-separated
FRB pairs; (ii) synchrotron maser in vacuum requires fine-tuned plasma
conditions where the B-field is nearly uniform (to within
an angle $\sim$$\gamma'^{-1}$) and particles' pitch-angle distribution is
narrowly peaked with spread $\lesssim 
\gamma'^{-1}$; (iii) synchrotron maser in plasma requires a
low radiation efficiency $f_{\rm r}\lesssim 5\times 10^{-9}\eps_{\rm
  B}^{-19/16}$ for FRBs detected at high frequencies $\sim$6
GHz; (iv) it is unclear how the population inversion condition
$\partial (N'_{\gamma'}/\gamma'^2)/\partial\gamma' > 0$ is
achieved\footnote{Note that our constraint on the radiation
  efficiency $f_{\rm r}$ is conservative, because the limit on
  the brightness temperature given by induced Compton scattering via
  eq. (\ref{eq:64}) will be stronger for mild population
  inversion with $\partial (N'_{\gamma'}/\gamma'^4)/\partial\gamma' <
  0$.
}; (v) during the maser amplification process, high-energy electrons
radiate faster than low-energy ones, so the population inversion
condition may be quickly destroyed; (vi) the maser due to bunching in
gyration phase at quasi-perpendicular shocks requires a low radiative
efficiency $f_{\rm r}\lesssim 4\times10^{-7}\mc{R}^{3/4}$, where
$\mc{R}$ is the luminosity ratio of the two colliding 
shells.

\section{The Antenna Mechanism}
In \S 4 and \S 5, we have explored many possible
maser mechanisms operating either inside the magnetosphere of a NS or
when a relativistic outflow undergoes internal/external dissipation. We
find that various maser mechanisms (in either vacuum or plasma)
proposed in the literature require unrealistic or extremely fine-tuned
plasma conditions. Thus, we are left with the antenna mechanism, which
will be described in this section. This mechanism requires
coherently moving charge bunches with sizes smaller than the
wavelength of emission $\lambda\sim 30\,$cm. This is only 
possible when there is large-scale ordered B-field lines, so we only
consider the plasma to be inside the magnetosphere of a NS. In the
situation where the dissipation of outflow energy occurs at large
distances from the central object, the B-field in the emitting plasma
is weak, and particles typically gyrate around B-field lines at random gyration
phases instead of forming coherent bunches. In \S 6.1,
we first go through the basic properties of bunches needed to produce FRB
luminosities, following \citet{2017MNRAS.468.2726K}. Then in \S 6.2,
we discuss possible bunch formation channels and show, for the first time,
bunches can form via two-stream instability in the twisted
magnetosphere of a magnetar.

\subsection{Properties of bunches}

Traditionally, coherent curvature radiation by bunches has long been
considered as a possible mechanism to explain pulsar radio
emission \citep{ 1975ApJ...196...51R}, but it suffers from a number of
critiques \citep[e.g.][]{1981IAUS...95..133M}. First, the growth  
time for bunches due to two-stream instability is too
long under the classical two-beam condition\footnote{However, it has
  also been proposed that using  
different particle distribution functions, the instability may grow on
much shorter time-scales \citep{1977ApJ...212..800C,
  1987ApJ...320..333U, 1992A&A...265L..33L}.} \citep[a primary beam
with $\gamma_{\rm 
  b}\sim 10^6$-$10^7$ interacting with a secondary pair plasma with
$\gamma_{\pm}\sim 10^2$-$10^3$, e.g.][]{1977MNRAS.179..189B,
  1999ApJ...521..351M}. Second,  the number of particles
per bunch required by the observed high brightness temperature leads to
too strong Coulomb repulsion and hence bunch dispersion
\citep[e.g.][]{1977ApJ...212..800C}. Third, none of the treatment in
the literature has included the formation of bunches and their coherent
radiation processes simultaneously. On the other hand, it has also
been proposed that charge bunches (solitons) could be produced by
modulational instability in a 
turbulent plasma provided that 
species of different charge signs have different effective mass
\citep[electrons and positrons with different streaming Lorentz factors do
have different mass, and there could also be some mixing of ions,
e.g.][]{1975PhyS...11..271K, 1986PhR...138....1S, 1990MNRAS.247..529A,
  2000ApJ...544.1081M}. In this paper, we do 
not attempt to unify pulsar radio emission with FRB radiation
mechanism, because the properties of the source plasma for FRBs is drastically
different from pulsars (as shown in \S 4.1).

\citet{2017MNRAS.468.2726K} considered coherent curvature emission by
the antenna mechanism in detail. If the local 
curvature radius of the B-field line is $\rho$ and an electron is
moving very close to $c$, the acceleration perpendicular to the
velocity is $c^2/\rho$. To produce curvature radiation at
frequency $\nu = \nu_9\rm\ GHz$, we require 
\begin{equation}
  \label{eq:37}
  \gamma \simeq 60\,\nu_9^{1/3}\rho_{6}^{1/3}.
\end{equation}
At such low Lorentz factors, the single-particle curvature power is
very small
\begin{equation}
  \label{eq:38}
  P_{\rm curv} \simeq {\gamma^4e^2c\over \rho^2}\simeq
  (8.5\times10^{-14})\,\mr{erg\,s^{-1}}\,\nu_9^{4/3} \rho_6^{-2/3}.
\end{equation}
When the particle is moving towards
the observer, the isotropic equivalent luminosity is
\begin{equation}
  \label{eq:39}
  \delta L_{\rm iso} \simeq \gamma^4 P_{\rm curv}\simeq
  (1.1\times10^{-6}\,\mr{erg\,s^{-1}})\,\nu_9^{8/3} \rho_6^{2/3}.
\end{equation}
We see that single-particle curvature radiation is extremely
inefficient. To produce the observed FRB luminosity $L_{\rm
  iso}\sim 10^{43}\rm\,erg\,s^{-1}$, electrons must form bunches and
radiate coherently.

The size of a bunch in the direction parallel to the line of sight
(hereafter {\it longitudinal} direction) must not significantly exceed
$\lambdabar=\lambda/2\pi= 4.8\,\nu_9^{-1}\,$cm. The radiation
formation length is $\rho/\gamma$, which corresponds to a time
$\rho/(\gamma^2c)$ in the 
electrons' comoving frame. Thus, the maximum size allowed by coherence
in the {\it transverse} direction is $\rho/\gamma^2\simeq
\gamma\lambdabar\simeq (2.9\times10^2\mr{\,cm})\,
\nu_9^{-2/3}\rho_6^{1/3}$. Therefore, the  
maximum number of particles in one 
coherent bunch is given by
\begin{equation}
  \label{eq:3}
  N_{\rm coh} \simeq \pi n\gamma^2\lambdabar^3\simeq
  1.3\times10^{24}\,n_{18} \nu_9^{-7/3} \rho_6^{2/3},
\end{equation}
where $n = 10^{18}n_{18}\rm\,cm^{-3}$ is the fiducial number
density in the NS frame. The isotropic equivalent luminosity from
one such bunch is
\begin{equation}
  \label{eq:169}
  L_{\rm iso}^{\rm bunch} \simeq N_{\rm coh}^2 \delta L_{\rm iso}.
\end{equation}
In reality, the transverse size of one local patch of
particles may be a factor\footnote{This factor was denoted as
  $\eta^{1/2}$ in \citet{2017MNRAS.468.2726K}.} of $\eta>1$ 
greater than $\gamma\lambdabar$ but coherence cannot be 
maintained due to causality, and then the total luminosity is the
incoherent sum of the emission from $\eta^2$ bunches
\begin{equation}
  \label{eq:160}
  L_{\rm iso} = \eta^2L_{\rm iso}^{\rm bunch} \simeq
  (1.8\times10^{42}\mr{\,erg\,s^{-1}})\, \eta^2 n_{18}^2
  \nu_9^{-2}\rho_6^2.
\end{equation}
% In case $\eta\gg1$, we can effectively consider the
% entire patch as $\eta^2$ separate coherent bunches radiating
% simultaneously.

Therefore, the observed FRB luminosity 
$L_{\rm iso}\sim 10^{43}\rm\,erg\,s^{-1}$ can be easily achieved by
tuning the parameters $n_{18}$, $\rho_6$ and 
$\eta$. Variations in these parameters can also lead to a large
range of FRB luminosities 
as observed. In the observer's frame, the emission from each bunch
lasts for $\nu^{-1}\sim 1\,$ns and it requires $10^{6}$ such bunches
to produce an FRB with intrinsic duration $\sim 1\,$ms. The beaming
factor of one bunch is $f_{\rm b}\sim \gamma^{-2}$, so the minimum
energy budget per FRB is $\sim\gamma^{-2}L_{\rm iso} t_{\rm
  FRB}\sim 3\times10^{36}\,{\rm erg}\, L_{\rm iso,43} t_{\rm FRB,-3}$,
for the Lorentz factor in eq. (\ref{eq:37}) and
  $\nu_9^{1/3}\rho_6^{1/3}\sim 1$.

In a man-made antenna, the kinetic energy of particles is
minuscule and the radiating power is supplied by the {\it ac}
electromotive force. Similarly, in the astrophysical antenna model of FRBs, the power
comes from a parallel E-field $E_{\para}$ which sustains the
Lorentz factor of electrons as required by eq. (\ref{eq:37}). Since
each electron is losing energy at a rate $N_{\rm coh}P_{\rm curv}$, we
obtain from energy conservation\footnote{Another way of thinking is
  that the radiation backreaction force operates on all particles
  within the coherent bunch, because the EM fields are nearly uniform 
  within the coherent region. Each particle experiences 
  a backreaction force due to all other radiating particles and this
  force must be balanced by the force from an external E-field
  $E_{\para}$.} $E_{\para}ec \simeq N_{\rm coh}P_{\rm curv}$, i.e.
\begin{equation}
  \label{eq:44}
  E_{\para}\simeq {\gamma^4e N_{\rm coh} \over \rho^2} \simeq
  (7.5\times10^{9}\mr{\,esu})\,n_{18}\nu_9^{-1}.
\end{equation}
% The bunch is made of particles of the same charge sign
% and hence the Coulomb field in the longitudinal direction is
% \begin{equation}
%   \label{eq:161}
%   E_{\rm Coul} \simeq {2\pi e N_{\rm coh} \over \pi \gamma^2
%     \lambdabar^2} \sim E_{\para}.
% \end{equation}
% Thus, a parallel E-field is also needed to prevent the charge
% bunches from dispersing or mixing with other bunches of the opposite
% charge sign. Bunch formation and survival will be discussed in the
% next subsection.

Such a strong, large-scale (length $\gtrsim
\rho/\gamma$) E-field may come from magnetic reconnection at small 
inclination angles in the magnetosphere
\citep{2017MNRAS.468.2726K}. We also note that the 
isotropic luminosity (eq. \ref{eq:160}) can be expressed in terms
of the parallel E-field $E_{\para} = 10^{10}E_{\para,10}\,$esu
\begin{equation}
  \label{eq:80}
  L_{\rm iso} \simeq \eta^2 E_{\para}^2\rho^2c \simeq
  (3.0\times10^{42}\mr{\,erg\, s^{-1}})\, \eta^2 E_{\para,10}^2\rho_6^2.
\end{equation}
The flow of the radiating particles (and counter-streaming particles
of the opposite charge sign) along the primary B-field leads to
a current $j = 2nce$ within a cylinder of radius $\eta \gamma\lambdabar$
(transverse length of the coherent bunch), which induces a strong
B-field in the transverse direction
\begin{equation}
  \label{eq:45}
  B_{\rm ind}\simeq {4\pi j\over c} {\eta \gamma \lambdabar\over 2} \simeq
  (4.0\times10^{12}\mr{\, G})\, L_{\rm iso,43}^{1/2} \nu_9^{1/3}
  \rho_6^{-2/3}. 
\end{equation}
This induced B-field leads to a torsion on the primary
B-field. Considering the fact that particles are locked in the lowest
Landau level and only move along local B-field lines, it is crucial
that the combined primary plus induced B-field still point towards
the observer within an angle of $\sim\gamma^{-1}$. This sets a lower
limit on the strength of the primary B-field
\begin{equation}
  \label{eq:47}
  \begin{split}
    B\gtrsim \gamma B_{\rm ind} &\simeq (2.4\times10^{14}\mr{\,G})\,
    L_{\rm iso,43}^{1/2} \nu_9^{2/3}\rho_6^{-1/3}.
  \end{split}
\end{equation}
This lower limit is stronger than the one from the energy
requirement in eq. (\ref{eq:58}), meaning the latter can be easily
satisfied.

% Pair production may be required during the magnetic reconnection if
% the plasma becomes charge starved. If $\theta_{\rm B}$ is the
% inclination angle between the B-fields of the two plasmas undergoing
% reconnection, charge-starvation corresponds to
% \begin{equation}
%   \label{eq:178}
%   B\sin\theta_{\rm B}/(\gamma \lambdabar) \gtrsim 4\pi n e.
% \end{equation}
% If the plasma inflow velocity is $v_{\rm in}$, the parallel E-field in
% the current sheet is $E_{\para} = B\sin\theta_{\rm B}v_{\rm in}/c$, and
% charge-starvation (eq. \ref{eq:178}) means $v_{\rm in}/c\lesssim E_{\para}/(4\pi
% \gamma \lambdabar n e) \simeq 5.8\times10^{-3} E_{\para,10} \nu_9^{2/3}
% \rho_6^{-1/3} n_{18}^{-1}$. Thus, for typical inflow velocity $v_{\rm
%   in}/c\sim 0.1$, there may not be pair production in the reconnection
% process.

\subsection{Formation of bunches and high-frequency FRB analogs} 

In the model of \citet{2017MNRAS.468.2726K}, it was shown that bunches
may form due to two-stream instability in the situation of counter-streaming 
electrons and positrons with Lorentz factor given by eq. (\ref{eq:37})
and density $n\gtrsim10^{17}\mr{\,cm^{-3}}$. However, since the
formation of bunches must occur simultaneously with the coherent
curvature radiation in their model, proper treatment of radiation
backreaction is needed. Moreover, the bunch formation length
$\sim\gamma^2\lambdabar$ is on the same order as the radiation formation
length $\sim\rho/\gamma$. It remains unclear whether bunches can form 
spontaneously (and survive the radiation formation process) during
magnetic reconnection in a plasma of quasi-uniform density.

Another possible way of bunching is a radiative instability proposed by
\citet{1971ApJ...170..463G}, where an initial uniform distribution of
particles moving along a thin circular ring spontaneously develops
bunches due to the backreaction of curvature radiation. However, since
the two non-linear processes --- the formation of bunches and coherent
emission by bunches --- were not treated together in a self-consistent
way, it is currently unclear whether radiation backreaction acts to
increase or decrease bunching \citep{1978ApJ...225..557M}.

In this paper, we propose a new mechanism for bunch formation. In
Appendix A, we show that, the plasma in
the twisted magnetosphere of a magnetar\footnote{Strong surface B-field
  is required by two independent arguments from (i) the energetics of
  the repeater FRB 121102 (eq. \ref{eq:58}), (ii) emission coherence
  (eq. \ref{eq:47}). These arguments imply that the progenitor is a
  magnetar rather than a normal radio pulsar.} is turbulent and
clumpy due to 
two-stream instability. When magnetic reconnection occurs, the
pre-existing density clumps may provide charge bunches for the antenna
mechanism to operate.
% In this section, we show that bunches of certain
% length-scales can survive the reconnection process and the curvature radiation
% formation process.

% In the following, we propose, for the first time, that bunches may
% already be in place before the reconnection occurs if there is significant twist
% in the NS magnetosphere before the reconnection trigger. \note{We also show that
% bunches of sizes exceeding a certain lower limit can survive the
% FRB radiation formation process.}

Guided by \S6.1, we use a fiducial number density $n =10^{18}
n_{18}\,\mr{cm^{-3}}$ and a fiducial B-field $B = 10^{14}B_{14}\,$G in
the following. This number density exceeds the G-J density
[$n_{\rm   GJ}=B/(ecP)$] by a factor of $\mc{M}\simeq 10^4n_{18}
B_{14}^{-1}P_{-1}$.
% and the minimum density required to sustain the
% magnetic twist induced currents $n_{\rm min} = |\nabla\times \b{B}|/(4\pi e)$
% by a factor of $\sim 10^2$. 
The twist-induced currents are supported by counter-streaming
particles of opposite charges from pair creation avalanches. In the
traditional model of pair creation above the polar cap of a rotating
NS \citep{1971ApJ...164..529S, 
  1975ApJ...196...51R}, $\gamma$-rays are produced by curvature
radiation of primary particles with Lorentz factor exceeding
$\sim10^7$. In the magnetar model, the creation of $\gamma$-rays is
mainly due to resonant scattering of ambient X-ray photons by initial
electrons pulled from the NS surface
\citep[e.g.][]{2007ApJ...657..967B}. The resonant scattering condition
is $\gamma_{\rm res}\eps_{\rm x} \simeq (B/B_{\rm QED})\me c^2$
(eq. \ref{eq:102}), i.e.
\begin{equation}
  \label{eq:78}
  \gamma_{\rm res}\simeq 1.2\times10^2\, B_{14} (\eps_{\rm
    x}/10\mr{\,keV})^{-1}.
\end{equation}
The scattered photons
have energy $\eps_{\rm IC}/\me c^2 \simeq 1.2\times10^2 B_{14}
(\eps_{\rm x}/10\mr{\,keV})^{-1}$ (eq. \ref{eq:103}) in the NS
frame. When $B\gtrsim B_{\rm QED}$, these $\gamma$-rays convert into
pairs within a propagation length $\rho\me c^2/\eps_{\rm IC}\simeq
(10^4\mr{\,cm})\, \rho_{6} B_{14}^{-1}(\eps_{\rm x}/10\mr{\,keV})$ when the pitch
angle becomes $\lesssim \me c^2/\eps_{\rm IC}$.

In the counter-streaming electron-positron plasma\footnote{
Note that the curvature antenna model may
not require a large amount of particle injection from an explosive
pair production process, so the pre-existing plasma could be made of
electrons and protons (instead of electrons and
positrons). Charge-to-mass ratio does
not make a difference in the curvature
radiation process, but the two-stream instability behaves
differently. It can be shown that our discussion is qualitatively
correct for an electron-proton plasma, as long as the parameter
$\gamma_{\rm c}$ in eq. (\ref{eq:162}) is redefined to include the
proton-to-electron mass ratio.}, two-stream 
instability naturally leads to density fluctuations with a broad
spatial power spectrum at wavenumbers smaller than a critical value
$k<k_{\rm max}\equiv \omega_{\rm p,eff}/c$, and the effective plasma
frequency $\omega_{\rm p,eff}$ depends on the distribution
function. In Appendix A, we show that, for a homogeneous
counter-streaming electron-positron plasma, the effective plasma
frequency is
\begin{equation}
  \label{eq:162}
  \begin{split}
      \omega_{\rm p,eff} &= \left( {4\pi e^2 n \over \gamma_{\rm c} m_{\rm
  e}}\right)^{1/2}\simeq (1.8\times10^{12}\mr{\,s^{-1}})\,
\left({ n_{18}\over \gamma_{\rm c,3}}\right)^{1/2},\\
\gamma_{\rm c}^{-1} &\equiv \langle\gamma^{-3}\rangle = \int_{-\infty}^{+\infty}
\gamma^{-3}f(u)\d u,
  \end{split}
\end{equation}
where $f(u)$ is the normalized one-dimensional distribution function of
particles' 4-velocities $u$. Thus, large density fluctuations with $\Delta
n/n\sim 1$ can develop on length-scales
longer than the effective skin depth
\begin{equation}
  \label{eq:65}
  \ell_{\rm skin}= c/\omega_{\rm p,eff}\simeq (1.7\times10^{-2}\mr{\,cm})\,
  \gamma_{c,3}^{1/2} n_{18}^{-1/2}.
\end{equation}
The distribution function of the counter-streaming pair 
plasma in the twisted magnetosphere is highly uncertain (depending on
$\b{B}$, $\rho$, $\eps_{\rm x}$, etc.), and hence $\gamma_{\rm c}$
could range from $\sim \langle \gamma\rangle$ (near Maxwellian
distribution) to $\sim \langle \gamma^3\rangle$ (near mono-energetic
distribution). Since typical Lorentz factors $\gamma<\gamma_{\rm
  res}\sim 10^2$ (eq. \ref{eq:78}), hereafter we take take $\gamma_{\rm c} =
10^3\gamma_{\rm c,3}$ as our fiducial parameter.
More work is needed to determine this quantity from the
distribution function and composition of the plasma in the
pre-existing twisted magnetosphere. As long as $\ell_{\rm
  skin}<\lambdabar = 4.8\nu_9^{-1}\mr{\,cm}$, fractional density
fluctuations of $\Delta n/n\sim 1$ over lengthscales
$\sim\lambdabar$ can be produced and our discussion is not
affected.

According to Appendix A, the growth rate of two-stream instability
scales linearly with wave number $k$ up
to $k\sim k_{\rm max}=\ell_{\rm skin}^{-1}$, but growth is impossible at $k>k_{\rm
max}$. Therefore, FRB analogs should exist at frequencies up to
$\omega_{\rm eff}/2\pi\simeq
(2.8\times10^{11}\mr{\,Hz})\,n_{18}^{1/2}\gamma_{\rm c,3}^{-1/2}$ (or
wavelength $\lambda\sim 1\,$mm). We
also note that both $n$ and $\gamma_{\rm c}\equiv
\langle\gamma^{-3}\rangle^{-1}$ have large  
uncertainties due to the unknown particle distribution function in the
magnetosphere of a magnetar. Since the
pair annihilation mean free path must be longer than the curvature
radius, i.e. $\gamma_{\rm
  c}/(\sigma_{\rm T}n)\sim (10^9\,\mr{cm})\, \gamma_{\rm
  c,3}n_{18}^{-1}\gtrsim (10^6\mr{\,cm})\,\rho_6$, i.e. $n_{18}^{1/2}\gamma_{\rm
  c,3}^{-1/2}\lesssim 10^{1.5}\rho_6^{-1/2}$, some very special
conditions may in principle allow FRB analogs at frequencies up to
$\sim$$10^{13}\,$Hz (wavelength $\sim$$30\,\mr{\mu m}$). We
encourage searching for millisecond transients in the mm up to far
infrared wavelengths.

In the following, we discuss the rate and luminosities of
FRB analogs at different frequencies in the curvature antenna
framework. We consider a charge bunch with longitudinal length $\ell_{\para} 
\gtrsim \ell_{\rm skin}$ (eq. \ref{eq:65}) and transverse length
$\ell_{\perp}$, which is moving with Lorentz factor
$\gamma$. To maintain coherence within the radiation formation time,
the longitudinal length is limited by the emission 
wavelength $\ell_{\para}\lesssim \lambdabar$. Note that, if particles
in the bunch have Lorentz factor 
spread $\Delta \gamma/\gamma\sim 1$, the bunch will disperse
after traveling a distance $\sim \gamma^2\ell_{\para}$. In this case,
to avoid bunch dispersion within the curvature radiation formation length
$\rho/\gamma$, we have $\ell_{\para}\gtrsim \lambdabar \simeq
\rho/\gamma^3$. However, smaller $\ell_{\para}<\lambdabar$ is allowed
if $\Delta \gamma/\gamma< 1$. In the following, we keep $\ell_{\para}$
as a free parameter within the range (0, $\sim$$\lambdabar$). Similar to
the situation in man-made antennas, the  
inertia of the charge carriers is minuscule and particles' motion is
controlled by the balance between power input (from a parallel
E-field) and output (due to radiation), i.e.
\begin{equation}
  \label{eq:194}
  \begin{split}
      E_{\para} e &\simeq \pi n\ell_{\para} \mr{min}[
  (\gamma\lambdabar)^2, \ell_{\perp}^2]\, P_{\rm curv}/c \\
&\simeq \pi n \ell_{\para} e^2 \mr{min}[1,
(\ell_{\perp}/\gamma\lambdabar)^2]
  \end{split}
\end{equation}
If the reconnecting B-field lines have inflow speed $\beta_{\rm in}c$
and inclination angle $\theta_{\rm B}$, then the parallel E-field is
roughly given by $E_{\para} \simeq B\sin\theta_{\rm B}\beta_{\rm
  in}$. Thus, we obtain
\begin{equation}
  \label{eq:191}
  B\sin\theta_{\rm B} \simeq \pi n\ell_{\para} e\beta_{\rm
    in}^{-1}\, \mr{min}[1, (\ell_{\perp}/\gamma\lambdabar)^2].
\end{equation}
The transverse size of the bunch $\ell_{\perp}$ is unknown. A natural
length-scale of the system is the thickness 
of the current sheet $\ell_{\rm cs}$ given by $\nabla\times B \simeq
B\sin\theta_{\rm B}/\ell_{\rm cs}\simeq 4\pi n e$, i.e.
\begin{equation}
  \label{eq:190}
  \ell_{\rm cs}\simeq B\sin\theta_{\rm B}/(4\pi n e)\simeq
  (1.7\times10^2\,\mr{cm})\, B_{14} \theta_{\rm B,-2} n_{18}^{-1}.
\end{equation}
Note that we have ignored the displacement current term, because
$c|\nabla\times \b{B}|(\partial E_\para/\partial 
t)^{-1} \sim B\sin\theta_{\rm B}/(\beta_{\rm in} E_{\para})\sim
\beta_{\rm in}^{-2}\gg 1$.
% We plug eq. (\ref{eq:191}) into eq. (\ref{eq:190}) and
% express the thickness as $\ell_{\rm cs}\sim
% \ell_{\para}/(4\beta_{\rm in})$. The physics of magnetic
% reconnection in the NS magnetosphere (with magnetization
% $\sim10^{15}$) has not been well studied, so the inflow speed is
% highly uncertain. For bunches of longitudinal sizes $\ell_{\para}\sim
% \lambdabar$, the inflow time across the current sheet should not
% exceed the duration of FRBs, so we require
% \begin{equation}
%   \label{eq:193}
%   \lambdabar/(4\beta_{\rm in}^2 c) \lesssim t_{\rm FRB}
%   \Longrightarrow \beta_{\rm in}\gtrsim 2\times10^{-4} t_{\rm
%     FRB,-3}^{-1/2} \nu_9^{-1/2}.
% \end{equation}
% Since $\gamma\lambdabar/\ell_{\rm cs}/ \sim 4\beta_{\rm
% in} \gamma \sim 0.24\beta_{\rm in, -3}\nu_9^{1/3}\rho_6^{1/3}$, it is
% unclear whether $\ell_{\rm cs}$ is greater or smaller than the
% allowed coherent length $\gamma\lambdabar$ in the transverse
% direction.
In the following, we take $\ell_{\perp}\sim \ell_{\rm cs}$ as given by
the magnetic reconnection physics (according to
eq. \ref{eq:190}). When $\ell_{\rm cs}\gtrsim \gamma\lambdabar$, which
applies to bursts at frequency $\nu\gtrsim (0.8\mr{\,GHz})\,
\rho_6^{1/2}\ell_{\rm  cs,2.5}^{-3/2}$, we make use of eqs. (\ref{eq:191}) and
(\ref{eq:190}) to eliminate $B\sin\theta_{\rm B}$ and obtain
$\beta_{\rm in}\simeq \ell_{\para}/4\ell_{\rm cs}\lesssim \ell_{\para}/(4\gamma
\lambdabar)$. The FRB luminosity should be dominated by bunches with
maximum longitudinal length allowed by coherence $\ell_{\para}\sim
\lambdabar$, so we find that the inflow speed of magnetic
reconnection is small
\begin{equation}
  \label{eq:200}
  \beta_{\rm in}\lesssim
(4\gamma)^{-1}\simeq4\times 10^{-3} \nu_9^{-1/3}\rho_6^{-1/3}.
\end{equation}
The isotropic equivalent luminosity
from the entire clump is given by the incoherent sum of $(\ell_{\rm
  cs}/\gamma\lambdabar)^2$ coherently 
radiating bunches of transverse area $\pi (\gamma\lambdabar)^2$, i.e. 
\begin{equation}
  \label{eq:195}
  \begin{split}
  &    L_{\rm iso}^{(1)} \simeq (\pi n\gamma^2\lambdabar^3)^2 P_{\rm curv}\gamma^4
  \left({\ell_{\rm cs}\over \gamma\lambdabar}\right)^2
\simeq {\pi^2 n^2 \rho^2\ell_{\rm cs}^2 e^2 c \over \gamma^2}\\
& \simeq (1.9\times10^{42}\mr{\,erg/s})\, n_{18}^2
\rho_{6}^{4/3}\ell_{\rm cs,2.5}^2 \nu_9^{-2/3} \mbox{, for $\ell_{\rm
    cs}\gtrsim \gamma\lambdabar$.}
  \end{split}
\end{equation}
On the other hand, when $\ell_{\rm cs}\lesssim \gamma\lambdabar$ (for
bursts at frequency $\nu\lesssim (0.8\mr{\,GHz})\, \rho_6^{1/2}\ell_{\rm 
 cs,2.5}^{-3/2}$), we again make use of eqs. (\ref{eq:191}) and
(\ref{eq:190}) and obtain $\beta_{\rm in}\simeq
\ell_{\para}\ell_{\rm cs}/(4\gamma^2 \lambdabar^2)\lesssim
\ell_{\para}/(4\gamma\lambdabar)$. Note that the constraint on the inflow
speed is the same as in eq. (\ref{eq:200}) for the case $\ell_{\rm
  cs}\gtrsim \gamma \lambdabar$. The isotropic luminosity
is given by the coherent radiation by all particles in the clump, i.e.
\begin{equation}
  \label{eq:199}
    \begin{split}
     & L_{\rm iso}^{(2)} \simeq (\pi n\lambdabar\ell_{\rm cs}^2)^2
      P_{\rm curv}\gamma^4 
\simeq{\pi^2 n^2 \gamma^2\ell_{\rm cs}^4 e^2 c}\\
&\simeq (2.5\times10^{42}\mr{\,erg/s})\, n_{18}^2 \rho_6^{2/3}
\ell_{\rm cs,2.5}^4 \nu_9^{2/3} \mbox{, for $\ell_{\rm
    cs}\lesssim \gamma\lambdabar$.}
  \end{split}
\end{equation}
Due to differences among reconnection events --- variations of particle
density $n$, curvature radius $\rho$, and current sheet
thicknesses $\ell_{\rm cs}$ --- we expect a wide range of FRB
luminosities. It is currently not possible to predict FRB
luminosities at a given frequency, due to the unknown plasma
conditions and magnetic reconnection physics in the magnetosphere of  
the NS. However, FRB analogs at higher frequencies are expected to
have lower burst rate due to narrower beaming angle ($\gamma^{-2}$),
i.e. the burst rate should decline with frequency as $f_{\rm
  b}\sim\gamma^{-2}\propto \nu^{-2/3}$. Moreover, we can see from
eq. (\ref{eq:195}) that FRB analogs at frequencies much higher than
$\sim$GHz are most likely dimmer, because the coherent volume
decreases faster than the luminosity gain from stronger beaming.

According to the antenna curvature model described here, the durations
of FRBs are controlled by the physics of magnetic
reconnection. 
% In fact, reconnection physics 
% in high energy density radiative plasmas, where radiative resistivity
% plays an important role, has been largely unexplored
% \citep{2016ASSL..427..473U}. In the FRB context, the radiative
% resistivity
% $E_{\para}/J\simeq \mr{min}[1,
% (\ell_{\perp}/\gamma\lambdabar)^2]\ell_{\para}/c$ (from
% eq. \ref{eq:194}) 
% depends on the level of bunching in the plasma. 
For quasi-uniform density distribution, the radiative resistivity is
negligible (incoherent curvature emission is
inefficient), and magnetic reconnection cannot proceed unless there is
another mechanism that can provide a much higher resistivity.
Charge bunches may flow into the current sheet in the longitudinal and
transverse directions. The former case has characteristic timescale $\pi
\rho/c\sim 100\rho_6\rm\,\mu s$. The
characteristic timescale in the transverse direction is $\ell_{\rm
  cs}/\beta_{\rm in}c\sim (10\mr{\,\mu s})\, \ell_{\rm
  cs,2.5}\beta_{\rm in,-3}^{-1}$. Both of these timescales are
much shorter than the typical duration of FRBs $t_{\rm FRB}\sim
1\,$ms. Therefore, the reconnection process may be unsteady and hence
FRBs may be made of multiple sub-bursts (each lasting for
$\sim$$10$-$100\rm\,\mu s$). 

The total FRB duration corresponds to the time
over which the accumulated stress in the reconnection region is
released. From eq. (\ref{eq:200}), we see that the reconnection inflow
speed is much smaller than the Alv{\'e}n speed ($\approx$$c$) in the
magnetosphere. This is because the energy inflow rate is limited by
the energy outflow rate in the form of coherent radiation. Thus,
individual FRBs do not require global reconnection on lengthscales of
the NS radius. The {\it maximum} size of the reconnection region in
the transverse direction is $t_{\rm FRB}\beta_{\rm in}c\sim (3\times
10^{4}\mr{\,cm})\, t_{\rm FRB,-3} \beta_{\rm in,-3}$. Without a
detailed model for the magnetic configuration and activity\footnote{ 
One possibility mentioned by \citet{2017MNRAS.468.2726K} is that the
buried magnetic flux emerges out of the NS surface and reconnects with
the pre-existing magnetospheric B-fields. This process occurs on a
time-scale $\Delta R/v_{\rm A}\sim (1\mr{\,ms})\, \Delta R_4
B_{14}^{-1} \rho_{0,13}^{1/2}$, where $\Delta R = 10^4\Delta
R_{4}\rm\,cm$ is the depth from which the flux emerges, $\rho_0 =
10^{13}\rho_{0,13}\,\rm g\,cm^{-3}$ is the mass density of the surface
layer, and $v_{\rm A} = B/(4\pi \rho_0)^{1/2}$ is the Alfv{\'e}n speed.
} near the surface of the magnetar, it is currently not possible to
predict FRB durations from first principles.

The typical time resolution of current FRB observations is $\sim$1~ms, which
is limited by the signal-to-noise ratio and intra-band dispersion (the
latter can be eliminated by coherent de-dispersion). Future
observations of brighter bursts or by   
more sensitive telescopes may be able to resolve the sub-burst structures
and provide valuable information on the reconnection physics.

To summarize the main results of this section, we find that the curvature
antenna model can reproduce the basic properties of FRBs provided that the B-field
strength of the NS is stronger than $\sim$$10^{14}\,$G and that bunches
with longitudinal size $\ell_{\para}\lesssim \lambdabar$ can form. We
propose that two-stream instability in the twisted magnetosphere of magnetars
(with persistent currents) provides a broad spectrum of density fluctuations at
length-scales larger than the plasma skin depth
(eq. \ref{eq:65}). Then, the coherent emission by charge clumps is
sustained by a strong E-field, which is produced by magnetic
reconnection. A prediction of this model is that,
since the initial two-stream instability leads to density
fluctuations on all length-scales $\gtrsim c/\omega_{\rm p,eff}$, FRB
analogs should exist at frequencies much higher than $\sim$GHz, up to
mm (or even far-infrared) wavelengths. We have also shown that FRB
analogs at frequencies $\gg\,$GHz are most likely dimmer (due to smaller
coherent volume) and that their rate may be lower (due to smaller
beaming angle).

\section{Comparison between FRBs and pulsar radio emission}

In this section, we briefly discuss some of the differences between
the mechanisms of FRBs and pulsar radio emission.

After many decades of debate, there is still no compelling answer
to the mechanism of pulsar radio emission. The basic reason behind the
debate is that only a tiny fraction ($\lesssim10^{-6}$ for the Crab)
of the pulsar's energy loss (via electromagnetic spin-down) goes to
radio photons, and therefore many maser or collective plasma emission
mechanisms are viable \citep{2016JPlPh..82c6302E, 2017RvMPP...1....5M}. 

However, constraining the radiation mechanism for FRBs may be much
easier than for pulsar radio emission. This is because, as explained
in the following, the total energy release in a FRB event may not be
much larger than the energy coming out in the radio band. In other
words, the radiation in the radio band is likely the dominant channel
of energy release in these transients. We have shown in previous sections
that the much higher isotropic luminosities of FRB compared to radio
pulsars (by more than 10 orders of magnitude) severely constrains the
radiation mechanism for FRBs. We have shown in \S 4.1 
and \S 5 that FRBs cannot be powered by the rotational energy of
NSs or internal/external dissipation of the free energy of
relativistic outflows from BHs or NSs. This leads us to conclude that FRBs 
are most likely produced by the dissipation of magnetic energy near the surface of
NSs. If the progenitor of FRB 121102 stays active for
$\gtrsim30\,$yr, then the cumulative energy output in the radio
band is $E_{\rm tot}\gtrsim10^{44}(f_{\rm b,tot}/f_{\rm r})\,$erg
(eq. \ref{eq:61}), where $f_{\rm b,tot}$ is the combined solid angle
of the radiation cones\footnote{If the radiation cones are not
    concentrated near the spin axis, we have $f_{\rm b,tot}\sim \Delta
    \theta$, where $\Delta \theta$ is the range of polar angles (with
    respect to the spin axis) occupied by the radiation cones of all
    bursts. The magnetic axis may be tilted 
    with respect to the spin axis of the progenitor at a large
    angle. If the emission is from the polar cap regions at the magnetic
    poles (as in pulsars), we have $f_{\rm b,tot}\sim \Delta\theta\sim
    (R_*/R_{\rm LC})^{1/2} \simeq 4.6\times10^{-2}P_{-1}^{-1/2}$. On
    the other hand, if the emission region is not concentrated near the
    magnetic poles (as implied by the non-detection of periodicity
    from FRB 121102), then $\Delta \theta$ is a fairly large angle,
    which means $f_{\rm b, tot}$ may be of order unity.
} of all bursts divided by $4\pi$ and 
$f_{\rm r}$ is the radio emission efficiency for each burst. Although these two
factors are not well constrained by observations, they tend to cancel
each other, and hence $E_{\rm tot}$ may be a significant fraction of
the total magnetic energy in the magnetosphere of a magnetar,
$\sim10^{45}B_{*,14}^2\,$erg.

When we consider collective plasma
emission (or plasma maser) within the magnetosphere of NSs, the
requirement that the magnetic energy density is much greater than
particles' kinetic energy density and the energy density of the FRB EM
waves provides a stringent lower limit on the
B-field strength in the emitting region (eq. \ref{eq:131}), and hence
the emission is generated close to the NS surface
(eq. \ref{eq:129}). The plasma
frequency must be much lower than the cyclotron frequency so that 
the B-field is strong enough to confine the emitting
plasma. Based on these constraints, it can be shown that various
beam instabilities proposed in the pulsar literature either do not
grow or have too small growth rates at $\sim$GHz frequencies, which
are many orders of magnitude lower than the plasma frequency and
cyclotron frequency. By the process of elimination, we arrive at a
unique solution for the FRB radiation mechanism --- the coherent curvature
emission. For pulsar radio emission, on the other hand,
whose isotropic equivalent luminosity is lower than typical FRB luminosity a
factor $\sim$$10^{10}$, these constraints on the plasma frequency and
cyclotron frequency become so weak that almost
any emission radius within the light cylinder is viable and many
beam instabilities may grow efficiently at $\sim$GHz frequencies.

We also note that the antenna mechanism for FRBs described in this
paper is significantly different from the antenna mechanism considered
in the radio pulsar literature. In the latter case, people have invoked a
fast primary beam colliding with a slower secondary plasma
\citep{1975ApJ...196...51R}. It is suggested that particle 
bunches form due to two-stream instability and 
then produce coherent curvature radiation. One of the major drawbacks with
this proposal is that the density contrast between bunches and
inter-bunch medium is small due to Coulomb repulsion within the bunch,
which severely limits the ability of this process to explain radio pulsar
emission \citep{1981IAUS...95..133M}.

The bunch formation process for FRBs we have proposed is a two-step
process. Step one is formation of roughly charge-neutral clumps
in the counter-streaming plasmas associated with strong current in the
twisted magnetosphere of a magnetar. And the second step is charge
separation of neutral clumps by the strong E-field inside the current
sheet associated with the magnetic reconnection process. This two-step
process avoids the well-known problem that Coulomb repulsion prevents
charge clumping. We also note that this two-step bunch formation
process may only work for transients like FRBs but not for regular
pulsars.

\section{summary}

In this paper, we have described the constraints on possible radiation
mechanisms for FRBs.

The extremely high brightness temperatures ($\gtrsim10^{35}\,$K) of FRBs
require that the EM fields radiated by individual
particles add up coherently. There are generally two classes of
such processes: maser and the antenna mechanism. We
consider collective plasma emission as a special type of maser
(named plasma maser). We use the observational properties of the
repeater FRB 121102 and general physical considerations to constrain
the plasma conditions 
needed for each of the coherent processes. We find that various maser
mechanisms require extremely fine-tuned plasma conditions or
unphysical parameters as summarized below; only the antenna curvature
mechanism operating near the surface of a magnetar is consistent with
the high isotropic luminosity of the repeater FRB 121102.

Plasma masers in the magnetosphere
of NSs can in principle operate when a beam of particles runs into
a target plasma and subluminal waves are excited and amplified due to
beam instabilities. However, we find that the cyclotron-Cherenkov (or
anomalous Doppler) resonance condition cannot be satisfied, because
the B-field must be strong enough to confine the motion of plasma
whose kinetic energy density must be at least as high as the FRB EM
waves. The Cherenkov resonance condition can be satisfied, but the
growth rate of the instability is too slow to be important for FRBs.

Vacuum curvature maser is possible only if the curved B-field is not
confined in a plane. This can be achieved when the B-field lines have
significant torsion, which could be caused by crustal motions at the
NS surface. However, the absorption cross-section is negative only
for O-mode waves propagating at an angle $\psi\lesssim \gamma^{-2}$ wrt
the momentum vector of the emitting particle. The absorption
cross-section becomes positive at larger angles, and for $\psi\sim
\gamma^{-1}$ the cross-section is larger by at least a  factor $10^5$
compared with the peak cross-section for wave amplification at $\psi
\sim \gamma^{-2}$. Due to the B-field curvature, photon
trajectories will unavoidably intersect with other nearby B-field lines
at larger angles, and then strong positive curvature
self-absorption will prevent the radiation from escaping.

Vacuum synchrotron maser is possible only
when the B-field is nearly uniform to within an angle
$\lesssim\gamma^{-1}$ and particles' pitch-angle distribution is
narrowly peaked with spread $\Delta \alpha\lesssim\gamma^{-1}$, where
$\gamma$ is the typical Lorentz factor of radiating particles in
the comoving frame of the source plasma. In order for the electron
cyclotron frequency to be in the radio band, the radiating plasma must
be at a large distance $\gtrsim 10^{14}\,$cm from the central
object. It is highly unlikely that the fine-tuned plasma conditions above
can be realized during internal/external dissipations of an outflow at
such large distances (much beyond the light cylinder of 
a NS).

For synchrotron maser near the effective Razin frequency of a 
relativistic plasma (due to internal/external dissipations of a
relativistic outflow), the brightness temperature is limited by
induced Compton scattering. To 
produce the observed brightness temperature $\gtrsim$$10^{35}\,$K, 
the radiation efficiency must be extremely low $f_{\rm r}\lesssim
5\times10^{-9}\eps_{\rm B}^{-19/16}$, where $\eps_{\rm B}$ is the fraction
of energy density in B-fields in the emitting region. Moreover, it is
unclear how the population inversion condition $\partial
(N_{\gamma}/\gamma^2)/\partial\gamma > 0$ can be 
achieved {\it and maintained} when electrons lose energy to radiation
(high-energy electrons radiate energy at a higher rate than low-energy ones and the 
population inversion is quickly destroyed unless it is actively
maintained by some unknown process). Bunching in gyration
  phase may occur due to coherent reflection of incoming particles at
  quasi-perpendicular shocks with high magnetization, and hence maser 
  emission may be produced when two consecutive shells ejected from
  the central engine collide. However, we show that the brightness
  temperature for such a maser mechanism is also limited by
  induced-Compton scattering. The observed FRB brightness temperature
  requires an extremely low radiation efficiency $f_{\rm r}\lesssim
  4\times10^{-7}\mc{R}^{3/4}$, where $\mc{R}$ is the luminosity ratio
  of the two colliding shells.

We find that the basic properties of FRBs are consistent with the antenna 
mechanism, where charge bunches with longitudinal sizes $\lesssim
\lambda$ move along the curved B-field lines with $B\gtrsim
10^{14}\,$G and produce coherent curvature emission. Similar to the
situation in man-made antennas, the kinetic energy of the radiating
particles is minuscule and the radiative power is supplied by an
E-field, which is produced by magnetic reconnection.
We find that bunches can form via two-stream instability in the twisted
magnetosphere of magnetars {\it before} the magnetic reconnection is
triggered. Electric currents flow along the strong B-field lines of a
magnetar whenever the field lines are twisted by crustal
motions. These currents are carried by counter-streaming electrons and
positrons. We have shown that two-stream instability leads to density
fluctuations on length-scales longer than the effective plasma skin
depth. A prediction of this curvature antenna model is that FRB
analogs should exist at frequencies much higher than $\sim$GHz, up to
mm or even far-infrared 
wavelengths (depending on the plasma distribution function). FRB
analogs at higher frequencies are  expected to be dimmer (due to
smaller coherent volume) and have a lower occurring rate (due to
smaller beaming angle).

The analysis presented in this paper, and the identification of the
most likely radiation mechanism for FRB 121102 (the curvature antenna 
mechanism) should apply to all those FRBs that have multiple outbursts
like FRB 121102.

Based on the calculations presented in this paper, the antenna
mechanism seems to be the most promising candidate for the radiation
process in FRBs. However, there are a number of technical issues that
require closer scrutiny and further study: (1) pair creation and
plasma distribution function in the twisted magnetosphere of
magnetars; (2) the physics of radiative
magnetic reconnection in the magnetosphere
of magnetars; (3) propagation of large amplitude radio waves (with
non-linearity parameter $a_0\gg 1$) through the magnetosphere of
magnetars.

With forthcoming telescopes such as
UTMOST \citep{2017MNRAS.468.3746C}, Apertif
\citep{2017arXiv170906104M}, CHIME \citep{2017arXiv170204728N}
and SKA \citep{2011PASA...28..299C}, the FRB sample is expected to grow
by 2-3 orders of magnitude. Some of these wide field-of-view
telescopes will be able to monitor 10-$10^2$ FRBs simultaneously and
hence will be better at finding repeaters (which can then be
localized). Observational search for analogs of FRBs at much higher
frequencies (mm to far-infrared) would provide very useful test of the
antenna mechanism and other radiation processes.

\section{acknowledgments}
We thank the referee, Jonathan Katz, for his careful
reading of the manuscript and for his numerous excellent comments and
suggestions that substantially improved the content and clarity of the
paper. We thank Don Melrose, George Smoot, Jon Arons, Anatoly
Spitkovsky, Bruce Grossan, Maxim
Lyutikov, Sterl Phinney, Eliot Quatart, Roger Blandford, Jing Luan, Paul Duffell,
Vikram Ravi, Cacey Law, Vishal Gajjar for useful discussions 
and comments. This research was funded by the Named Continuing
Fellowship and David Alan Benfield Memorial Fellowship at the
University of Texas at Austin.

% \label{lastpage}
% \end{document}

\appendix
\section{Two-stream instability}
% In this appendix, we first consider the initial growth of Langmuir
% waves in a homogeneous counter-streaming system and then estimate the
% maximum density fluctuations when the wave growth has saturated.
% \subsection{Growth of Langmuir waves}
We consider counter-streaming electrons and positrons with 1-D
distribution function $f(u)$, where $u=\gamma \beta$ is 
the four velocity ($u>0$ for electrons and $u<0$ for positrons). We
assume $f(u)$ to be symmetric $f(u) = f(-u)$ and normalized
$\int_{-\infty}^{+\infty} f(u)\d u = 1$. The two streams have
identical number densities $n_-=n_+=n/2$, where $n$ is the total number
density in the lab frame. We take $c=1$ and define the
non-relativistic plasma frequency as
\begin{equation}
  \label{eq:20}
    \omega_{\rm p} = (4\pi e^2 n/m_{\rm e})^{1/2}.
\end{equation}
The following derivation is valid from non-relativistic to
ultra-relativistic distribution functions.

\begin{figure*}
  \centering
\includegraphics[width = 0.8 \textwidth,
  height=0.3\textheight]{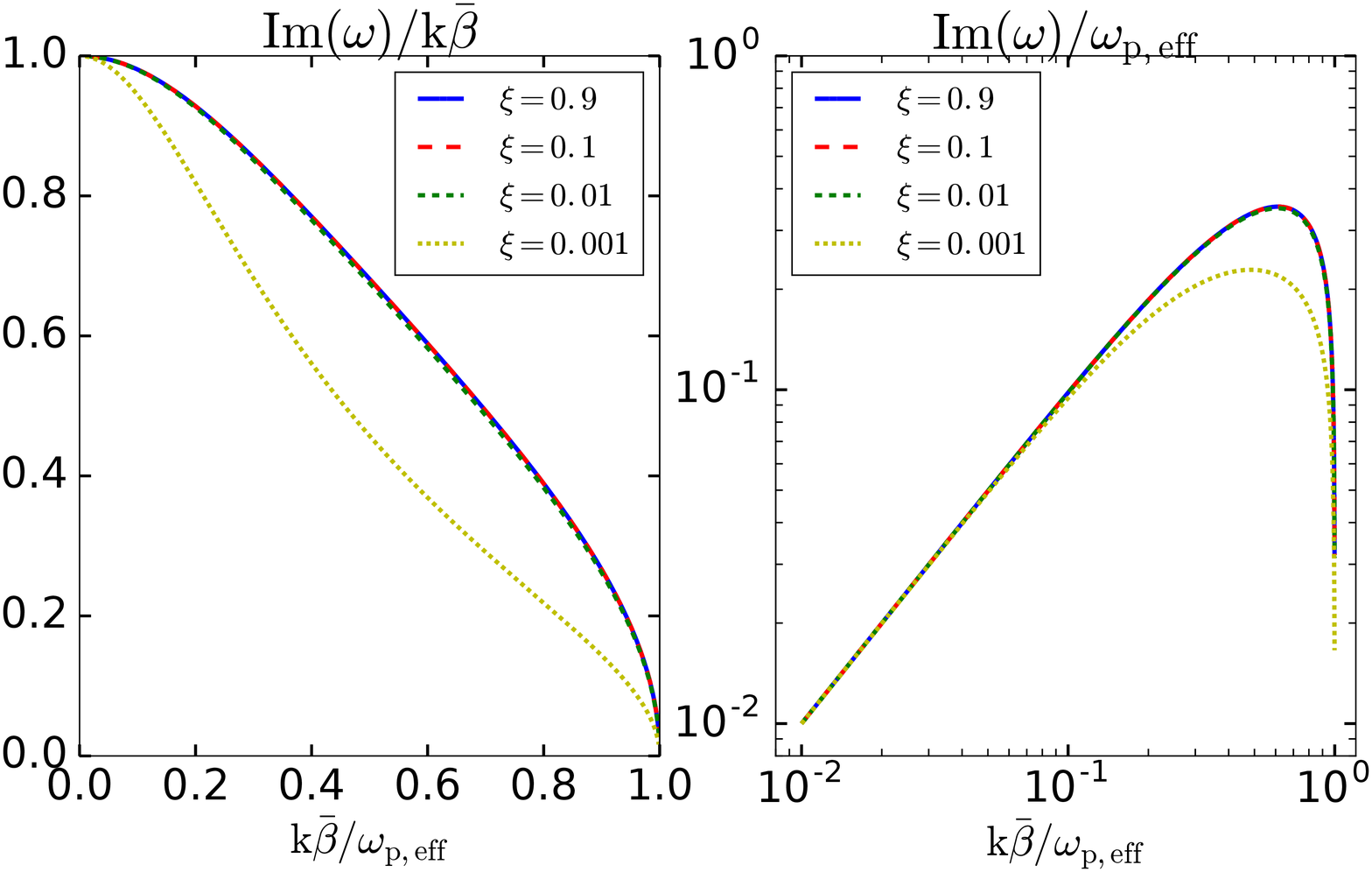}
\caption{The growth rate of two-stream instability for counter-streaming
  $e^\pm$ with ``waterbag'' (flat) distribution function between $\xi u_0$ and
  $u_0$. Here we use $ u_0 = 100\, (\approx \gamma_0)$. The geometric mean 
  speed is defined 
  as $\bar{\beta} = \sqrt{\beta_{\rm min}\beta_{\rm max}}$. The three
  cases with $\xi = 0.9, 0.1, 0.01$ are relativistic in that
  $\bar{\beta}\approx 1$ and the case with $\xi = 0.001$ has a
  fraction ($\sim u_0^{-1}$) of non-relativistic particles and
  $\bar{\beta}\sim 10^{-1/2}$. According
  to eq. (\ref{eq:28}), the maximum unstable wave-numbers for the four
  cases are $k_{\rm max}/(\omega_{\rm p} \gamma_0^{-3/2}) = 1.08$ ($\xi =
  0.9$), $7.4$ ($\xi = 0.1$), $65$ ($\xi = 0.01$), $3.0\times10^2$
  ($\xi = 0.001$). The fact that the curves overlap for the three
  relativistic cases ($\xi = 0.9,\, 0.1,\, 0.01$) do not mean the
  growth rates are the same, because $k_{\rm max}$ (or $\omega_{\rm
    p,eff}$) are different in these cases.
  The $\xi = 0.9$ case is close to mono-energetic
  distribution and the $\xi=0.001$ case is close to
  Maxwell-J{\"u}ttner distribution with a low $u$ cut-off.
}\label{fig:growth_rate}
\end{figure*}

The dispersion relation for longitudinal Langmuir waves of real
wave-number $k$ and complex frequency $\omega$ is given by
\begin{equation}
  \label{eq:21}
  1 = \int_{-\infty}^{+\infty} {\omega_{\rm p}^2 \over \gamma^3}
  {f(u)\over (\omega - \beta k)^2} \d u = \int_{-1}^{+1}
  \omega_{\rm p}^2 {f(\beta)\over (\omega - \beta k)^2} \d \beta,
\end{equation}
which can be solved for any given distribution function $f(u)$.
We plug the complex phase velocity $\omega/k = z + iy$ into
eq. (\ref{eq:21}) and obtain two separate equations for two real
unknowns $z$ and $y$ (from the real and imaginary parts of
eq. \ref{eq:21} respectively)
\begin{equation}
  \label{eq:222}
  \begin{split}
          {k^2\over \omega_{\rm p}^2} &= \int_{-1}^{+1} {(z- \beta)^2 - y^2
    \over [(z+\beta)^2 + y^2]^2} f(\beta)\d \beta, \\
  0 &= \int_{-1}^{+1} {(z- \beta)y
    \over [(z+\beta)^2 + y^2]^2} f(\beta)\d \beta.
  \end{split}
\end{equation}
We are interested in $y> 0$ solutions which correspond to wave
growth. If we make use of the symmetry $f(\beta) = f(-\beta)$, it can
be shown that the second expression above can only be satisfied when
$z = 0$. This means that only (purely imaginary or
electrostatic) standing waves can grow via  
 two-stream instability. Then the first expression in eq. (\ref{eq:222})
becomes
\begin{equation}
  \label{eq:24}
   {k^2\over \omega_{\rm p}^2} = 2\int_0^{1} {\beta^2 - y^2
    \over [\beta^2 + y^2]^2} f(\beta)\d \beta.
\end{equation}
Our goal is to solve this equation for the growth rate
$\mr{Im}(\omega) = yk$ under different distribution functions
$f(\beta)$.

A simple but interesting case is the ``waterbag'' (or step-function)
distribution given by two 
parameters $u_0> 0$ and $0\leq\xi<1$
\begin{equation}
  \label{eq:25}
  f(u) = 
  \begin{cases}
    {1 \over 2(1 - \xi) u_0},\ &\mr{if}\ \xi u_0 < |u|< u_0,\\
    0, \ &\mr{otherwise}.
  \end{cases}
\end{equation}
This distribution has the property that, when $\xi\rightarrow
0$, it resembles the relativistic Maxwell-J{\"u}ttner distribution
of temperature $T\sim u_0$ (in unit of $mc^2/k_{\rm B}$)
\begin{equation}
  \label{eq:1126}
  f_{\rm MJ} = {1\over 2K_1(T^{-1})} \mr{e}^{-\gamma/T},
\end{equation}
where $K_1(z)$ is the modified Bessel function of order
one. On the other hand, when $\xi\rightarrow 1$,
the distribution in eq. (\ref{eq:25}) describes two mono-energetic
beams running against each other. The particle injection in the
twisted magnetosphere of magnetars may be somewhere in the middle of
these two extreme cases ($0<\xi< 1$).

We define the maximum and minimum Lorentz factors as $\gamma_{\rm
  max} \equiv \sqrt{1 + u_0^2}$ and $\gamma_{\rm min}
\equiv \sqrt{1 + \xi^2u_0^2}$ and the corresponding speeds as
$\beta_{\rm max} \equiv  u_0/\gamma_{\rm max}$ and $\beta_{\rm min}
\equiv \xi u_0/\gamma_{\rm 
  min}$. The analytical solution to eqs. (\ref{eq:24}) and (\ref{eq:25}) is
\begin{equation}
  \label{eq:1127}
\left({\mr{Im}(\omega)\over k\beta_{\rm max}}\right)^2 = { \sqrt{ (1 +
    r^2 + a)^2 + 4r(a-r) } - (1 + 
    r^2 + a) \over 2},
\end{equation}
where 
\begin{equation}
  \label{eq:28}
  \begin{split}
      &\omega_{\rm p,eff} \equiv \omega_{\rm p}
  \left[ \beta_{\rm max} - \beta_{\rm min} \over (1 - \xi) u_0
\right]^{1/2} =\omega_{\rm p} \langle \gamma^{-3}\rangle^{1/2},\\
&\langle \ldots\rangle = \int_{-\infty}^{+\infty} (\ldots)f(u)\d u,\ r \equiv
{\beta_{\rm min}\over\beta_{\rm max}},\ a \equiv {\omega_{\rm p,eff}^2 
    \over k^2 \beta_{\rm max}^2},\\
&k<k_{\rm max}\equiv
{\omega_{\rm p,eff}\over \bar{\beta}},\ 
\bar{\beta} = (\beta_{\rm min}\beta_{\rm max})^{1/2}.
  \end{split}
\end{equation}
Unstable Langmuir waves are possible only when $a>r$, which is
equivalent to $k<k_{\rm max}$. As can be seen in
Fig. \ref{fig:growth_rate}, the growth rate $\mr{Im}(\omega)$ at very
small wave-number $k\ll k_{\rm max}$ equals to $k\bar{\beta}$; the
growth rate near the maximum wave-number $k\sim k_{\rm max}$ roughly
equals to $\omega_{\rm p,eff}/\bar{\beta}$. In the limiting case of
$\xi\rightarrow0$ (or Maxwell-J{\"u}ttner distribution), we have
$\bar{\beta} = 0$ ($k_{\rm max}\rightarrow 0$), so the growth rate at
any wave-number $k$ is zero, which is expected because the system is
already in equilibrium.

The effective plasma frequency is $\omega_{\rm p,eff} = \omega_{\rm p}
\langle \gamma^{-3}\rangle^{1/2}$ and the effective skin depth is
$\ell_{\rm skin,eff} \equiv k_{\rm max}^{-1} = \bar{\beta}/\omega_{\rm p,eff}$.
For mono-energetic ultra-relativistic case ($\xi \approx 1$ and
$\gamma_0\approx u_0\gg 1$), we have $\omega_{\rm p,eff} \approx
\omega_{\rm p} \gamma_0^{-3/2}$. For a broad but ultra-relativistic
distribution ($\xi\ll 1$ and $\gamma_{\rm min} \approx \xi u_0 \gg
1$), we have $\omega_{\rm p,eff} \approx \omega_{\rm p}(2\gamma_{\rm
  min}^2\gamma_0)^{-1/2} = \omega_{\rm p}\gamma_0^{-3/2}/(\sqrt{2}\,
\xi)$, which is significantly greater (by a factor of $\sim\xi^{-1}$)
than that in the mono-energetic case.

% Consider a system which starts with mono-energetic counter-streaming
% electrons and positrons. Initially, only wave-modes with $k\lesssim
% \omega_{\rm p}\gamma_0^{-3/2}$ can grow (the growth of waves with
% $\omega_{\rm p}\gamma_0^{-3/2}< k \lesssim \omega_{\rm p}
% \gamma_0^{-1/2}$ is prohibited). The modes with $k\sim \omega_{\rm
%   p}\gamma_0^{-3/2}$ grow the fastest. As the
% distribution broadens (and $\xi$ decreases), wave
% modes with higher and higher $k$ begin to grow. After some time ($\gtrsim 10
% \gamma_0^{3/2} \omega_{\rm p}^{-1}$), the distribution becomes very
% broad $|u|\in (\sim 1, \sim u_0)$, then all modes with $k \lesssim
% \omega_{\rm p} \gamma_0^{-1/2}$ start growing. After a sufficiently long time
% (the relaxation time), the distribution
% function is expected to become relativistic Maxwell-J{\"u}ttner. 

\section{Dynamics of external shocks}

% \note{[We note that for a NS
%   progenitor, the CSM is ultra-relativistic pulsar wind and
%   the outflow (launched along the open field lines) propagates like
%   in vacuum until it runs into the wind termination shock at large
%   radius $\gtrsim 10^{16}\,cm$. The timescale is long unless the
%   outflow has a very large Lorentz factor.]}

When a relativistic outflow of isotropic equivalent energy $E_{\rm iso}$ and
Lorentz factor $\Gamma$ ploughs its way through the
cold circumstellar medium (CSM), two shocks  
form in this process, the forward shock going into the CSM and the
reverse shock going into the ejecta. The bolometric emission from the
outflow peaks roughly when the ejecta reaches the 
deceleration radius $r_{\rm dec}$, which is given by
\begin{equation}
  \label{eq:30}
  E_{\rm iso} \sim \Gamma^2 m_p c^2\int^{r_{\rm dec}} 4\pi r^2
  n_{\rm CSM}(r)\d r.
\end{equation}
In the observer's frame, the typical emission timescale is the
deceleration time $t_{\rm dec} \simeq r_{\rm 
  dec}/(2\Gamma^2c)$. 

In the following, we assume a
power-law density profile $n_{\rm CSM}\propto r^{-k}$ (our discussion
can be generalized to an arbitrary density profile). After the 
deceleration time, the Lorentz factor of the forward shock decreases
with radius as a power-law 
$\Gamma(r/r_{\rm dec})^{(k-3)/2}$ \citep{1976PhFl...19.1130B} before it
decelerates to Newtonian speeds. The emission from the forward-shocked
region decreases
with time as a power-law. For some FRBs without significant scattering
broadening, we usually see a sharp cuf-off instead of a power-law at
the end of the burst. On the other hand, the emission from the reverse
shock may have a sharp cut-off due to adiabatic sideway expansion after the
reverse shock crosses the ejecta. 

Now we consider a second outflow with
a similar initial Lorentz factor $\sim\Gamma$ but launched with a delay of
$t_{\rm dec}\ll t_{\rm delay}\ll r_{\rm dec}/c$ wrt the first
outflow. We note that the time intervals between
some burst pairs from FRB 121102 can be as short at $\sim30\,$ms to $40\,$ms
\citep{2017ApJ...846...80S, 2017MNRAS.472.2800H}, so the CSM does not
have time to recover to its original undisturbed state 
after the first outflow passed by. When the second outflow
catches up with the first one, if the first one is still
ultra-relativistic, then the catch-up radius $r_{\rm c}$ is can be
estimated by
\begin{equation}
  \label{eq:7}
  2ct_{\rm delay}\Gamma^2(r_{\rm c}/r_{\rm dec})^{k-3} \simeq r_{\rm c} -
  ct_{\rm delay}\simeq r_{\rm c},
\end{equation}
i.e.
\begin{equation}
  \label{eq:20}
  r_{\rm c}\simeq r_{\rm dec} (t_{\rm delay}/t_{\rm dec})^{1/(4-k)},
\end{equation}
This means that the second FRB will have a longer duration than the first
one by a factor of $(t_{\rm delay}/t_{\rm dec})^{1/(4-k)}$, where $t_{\rm dec}$ is
the duration of the first FRB. If the first
outflow has decelerated to a Newtonian speed when it
is caught up by the second one, the duration of the second FRB is
determined by the radius where the first outflow becomes Newtonian
$r_{\rm N}\, (\gg r_{\rm dec})$ divided by $2\Gamma^2c$. Still, the duration of the second
FRB should be much longer than the first one. This is inconsistent
with observations. For instance, the two bursts detected by
\citet{2017ApJ...846...80S}, ``GBT-1'' and ``GBT-2'', were separated by $\sim
40\,$ms and they have very similar durations of $\sim 2\,$ms.

\section{Induced Compton Scattering}
We consider a spatially uniform electron-radiation mixture. The
electron distribution function is isotropic and mono-energetic with
Lorentz factor $\gamma$ and number density
$n_{\rm e}$ (a distribution of Lorentz factors will be considered
later). The radiation field is also isotropic with intensity 
$I_{\nu}$ only a function of frequency $\nu$. In terms of the
distribution function of the radiation field $f(\b{r}, \b{p}, t)$, the
photon occupation number $\t{f}$ is defined as the number of photons
in a phase-space 
volume $h^3$ ($h$ being the Planck constant) and we have
\begin{equation}
  \label{eq:2}
  \t{f} \equiv h^3f(\b{r}, \b{p}, t) = {I_{\nu} c^2 \over 2 h\nu^3}.
\end{equation}
We are interested in the time evolution of the photon occupation number
$\t{f}_0\equiv \t{f}(\nu_0, \b{\Omega}_0)$ at a given frequency $\nu_0$
along a given direction $\b{\Omega}_0$ due to induced (or stimulated)
Compton scattering in the regime $\t{f}\gg 1$.

\begin{figure}
  \centering
\includegraphics[width = 0.47 \textwidth,
  height=0.2\textheight]{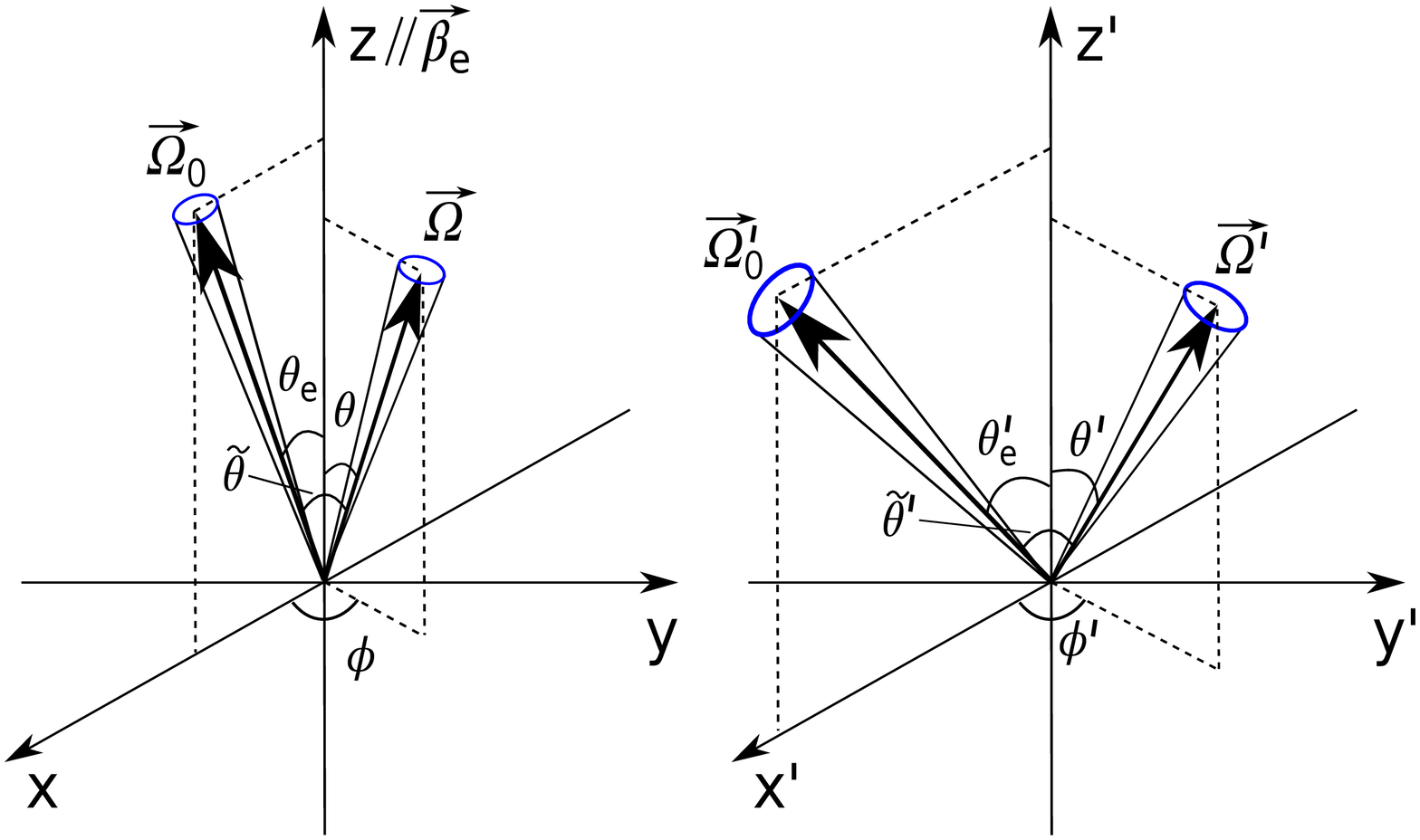}
\caption{Geometry for induced Compton scattering. In the lab frame
  (left panel), we
  consider the time evolution of the intensity in the given direction
  $\b{\Omega}_0$ due to induced Compton scattering from and into other
  directions $\b{\Omega}(\theta,\phi)$. The polar coordinate system
  has polar axis aligned with electrons' velocity vector
  $\b{\beta}_{\rm e}$. We put $\b{\Omega}_0$ in the x-z plane. The
  angle between $\b{\Omega}_0$ and 
 $\b{\beta}_{\rm e}$ is denoted as
  $\theta_{\rm e}$ and the angle between $\b{\Omega}_0$ and $\b{\Omega}$
  is $\t{\theta}$. The calculation is done in the comoving frame of
  electrons (right panel), where all quantities are denoted with a
  prime ($'$).
}\label{fig:compton}
\end{figure}

We consider a subgroup of electrons moving in the $\hz$ direction
within an infinitesimal solid angle $\d\Omega_{\rm e}$. The angle between
$\b{\Omega}_0$ and $\hz$ is denoted as $\theta_{\rm e}$. The 
number density of these electrons in their comoving frame is $\d n_{\rm e}' =
n_{\rm e}\d \Omega_{\rm e}/(4\pi \gamma)$. Hereafter, all quantities in the
electrons' comoving frame are denoted with a prime ($'$) and the
unprimed quantities are measured in the lab frame.

As shown in the right panel (comoving frame) of
Fig. \ref{fig:compton}, photons originally moving near the
$\b{\Omega}_0'$ direction can be scattered to  an arbitrary 
direction $\b{\Omega}'$ and there will also be photons scattered from
$\b{\Omega}'$ back to $\b{\Omega}_0'$. The direction $\b{\Omega}'$ is
given by two angles ($\theta',\phi'$) in a polar coordinate system
with polar axis $\b{z}$ aligned with electrons' velocity vector
$\b{\beta}_{\rm e}$ in the lab frame. In this coordinate system, the
direction $\b{\Omega}_0'$ is taken to be in the $x'$-$z'$ plane, given
by ($\theta' = \theta_{\rm e}',\phi'=0$). The angle between
$\b{\Omega}'$ and $\b{\Omega}_0'$ 
is denoted as $\t{\theta}'$ (and $\t{\mu}'\equiv \cos\t{\theta}'$).

For convenience, we define two Doppler factors
\begin{equation}
  \label{eq:2228}
  \begin{split}
      &\mc{D} = \gamma(1 + \beta \mu'),\ \mc{D}_{\rm e} = \gamma(1 +
  \beta\mu_{\rm e}'),\\
&\mu' \equiv \cos\theta',\ \mu_{\rm e}' \equiv
  \cos\theta_{\rm e}',
  \end{split}
\end{equation}
and then the Lorentz transformation of frequencies and angles are
given by 
\begin{equation}
  \label{eq:29}
  \begin{split}
     & \nu' = \nu/\mc{D},\ \nu_0' = \nu_0/\mc{D}_{\rm e},\ \phi' =
     \phi,\\
     & \mu' = {\mu - \beta \over 1 -
    \beta\mu},\
\mu'_{\rm e} =  {\mue - \beta \over 1 -
    \beta\mue},\ \d \mue' = \mc{D}_{\rm e}^2\d \mue.
  \end{split}
\end{equation}
The photon occupation number $\t{f}$ is Lorentz
invariant. The rate of change in $\t{f}_0 \equiv \t{f}(\nu_0',
\b{\Omega}_0')=\t{f}(\nu_0,\b{\Omega}_0)$ in the comoving frame is given
by \citep{1982MNRAS.200..881W}
\begin{equation}
  \label{eq:35}
  {\d (\mr{ln}\t{f}_0) \over \d t'} = {3hc \sigma_{\rm T}\over
    8\pi\me c^2} \d n_{\rm e}' \int \d \Omega' (1 +\t{\mu}'^2)(1 -
  \t{\mu}') \left. {\partial (\nu'^2 \t{f}) \over
    \partial \nu'}\right|_{\nu_0'},
\end{equation}
where $\sigma_{\rm T}$ is the Thomson cross-section. 

From $\d\nu' = \d \nu/\mc{D}(\mu')$ (for a given direction $\b{\Omega}'$),
we obtain
\begin{equation}
  \label{eq:110}
  \begin{split}
      \left. {\partial [\nu'^2 \t{f}(\nu', \b{\Omega}')] \over
    \partial \nu'}\right|_{\nu_0'} &= {1\over \mc{D}} 
\left. {\partial [\nu^2 \t{f}(\nu)] \over
    \partial \nu}\right|_{\nu=\nu_0'\mc{D} = \nu_0\mc{D}/\mc{D}_{\rm 
  e}}\\
&= {c^2 \over 2h\mc{D}}  \left. {\partial (I_{\nu}/\nu) \over
    \partial \nu}\right|_{\nu= \nu_0\mc{D}/\mc{D}_{\rm
  e}}.
  \end{split}
\end{equation}
To compute the time evolution of $\t{f}_0$ due to the scattering of
all electrons, we need to integrate eq. (\ref{eq:35}) over the
electrons' angle distribution
\begin{equation}
  \label{eq:91}
  \int \d n_{\rm e}'(\ldots) = {n_{\rm e}\over 4\pi \gamma} \int \d\Omega_{\rm
    e} (\ldots) = {n_{\rm e}\over 2 \gamma} \int_{-1}^{1} \d\mu_{\rm
    e}' \mc{D}_{\rm e}^{-2} (\ldots),
\end{equation}
where we have made use of the fact that $(\ldots)$ has no $\phi_{\rm
e}$ dependence (due to the symmetry of the system) and $\d\mue = \d
\mue'/\mc{D}_{\rm e}^2$. Then, we use $\d t' = \d t/\gamma$ and write out
the full integral explicitly
\begin{equation}
  \label{eq:93}
  \begin{split}
     & {\d (\mr{ln}\t{f}_0) \over \d t}  = {c^2 \over 2\me c^2}
     {3cn_{\rm e} \sigma_{\rm T} \over 16\pi \gamma^2}
\int_{-1}^{1}\d \mue' \mc{D}_{\rm e}^{-2} \int_{-1}^{1}\d \mu' \mc{D}^{-1}\\
&{\hspace{0.3cm}}\cdot 
\int_0^{2\pi}\d \phi' (1 + \t{\mu}'^2)(1 - \t{\mu}')
\left. {\partial (I_{\nu}/\nu) \over
    \partial \nu}\right|_{\nu= \nu_0\mc{D}/\mc{D}_{\rm
  e}},
  \end{split}
\end{equation}
where
\begin{equation}
  \label{eq:98}
  \t{\mu}' = \sin\theta_{\rm e}'\sin\theta' \cos\phi' + \mue'\mu'.
\end{equation}
In the limit where electrons are at rest ($\beta=0, \gamma=1,
\mc{D}=\mc{D}_{\rm e}=1$), we have $\int\d\mu'\int \d \phi' (1 +
\t{\mu}'^2)(1 - \t{\mu}')=16\pi/3$ and the classical result is
recovered
\begin{equation}
  \label{eq:111}
  {\d (\mr{ln}\t{f}_0) \over \d t}  = {c^2 \over 2\me c^2}
     {cn_{\rm e} \sigma_{\rm T}} \left. {\partial (I_{\nu}/\nu) \over
    \partial \nu}\right|_{\nu_0}.
\end{equation}
In the ultra-relativistic regime $\gamma\gg 1$, the factor $\int\d \phi'(1 +
\t{\mu}'^2)$ roughly gives $8\pi/3$ (with error of order unity) and
is hence not important in our order-of-magnitude estimate. In the
following, we also drop the 
term $\sin\theta_{\rm e}'\sin\theta' \cos\phi'$ in $\t{\mu}'$ because
it is an odd function of $\cos\phi'$ and $\int \d
\phi'\cos\phi' = 0$.

If the system has length-scale $\ell$, the Thomson optical
depth is $\tau_{\rm T} = \ell \sigma_{\rm T}n_{\rm e}$. After a
light-crossing time $\ell/c$, the photon occupation number changes by
a factor of $\mr{e}^{\tau_{\rm eff}}$, where the effective optical
depth is roughly given by
\begin{equation}
  \label{eq:106}
  \begin{split}
      \tau_{\rm eff} \sim {c^2 \tau_{\rm T}\over 4\gamma^2\me c^2}
      \int_{-1}^{1} {\d \mue'\over \mc{D}_{\rm e}^{2}} \int_{-1}^{1}
      {\d \mu' (1 - \mue'\mu') \over \mc{D}}
\left. {\partial (I_{\nu}/\nu) \over
    \partial \nu}\right|_{{\mc{D}\nu_0\over \mc{D}_{\rm
  e}}},
  \end{split}
\end{equation}
We note that, due to induced Compton scattering,
the radiation field $I_{\nu}(\nu_0,\b{\Omega}_0)$ is in principle
coupled with the derivative of $I_{\nu}/\nu$ in all other directions
and at a wide range of frequencies $\mc{D}\nu_0/\mc{D}_{\rm e}\in
(\gamma^{-1}\nu_0, \gamma \nu_0)$. Thus, the effective optical depth
is sensitive not only to $\gamma$ but also to the spectral broadness.

In the following, we discuss two extreme
cases: (i) a narrow step-function or Gaussian spectrum with
$\Delta\nu/\nu_0\sim 1$ and (ii) a broad power-law spectrum in the
range  $\nu/\nu_0\in (\gamma^{-1},\gamma)$.

In case (i), the integral in eq. (\ref{eq:106}) is non-zero only when
$\mc{D}\sim \mc{D}_{\rm e}$. In the range $\mue'\in (-1, -1
+\gamma^{-2})$, we can estimate $\mc{D}\sim \mc{D}_{\rm e}\sim
\gamma^{-1}$, $\Delta \mu'\sim \gamma^{-2}$, $1 - \mue'\mu'\sim
\gamma^{-2}$, and hence the contribution to the integral is roughly
$\sim \gamma^{-3} I_{\nu_0}/\nu_0^2$. In the range where $\mue'$ is
far from $-1$ ($\Delta\mue'\sim 1$), we can estimate $\mc{D}\sim
\mc{D}_{\rm e}\sim \gamma$, $\Delta \mu'\sim 1$, $1 -
\mue'\mu'\sim 1$, and hence the contribution to the integral is
$\sim \gamma^{-3} I_{\nu_0}/\nu_0^2$. We define the brightness
temperature $T_{\rm B}$ at $\nu_0$ as 
\begin{equation}
  \label{eq:170}
  I_{\nu_0} = {2\nu_0^2 k_{\rm B}T_{\rm B} \over c^2}\
  \Longrightarrow\ \
  k_{\rm B}T_{\rm B} = {c^2I_{\nu_0}\over 2\nu_0^2}.
\end{equation}
Then the effective optical depth is roughly
\begin{equation}
  \label{eq:171}
  \tau_{\rm eff} \sim {k_{\rm B}T_{\rm B} \over \me c^2} {\tau_{\rm
      T}\over \gamma^5}.
\end{equation}

In case (ii), we assume that the spectrum is a single power-law
$I_{\nu}\propto \nu^{p}$ ($p\neq 1$) with normalization given by the
brightness temperature at $\nu_0$ as in eq. (\ref{eq:170}). Then we
have
\begin{equation}
  \label{eq:113}
\left. {\partial (I_{\nu}/\nu) \over
    \partial \nu}\right|_{\mc{D}\nu_0/\mc{D}_{\rm
  e}} = (p-1) {2k_{\rm B}T_{\rm b}(\nu_0) \over c^2}
\left({\mc{D} \over \mc{D}_{\rm e} 
}\right)^{p-2},
\end{equation}
and the effective optical depth becomes
\begin{equation}
  \label{eq:120}
  \begin{split}
            &\tau_{\rm eff} \sim {(p-1)\over 2} {k_{\rm B}T_{\rm b}(\nu_0)\over
        \me c^2} {\tau_{\rm T}\over \gamma^2}\cdot Q,\\
    &Q\equiv \int_{-1}^{1}\d \mue' [\mc{D}_{\rm e}(\mue')]^{-p}
    \int_{-1}^{1}\d \mu' (1 - \mue'\mu') [\mc{D}(\mu')]^{p-3}.
  \end{split}
\end{equation}
When $p < 1$, we have $\tau_{\rm eff}<0$ and $\t{f}_0$ decreases
with time exponentially; when $p > 1$, we have $\tau_{\rm eff}>0$ and
$\t{f}_0$ increases with time exponentially. 
We integrate eq. (\ref{eq:120}) analytically and obtain
\begin{equation}
  \label{eq:109}
Q\sim 
  \begin{cases}
  \gamma^{1-2p},\ &\mr{if}\ p < 1.5,\\
  \gamma^{2p-5},\ &\mr{if}\ p > 1.5.
\end{cases}
\end{equation}
For $p = 1.5$ (a rising spectrum), the integral reaches the minimum
$Q_{\rm min}\sim \gamma^{-2}$, which means $|\tau_{\rm eff}| \sim
\gamma^{-4}\tau_{\rm T} 
(k_{\rm B}T_{\rm b}/ \me c^2)$. For a flat spectrum $I_{\nu}\propto
\nu^0$ (or $p = 0$), we have $Q\sim \gamma$ and hence $|\tau_{\rm
  eff}| \sim \gamma^{-1}\tau_{\rm T} (k_{\rm B}T_{\rm b}/ \me c^2)$.
Above the peak of the spectral energy
distribution $p < -1$, we have $Q\gtrsim \gamma^3$ and
$|\tau_{\rm eff}| \gtrsim \gamma\tau_{\rm T} (k_{\rm B}T_{\rm b}/ \me
c^2)$, which means the photon occupation number (and hence flux)
$\t{f}_0$ drops even more quickly than in the flat-spectrum case. The net
effect is that the peak of the spectrum moves towards lower and lower
frequencies and the radiation energy is transferred to the kinetic
energy of electrons.

Observationally, the spectrum of FRBs may or may not be broad. We note
that co-detections at multiple telescopes operating at different
frequencies (e.g. 1.4 and 3 GHz) have been reported by
\citet{2017ApJ...850...76L} and \citet{2017ApJ...846...80S}. In this
paper (eq. \ref{eq:64}), we take the most conservative limit $Q\sim
\gamma^{-3}$ (for a narrow spectrum with $\Delta \nu/\nu_0\sim 1$) and
hence $|\tau_{\rm eff}| 
\gtrsim \gamma^{-5}\tau_{\rm T} (k_{\rm B}T_{\rm b}/ \me c^2)$. For
any spectra broader than $\Delta \nu/\nu_0\sim 1$, the effective
optical depth $|\tau_{\rm eff}|$ will be larger and hence induced
Compton scattering will be more efficient.

Finally, we integrate eq. (\ref{eq:120}) over the (normalized) Lorentz factor
distribution of electrons $\t{N}_\gamma\equiv \d \t{N}/\d \gamma$
($\int_1^{\infty} \t{N}_{\gamma}\d \gamma = 1$) and obtain the total
effective optical depth
\begin{equation}
  \label{eq:180}
  \tau_{\rm eff,tot}\gtrsim {k_{\rm B} T_{\rm b}(\nu_0) \over \me c^2}
  \tau_{\rm T} \int_1^{\infty}\gamma^{-5} \t{N}_{\gamma} 
   \d \gamma.
\end{equation}
We see that the contribution from high-energy electrons are strongly
suppressed by the $\gamma^{-5}$ factor. In realistic dissipations
caused by shocks or magnetic reconnection, the distribution function
is usually a power-law 
$\t{N}_{\gamma}\propto \gamma^{-q}$ ($q>1$) above the peak Lorentz
factor, which is also roughly the mean Lorentz factor
$\bar{\gamma}$, but the part below the peak 
Lorentz factor may be uncertain. In the case of
a Maxwell-J{\"u}ttner distribution, we have $\t{N}_{\gamma}\propto
\gamma^2$ and hence most of the contribution comes from electrons near
the lowest Lorentz factors $\sim 1$ and $\int
\gamma^{-5}\t{N}_{\gamma}\d \gamma \sim \bar{\gamma}^{-3}$. On the
other hand, in the case of an infinitely sharp cut-off below the peak
Lorentz factor, we have $\int \gamma^{-5}\t{N}_{\gamma}\d \gamma \sim
\bar{\gamma}^{-5}$. In this paper, we use the result in the
latter (most conservative) case.

\label{lastpage}
\end{document}